\documentclass[12pt, apj, twocolumn]{openjournal}

\usepackage{lipsum}

\usepackage{xcolor}
\usepackage{textgreek}
\usepackage[utf8]{inputenc}
\usepackage[english]{babel}
\usepackage{parskip}
\usepackage{hyperref}
\hypersetup{
    colorlinks=true,
    linkcolor=blue,
    filecolor=blue,      
    urlcolor=blue,
    citecolor=blue,
    urlcolor=blue,
}
\usepackage{color}
\definecolor{linkcolor}{rgb}{0.0,0.3,0.5}
\usepackage{tensind}
\tensordelimiter{?}
\DeclareGraphicsExtensions{.bmp,.png,.jpg,.pdf}
\usepackage{verbatim}
\usepackage[normalem]{ulem}
\usepackage{orcidlink}
\usepackage{soul}
\usepackage{graphicx}
\usepackage{subcaption}
\usepackage{float}
\usepackage{booktabs}
\newcommand{\gt}{\ensuremath{>}}

\graphicspath{ {./figs/} }

\begin{document}
\title{Astrophysics Wrapped 2025: Year-in-Review of Every Astrophysics arXiv Paper from 2025}
\author{\vspace{-1.3cm}
Rommulus Francis Lewis\,\orcidlink{0009-0004-2970-6805}$^{1,2, \star}$,
Hetansh Shah\,\orcidlink{0009-0007-8431-1322}$^{3}$,
Amruth Alfred\,\orcidlink{0000-0003-1276-1248}$^{1,2}$,
and The Astrophysics Community{$^\dagger$}}
\thanks{$^{\star}$E-mail: \url{rommulus.astrophysics@gmail.com}}

\affiliation{
\vspace{0.1cm}
$^{1}$Department of Physics, University of Hong Kong, Hong Kong, Hong Kong SAR\\
\vspace{0.1cm}
$^{2}$Hong Kong Institute for Astronomy and Astrophysics, University of Hong Kong, Hong Kong, Hong Kong SAR\\
\vspace{0.1cm}
$^{3}$Department of Computer Science, University of Massachusetts, 300 Massachusetts Ave, Amherst, MA 01003, United States\\
\vspace{0.1cm}
$^{\dagger}$ Everyone who reached out to inform us about improvements and corrections, Worldwide\\
}

\begin{abstract}
Astrophysics has experienced an overwhelming increase in research output, as is evident from the year-over-year increase in the number of research papers submitted to the online repository arXiv. As a result, keeping up with progress happening outside our respective sub-fields can be exhausting. While it is impossible to be informed on every single aspect of every sub-field, this paper aims to be the next best thing. We present a summary of statistics for every paper uploaded onto the Astrophysics arXiv over the past year - 2025. We analyse a host of metrics like the most used keywords, subfields and telescopes, the distribution of journals, the most studied astrophysical objects like GW, GRB, FRB events, exoplanets and much more. We also indexed the authors’ affiliations to put into context the global distribution of research and collaboration. Combining this data with the citation information of each paper allows us to understand how influential different papers have been on the progress of the field this year. We also present a first of its kind Astrophysical Spectral Fingerprint showing the distribution of research across the electromagnetic spectrum as well as the distribution of research by redshift. Overall, these statistics highlight the general current state of the field, the hot topics people are working on and the different research communities across the globe and how they function. We hope that this is helpful for both students and professionals alike to adapt their current trajectories to better benefit the field.
\end{abstract}


\maketitle

\section{Introduction}
\label{sec:intro}

Astronomy is widely regarded as the oldest science and has played a central role in the development of human civilisation. From Galileo's use of the telescope to challenge the geocentric model of our Solar System, to Newton's proposal of the laws of motion and the invention of calculus to explain the orbits of planets, and more recently, to provide the first evidence for Einstein's theory of general relativity through observations of the solar eclipse in 1919. These are not just discoveries about our Universe, but crucial pillars on which modern technology is rooted. In the present era, humanity has launched unbelievably advanced devices like the Hubble Space Telescope and the more recent James Webb Space Telescope, among many others. The increasing influx of data has enabled us to disprove, refine and discover fundamental theories that explain the workings of our Universe, typically compiled in the form of research articles. While in the near past, one had to find (if lucky!) and read physical prints of research papers in libraries, the explosion of the internet and computers has resulted in vast online repositories. Although several of these exist for astrophysics research, arXiv\footnote{\url{https://info.arxiv.org/about/index.html}} is one of the most widely used platforms for researchers to share their work as preprints before publication, at no cost. The ease of use and popularity of this system means that almost every astrophysics paper, irrespective of discipline, at some point, ends up on the arXiv.

In this so-called ``Golden Age of Astronomy", given the sheer scale of how much progress we make in a day, evident from the number of papers submitted every day, it is easy to lose sight of what each of us is doing in our respective sub-fields. Today, it is next to impossible, if not impossible, to read all the papers uploaded on any particular day. Sometimes, keeping up with our own sub-fields can be a challenge. While we might not all agree on which papers on the arXiv are ``good papers", we personally believe that all these papers tell us something useful.

Here, we present the first-ever analysis, to our knowledge, of all papers released on the Astrophysics arXiv in the past year 2025. The most similar work to this is that of Virginia Trimble and collaborators who carried out an incredible annual summarization of all the major events and research for every year in astrophysics from 1996 to 2006 \citep{trimble2006, trimble2005, trimble2004, trimble2003, trimble2002, trimble2001, trimble2000, trimble1999, trimble1998, trimble1997, trimble1996}. The difficulty in doing something like \cite{trimble2006} today lies in how vast and inhomogeneous the dataset is. Since 2006, the total annual number of papers released on the arXiv has more than doubled. In addition, research is now much more diverse and covers highly specialised aspects of the field known to a select few, making a summary similar to \cite{trimble2006} require a deep understanding of every field. However, our work addresses this by capturing both broad and niche areas of the field without requiring in-depth knowledge of every topic. This allows all readers, whether within or outside the field, to gauge the general state of astrophysics in 2025 without encountering excessive technical jargon.

We would like to stress that this paper is not meant to incite unhealthy competition among the community, nor is it meant to be a cheat-sheet to maximising metrics. Rather, it is an appreciation or `pat on the back' for the incredible efforts we have collectively put in to try and make sense of the Universe. We hope that this work reveals the many interesting statistics we would all like to know and helps us gauge what we as a community should be working towards in the long run.

In the following, we present our results across a broad range of metrics calculated for all new papers released on each day of 2025. With this data, we are able to show metrics in relation to paper topics, keywords and citations. From each paper's title and abstract, we can track what the community has focused on and where we are headed. Furthermore, we collected the affiliations of all authors and show geographical distributions of the aforementioned metrics. Using the author affiliations, we can quantify the extent of collaboration across groups and subfields. We will demonstrate that the patterns we uncover are valuable in informing crucial decisions regarding the future of astrophysics on the planet.

In section \ref{methods} we will elaborate on the methodology and the extensive data collection and extraction involved in this work, followed by a section \ref{sec:results} where we present all the statistics we collected as a plethora of plots and tables. Section \ref{sec:honors} highlights some unconventional achievements and finally \ref{sec:conclusions} provides an executive summary. \\

\section{Methods} \label{methods}
In this section, we outline how we collected our data and the various metrics we utilise in the rest of this paper. We also define some useful indices which we will reference later on in our analysis.

\subsection{Data} \label{subsec:data_collection}
The data used in this analysis consists of all 18660 papers uploaded to the arXiv in 2025 under Astrophysics. This does not include replacement submissions, as replacement submissions usually only involve changes in paper content, which we do not collect. The dataset also does not include cross-submissions, so as to confine the dataset to only pure Astrophysics papers. Each of these papers is assigned a Primary Subject by the author out of a list of seven categories - Astrophysics of Galaxies, High Energy Astrophysical Phenomena, Solar and Stellar Physics, Cosmology and Nongalactic Astrophysics, Earth and Planetary Astrophysics, Instrumentation and Methods for Astrophysics and a final general category, Astrophysics. A submission also includes comments where authors typically mention the number of pages, tables and figures as well as publication information. This metadata, along with the title and abstract of each paper, was obtained by scraping the new submissions of the arXiv everyday over the past year, a practice which also enabled us to track the exact date a paper was listed. We also utilise a series of other metrics which are not inherently reported on the arXiv owing to which we rely on other sources.

Citation information was collected using the NASA Astrophysics Data System (NASA ADS\footnote{\url{https://ui.adsabs.harvard.edu}}), which tracks both the total number of citations as well as the number of citations excluding self-citations. While these metrics are interesting in themselves, any statistics calculated from them are sensitive to the fact that not all papers on the arXiv are published in a journal. This is why we develop our own citation indices to remove this bias and highlight citation metrics specific to those papers published in journals. We define the citation indices as follows:

\begin{enumerate}
    \item All Articles Citation (AAC): Under a certain category, the AAC is defined as the average number of citations for all articles under that category, including articles that have not been published in a journal. This number includes self-citations.
    \item Journal Articles Citation (JAC): The JAC of a category is the average number of citations per paper, for papers under that category that are published in a journal. We do not count arXiv e-prints as journal papers, and hence citations to arXiv papers are not included in the JAC. 
    \item Excluding Self All Articles Citation (EAAC): The same as the ACC, but only considering citation count excluding self-citations. 
    \item Excluding Self Journal Articles Citation (EJAC): Same as the JAC, but only considering citation count excluding self-citations. 
\end{enumerate}

Although our analysis mainly focuses on AAC, we have included extended statistics for all indices in Supplementary Information.

The calculation of citation indices requires us to flag whether a paper has been published in a journal. However, considering how involved the act of publishing is with the field we also thought it was necessary to track which journal each paper was published in, information that we use in our analysis of journals and journal costs. To determine which journal a paper was published in, we once again turned to NASA ADS, which filled in journal information for a substantial number of papers. We note that this process was done in the third week of December (as late as possible) to allow sufficient time for papers to go through the publication process. However, it is not all encompassing and misses out on papers that have been accepted for publication but are not yet published. Thus, we turn to the arXiv comments left by authors, which more often than not confirm a publication in the form of phrases like `accepted for publication in \textit{insert journal}'. We search for the word `published' and its other forms, and if found, then look for a journal name. Terms like `submitted' disqualify the paper from being counted as a journal paper. We also use the comments to our advantage to fill in the number of pages, tables and figures where available, as a secondary method to an automatic script which otherwise counts these values by looking through the PDF of the paper. As a final means to ensure completeness, we also include journal names from the journal reference arXiv metadata, which is filled in for a very small number of papers.

Geographic analysis is a big part of this paper, so we ensured to capture the geographic information of as many papers as possible. This was two-fold in that we separated the general country or region and the specific institution and used the two independently in our analysis. The maps and general country or region related statistics have been constructed by searching for country or region names in the latex files (available for download from the arxiv) of each paper from a list of 245 countries and regions all over the world. For institutions on the other hand, we search for more specific words like `University', `Centre' and their Spanish, French and German translations, arranged according to a priority, with `University' having the highest priority. This was done to ensure that phrases like `Department of Physics' or `Institute of Science', which do not include sufficient specificity for normalisation, are not caught. If there is no `University' mention then we go to the subsequent lower priority words. Not all papers have parse-able LaTeX, and not all affiliations list their country or region at the very end. For papers were the script fails at finding an affiliation or region, we again turn to the NASA ADS and retrieve the affiliations for the authors on that paper. Since authors rarely fill in or update their affiliations regularly on the ADS, this is used as a last resort to fill in whatever is missing, with the bulk of papers taken care of by LaTeX searching.

The combination of journals and geolocation enabled us to improve the accuracy of the calculation of the total publishing costs for the year (presented in \ref{sec:journal_costs}) as we were able to account for discounted publication costs for authors from certain regions for certain journals. The number of pages were used similarly to account for journals that have per-page pricing or set pricing for all articles within a certain page limit.

Importantly, the journals, affiliations and institutions all have a certain normalisation applied to them since these metrics have been indexed from different sources which may have different formats. For example, the journal `Astronomy and Astrophysics' can also be referred to as `A\&A'. Normalisation ensures that these two entries are counted as the same journal.

\begin{figure*}[t]
	\includegraphics[width=\textwidth]{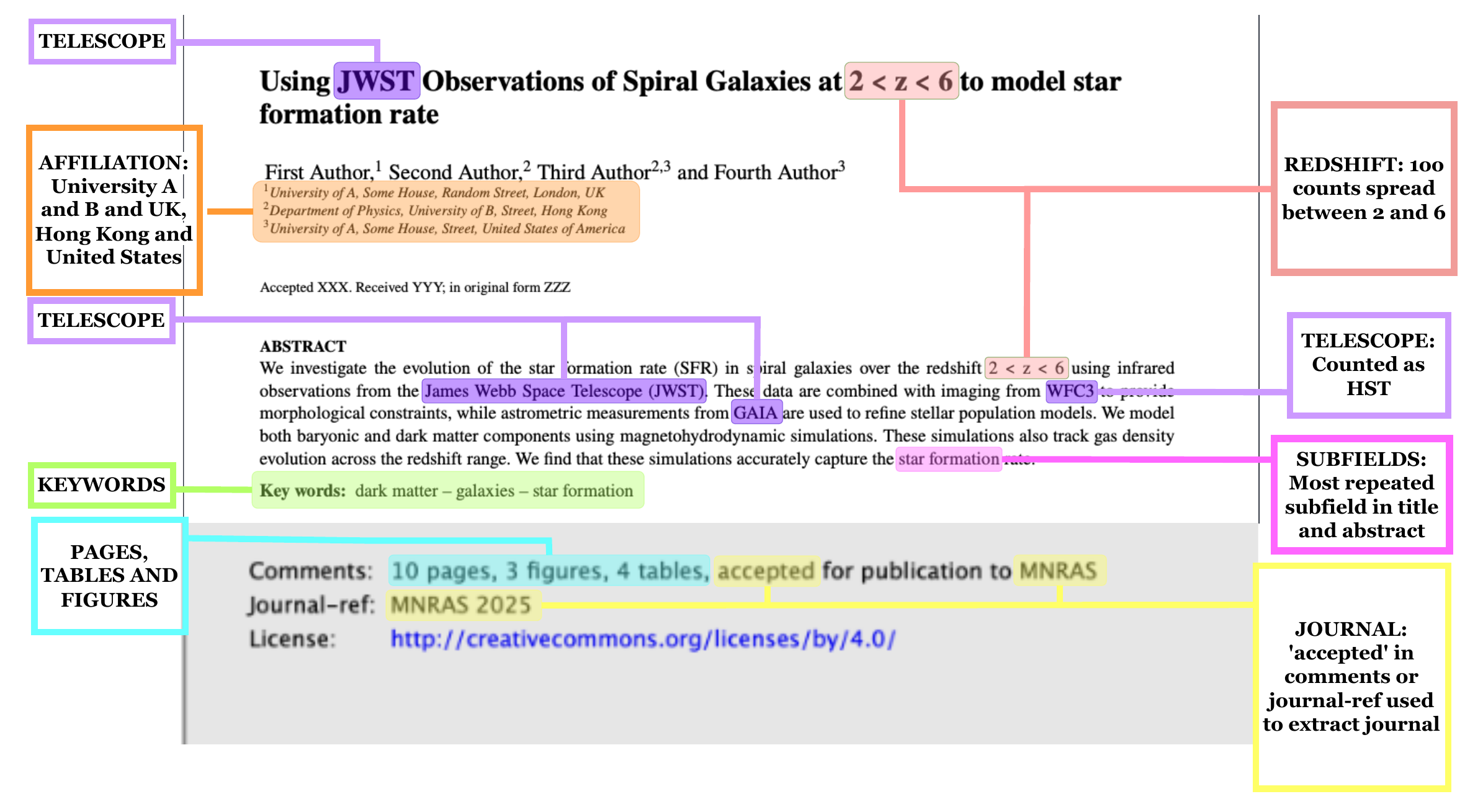}
    \caption{Example Paper showing all Extracted Metadata and Metrics. Using the paper LaTeX, PDF and arXiv metadata, a robust extraction script picks up all aspects of the paper. The demo serves to show how much information we extract from a single paper. For telescopes, we extract WFC3 as an instrument of HST and the telescope as HST, among others. However, for the spectral fingerprint plot, the WFC3 operating wavelength range is counted instead of the full HST wavelength window, along with the ranges of the other telescopes. Since the redshift studied is a range, 1 count for this paper is split across 50 equally distributed points within the range, and those points contribute towards the redshift plot.}
    \label{fig:figure_wrapped_demo}
\end{figure*}

\subsection{Collaborative Indices and Terms} \label{subsec:collab_indices}
Collaboration has become an important aspect of how research is conducted in the modern era. In order to better represent how collaborative an author, paper or region is, we define the following indices and terms:

\begin{enumerate}
    \item Local Collaborative Index (LCI): We define the Local Collaborative Index of a paper as the number of author affiliations from the same country or region as the first author. This term represents how locally collaborative or domestically centred an author or paper is. 
    \item Local Collaborative Ratio (LCR): Similar to the LCI, it is the ratio of the number of author affiliations from the same country or region as the first author to the total number of authors on a paper (including the first author). The LCR tells us the percentage split of authors who are local versus international. 
    \item Global Collaborative Index (GCI): This is the number of author affiliations from a country or region different from that of the first author and thus represents how many international collaborators are present on a paper. 
    \item Non-repeated Global Collaborative Index (NGCI): Lastly, we define the NGCI as the number of unique countries or regions on a paper other than the first author's region. 
\end{enumerate}

To illustrate the difference in these indices, we provide an example- a paper with a total of 5 authors, 2 from Hong Kong (including the first author), 2 from India, and 1 from the United States would have an LCI = 2, a GCI = 3 and an NGCI = 2. Here, we make an important distinction in that the GCI takes into account multiple authors from the same region (as in the case of the example, both authors from India are considered), to maintain symmetry with the way we define the LCI. The NGCI counts multiple authors from the same region as one. The LCR in this example would be 0.4, implying 40\% of the authors are local, and 60\% are global. In summary, the LCI is the number of local collaborators, the GCI is the number of global collaborators, the NGCI is the number of countries or regions collaborated with and the LCR is the percentage of the total authors who are local.

\subsection{Extracting Content-related Information}
We did not want to limit this work to reporting number statistics but instead wanted to reflect on the specific topics and areas of interest that the community has focused on. This required us to go beyond readily-available and indexable metrics and look through the actual content that papers are written about. Details like the telescope used in a study or the sub-field referenced in a paper are buried in the paper content. To extract these metrics, we developed a general methodology of searching for words and phrases related to the metric that are present in the text of the title and abstract of a paper. The title and abstract are the first two, if not only, sections of a paper that a reader skims through before deciding if the paper is interesting enough to warrant a proper perusal. Knowing this, authors should only include the most relevant and important pieces of information in the title and abstract, which is why we believe it is justified to conduct our extraction only using these two sections of the paper. As an added bonus, these sections do not permit referencing, which limits authors to focus on their work only, rather than prior literature, preventing our extraction from picking up on referenced work.

We use very specific searching phrases for each metric to ensure we only extract information that is relevant to the metric. We simultaneously search for a series of different phrases and variations of phrases to account for differences in the way authors write. In the following paragraphs, we highlight some metric specific searching techniques we employ.

Redshifts were extracted by searching for any text that included the lower case letter $z$ either preceded or followed by an equality or inequality symbol and a numerical value. A redshift phrase with a discrete numerical value is counted as a single mention in \ref{fig:redshift}. For phrases that include a range, the range is divided into 50 equally weighted mentions that add up to 1. For example, a paper mentioning the phrase `$2 < z < 6$' will contribute a distribution of 50 down-weighted mentions between 2 and 6, which sum up to 1 mention over the entire range. This ensures papers that study a particular range do not only contribute two numerical mentions, which would undervalue that study's impact.

In a similar way, we find the telescopes used in a paper by searching for the names of telescopes in the title and abstract from a set list we create. The list includes about 60 space and ground based telescopes with a preference for currently operational and future facilities. We look for both the full names and abbreviations of telescopes keeping case sensitivity in mind for the latter, to avoid matching to parts of common words. For certain prominent telescopes, we also include the names of instruments on board like MIRI for JWST and WFC3 for HST, because some papers tend to report the instrument on the premise of the associated telescope being common knowledge.

Building on our telescope extraction, we looked for mentions of wavelength, frequency and energy related phrases by searching for their respective units. For wavelength, we included mentions such as `nm', `$\AA$', but avoided meters owing to the possibility that it identifies telescope sizes and the fact that wavelengths are rarely reported in meters. The same was done for frequencies (searched for Hz, GHz, etc) and energy (search terms included eV, GeV, etc), after which all numerical values associated were converted into wavelength. Papers that mentioned a telescope but no wavelength, frequency or energy-related term were set to be representative of the general wavelength range of the telescope. If specific instruments of the telescope were mentioned, then the operational range of those instruments was considered, though this was done only for some major telescopes like JWST and HST. We also accounted for mentions of specific spectral lines and transitions such as $H\alpha$, the 21 cm line, Calcium H and K lines and a few more. The same weighting technique that we implemented for redshift was applied for all cases where a range of numerical values was reported instead of a discrete value.

Finally, for keywords and subfields, we apply the same text extraction we have done everywhere else. Some papers already have three keywords assigned to them by the authors. For papers without keywords, we assigned a set of three by matching the most commonly occurring words in the titles and abstracts to a dataset of approximately 1000 of the most used keywords from papers that were assigned keywords. We believe this is justified as about 68\% of papers in the dataset already have keywords assigned, meaning that this list of top keywords represents a good proportion of what is used in the field. We realised that a few trends change when we include our self-assigned keywords, so we report statistics both with and without them. Sub-fields, though similar in principle is assigned as one per paper from a pre-defined list of around 100 words that we made up. The list is meant to be distinct from keywords, avoiding some of the most frequently used keywords, and covers a broad range of topics and more general method-specific terms, such as simulations.

In general, every content-related metric that we search for has a series of carefully throughout out terms used for searching that are narrow enough to trigger on the phrases we are interested in but broad enough or extensive enough to account for the disparity in the way the same term can be reported across the literature.

\subsection{Fidelity, Completeness and Caveats}
Science must constantly wrestle with the fact that all conclusions are subject to uncertainty. So is the case with this paper. While we have the titles and abstracts and basic metrics like citations for all papers and report 100\% completeness, metrics which require more complex extraction like the affiliations and content-related metrics, are subject to uncertainties. These uncertainties are not in the form of incorrect information but rather incomplete information. Since our extraction is strict and is based on text taken directly from the arXiv, the possibility of having errors or false positives is very low. Especially for the content related metrics, the search only trigger if the search term is found. For a term like `JWST', there is only one subject this term could be referenced in relation to. Thus, false positives are extremely rare.

We understand that this is not very convincing, which is why we manually checked the accuracy and completeness of these metrics for 100 randomly selected papers and report fidelity metrics based on these papers. We find that the most unreliable metrics are the pages, tables and figures which, although 100\% complete, are around 85\% accurate. This is only because of the fact that authors inconsistently report these metrics inclusive or exclusive of the appendix, in the arXiv comments. Nevertheless, this does not affect any of the conclusions in this paper, and the numbers differ by small numerical values. For citations, we report 100\% completeness as NASA ADS keeps track of preprint paper citations as well. Out of 100 papers, 97 of them had journal information complete and correctly reported, with the three being incorrect only due to the fact that our normalisation does not account for very niche journal names. As for our content extraction, we report 100\% completeness and fidelity with all papers having redshift-related phrases correctly extracted. We did, however, notice one paper with a telescope mentioned that was not in our set list of searchable telescopes, which is why our extraction missed it. Affiliations and geolocations are also well filled, with 96.4\% of papers having at least one extracted affiliation or geolocation. Out of these, 95\% of affiliations and locations are correct, with the 5\% of incorrect entries coming from matching to country names associated with other sections of the paper, like the acknowledgements instead of the affiliations. For the 96\% of papers that we have extracted affiliations for we report a 71\% completeness owing to extreme differences in LaTeX formatting. It is important to keep in mind that this completeness is not reflective of the number of affiliations missed on a paper and hence does not mean we only account for 71\% of authors. Rather, its interpretation is 71\% of papers have at least 1 author affiliation missing. We note that during our manual checking a vast majority of the time, the code misses 1 or 2 author affiliations and rarely does it completely fail on a paper. To put some form of a number statistic to this, assuming on average an author has only one affiliation (which is a reasonable assumption considering having multiple affiliations on a single paper is not very common), 97.8\% of authors in our dataset have an associated affiliation. All in all, our location and affiliation statistics are near complete, given how difficult it is to obtain a researcher's affiliation, let alone extract it from a set of non-conformal research papers.

A large language model approach for extracting metrics was considered and tested, especially for affiliations, but the cost involved for processing such a large dataset and the possibility of hallucinations convinced us otherwise. In fact, during a preliminary test, the LLM hallucinated an author affiliation within the first five papers checked. While this is presently unfeasible, with more experimenting and resources it might be useful in the future.

This uncertainty and incompleteness was not easy to deal with since it almost always originates either from the fact that a paper has a slightly different format that makes it difficult to parse or from a search term that does not cover the 0.1\% of papers that include a rarely used niche nomenclature. Hence, we must lay out some caveats. The number statistics that we report are accurate and precise, but do not have perfect precision. If we report that a certain telescope has an ACC of 5.3, we expect readers to inherently associate an uncertainty with this number. The focus should be more on comparing statistics (telescope A has a higher AAC than telescope B), rather than exact numerical values. This is the inherent uncertainty that we must accept when working with such inhomogeneous data. We also caveat this analysis with the fact that parts of this paper use text matching, which is sensitive to the terms we search for. These search terms have been extensively constructed to account for diversity in writing and have aimed to be as inclusive as possible. However, we cannot include everything. We hope that the completeness statistics we reported in the previous paragraph have convinced you of the effort to be inclusive and of the overwhelming majority of papers that are correctly indexed and analysed in this work. Finally, this work is based on arXiv papers. Different communities use the arXiv in different ways, which is why it is possible that certain communities are under represented in our original sample.

Despite these small but tenacious caveats, we believe this work accurately captures the general state of astrophysics for the year 2025. We believe this work has successfully picked up on the overarching trends and has given as a good compass for determining the past, present and future of the field. As we continue to write these papers year after year, we learn from the papers we read and past year's experiences and will implement improvements to our analysis that reduce uncertainty and account for the small percentage of papers that we miss. For the time being, we accept these uncertainties as we do with all other fields, but pride ourselves in the fact that we have ambitious captured what astrophysics was like in 2025. While it is impossible to calculate an exact number, we hope this detailed methodology and our fidelity checks have convinced readers that this paper covers somewhere between 90 and 95\% of the field with minimal uncertainty, which, for a first-time experiment and analysis, is extremely high. \\

\section{Results and Discussion} \label{sec:results}
In this section, we lay out all the statistics we have collected. Since we are presenting these statistics from different perspectives or relative to different base metrics, we split up this section such that each subsection corresponds to a different base metric. \\

\subsection{General Statistics} \label{sec:general_stats}

\begin{figure*}
	\includegraphics[width=\textwidth]{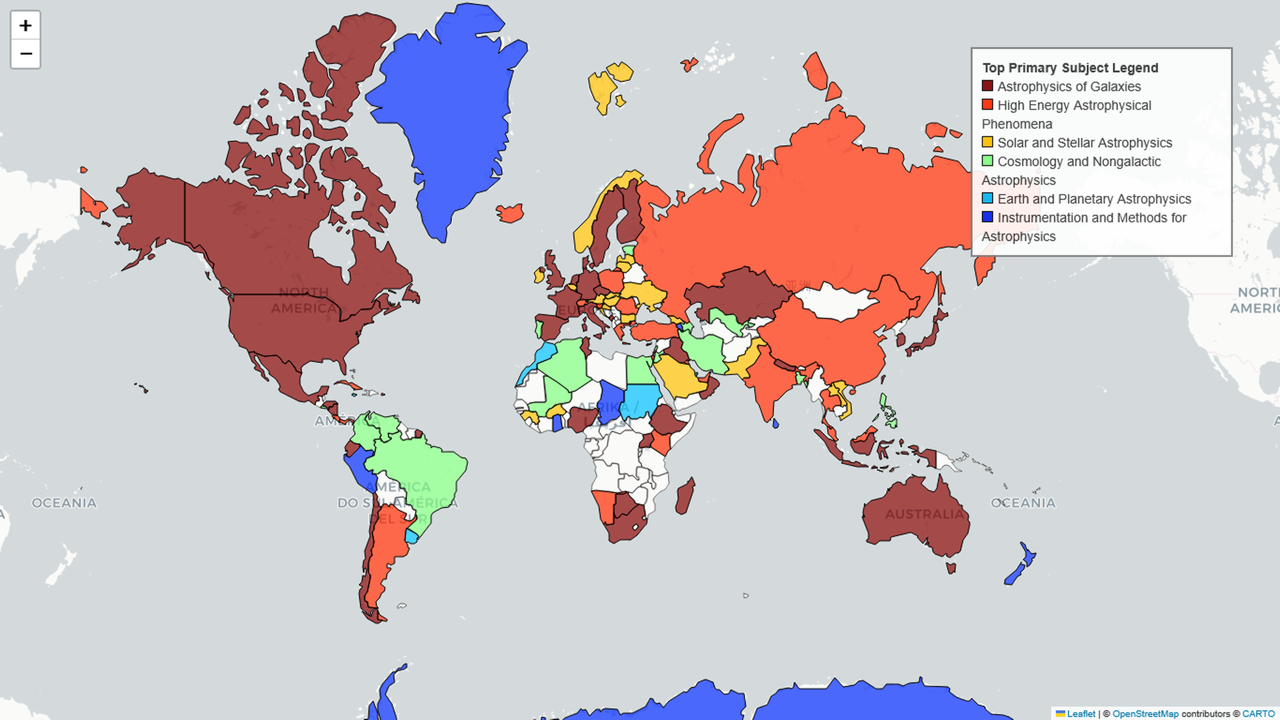}
    \caption{Top Primary Subject by First Author Country or Region. First authors with multiple affiliations have each region counted individually. For example, a first author affiliated with China and the US under `Astrophysics of Galaxies' will have one count for the category going to China and one going to the US. All countries or regions in white had no papers this year or were missed by our affiliation extraction.}
    \label{fig:p_sub_map}
\end{figure*}

\begin{figure}
	\includegraphics[width=\columnwidth]{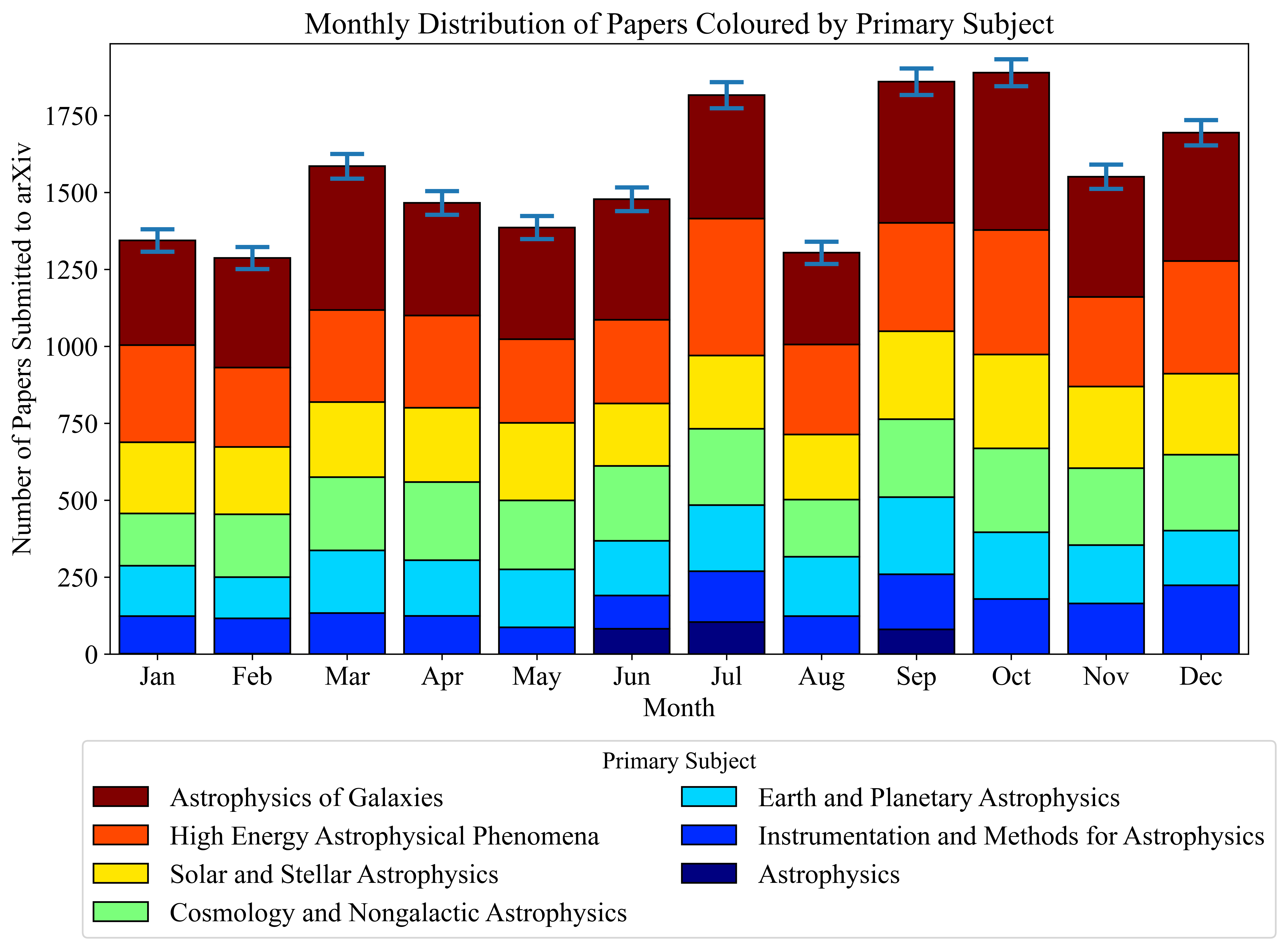}
    \caption{Monthly Distribution of Papers coloured by the arXiv Primary Subject they were submitted under with Poisson error bars. The plot also shows us that October, September and July were the months with the most number of papers added to the arXiv.}
    \label{fig:figure1}
\end{figure}

We begin with some general, overarching statistics from the year. As mentioned before, there were 18660 research papers published this year on the arXiv under the Astrophysics category in comparison to 16333 articles published in 2024. On average, there were 1555 papers per month, or about 72 papers per day, excluding weekends and accounting for the fact that arXiv uploads 5 days a week. This changes to 51 papers per day including weekends and days the arXiv does not publish. The most active months of the year by far (in terms of total number of articles published) were October, September and July in descending order, with the 4th most active month, December, being 0.56$\sigma$ away from July. Accounting for the number of days on which papers were published each month, September was the most active with 62 papers/day, followed by October at 61 papers/day. The least active month was August with an average of 42 papers/day. 

\noindent The AAC for all papers in the dataset is 2.53, and the EAAC is 1.77. Only considering the articles published in a journal, the JAC is 2.84 and the EJAC is 1.97, implying that a paper published in a journal is self-cited an average of 0.87 times. More journal-specific discussion will be provided later in Section \ref{sec:journals}. 

\noindent All arXiv papers are submitted with a Primary Subject metadata category consisting of 7 categories- Astrophysics of Galaxies, High Energy Astrophysical Phenomena, Solar and Stellar Astrophysics, Cosmology and Nongalactic Astrophysics, Earth and Planetary Science, Instrumentation and Methods for Astrophysics and finally, Astrophysics. The monthly distribution of papers coloured according to their Primary Subject is given in Figure \ref{fig:figure1}. From Figure \ref{fig:figure1} we see that every month, Astrophysics of Galaxies is consistently the highest submitted primary subject category with 4761 papers submitted over the year and an average of 397 papers per month under this category. The fluctuations in submission approach a consistent peak just after the summer holidays, presumably the culmination of summer projects carried out by students. A complete set of Primary Subject Statistics including citation indices are available in Table \ref{tab:table1}. Interestingly, `Cosmology and Nongalactic Astrophysics' papers have the highest number of citations across all indices, by some margin, despite having a lower number of papers submitted under this category. This tells us that while there might be fewer papers under this category, these papers are cited more on average, than the papers from any of the other categories.

We will go deeper than this broad primary subject classification and discuss the statistics of sub-fields in more detail later in sub-section \ref{sec:subfields}. With author affiliations we can show the global distribution of the most published primary subject category for each country or region based on the country or region indexed for the first author on the paper. This map is given in Figure \ref{fig:p_sub_map}. A more detailed description of the author affiliations will be given later in Section \ref{sec:location_and_collab}. 

\begin{table*}
\centering
\resizebox{\textwidth}{!}{%
\begin{tabular}{lcccccccc}
\hline
Primary Subject                              & Annual Count & Average Per Month & Average Per Day* & Highest Month Count & AAC$^{\dagger}$ & EAAC$^{\dagger}$ & JAC$^{\dagger}$ & EJAC$^{\dagger}$\\ \hline \hline
Astrophysics of Galaxies                     & 4761  & 397   & 18  & October   &  2.78 & 1.75 & 2.93 & 1.86 \\
High Energy Astrophysical Phenomena          & 3869  & 322   & 15  & July      &  2.29 & 1.66 & 2.58 & 1.81 \\
Solar and Stellar Astrophysics               & 2958  & 246   & 11  & October   &  1.41 & 0.93 & 1.62 & 1.07 \\
Cosmology and Nongalactic Astrophysics       & 2789  & 232   & 11  & October   &  4.83 & 3.86 & 7.02 & 5.57 \\
Earth and Planetary Astrophysics             & 2293  & 191   & 9   & September &  2.25 & 1.47 & 2.55 & 1.72 \\
Instrumentation and Methods for Astrophysics & 1722  & 144   & 7   & December  &  1.06 & 0.59 & 1.54 & 0.91 \\
Astrophysics                                 & 268   & 54    & 2   & July      &  2.06 & 1.47 & 2.65 & 1.93 \\ \hline
\end{tabular}}
\caption{Primary Subject Statistics ($\dagger$ - per Paper, * - excluding weekends calculated with 5 days a week}
\label{tab:table1}
\end{table*}

\noindent In our dataset, 8684 papers also include a secondary subject. Table \ref{tab:table2} shows the top-15 most commonly reported secondary subjects. Once again, following the trends from Primary Subjects, Astrophysics of Galaxies is also the most common secondary subject. We see further similarities between the most common primary and secondary subjects.

\begin{table}
\resizebox{\columnwidth}{!}{%
\begin{tabular}{lc}
\hline
Secondary Subject                            & Annual Count \\ \hline \hline
Astrophysics of Galaxies                     & 2049         \\
Solar and Stellar Astrophysics               & 1632         \\
Instrumentation and Methods for Astrophysics & 1224         \\
General Relativity and Quantum Cosmology     & 1205         \\
High Energy Astrophysical Phenomena          & 983          \\
Cosmology and Nongalactic Astrophysics       & 892          \\
High Energy Physics - Phenomenology          & 831          \\
Earth and Planetary Astrophysics             & 605          \\
High Energy Physics - Theory                 & 340          \\
Space Physics                                & 263          \\
Plasma Physics                               & 223          \\
Machine Learning                             & 191          \\
Nuclear Theory                               & 149          \\
High Energy Physics - Experiment             & 111          \\
Computational Physics                        & 80           \\
\end{tabular}}
\caption{Top 15 Secondary Subject Paper Categories. Not all papers have a secondary subject assigned so the counts do not add up to 18660. With the exception of `Astrophysics of Galaxies', there is some shuffling between the most submitted categories when comparing primary and secondary subjects.}
\label{tab:table2}
\end{table}

\subsection{Redshift and Spectral Fingerprint} 
\label{sec:redshift_and_sed}

\begin{figure*}

 \begin{subfigure}{\textwidth}
       
	\includegraphics[width=\linewidth]{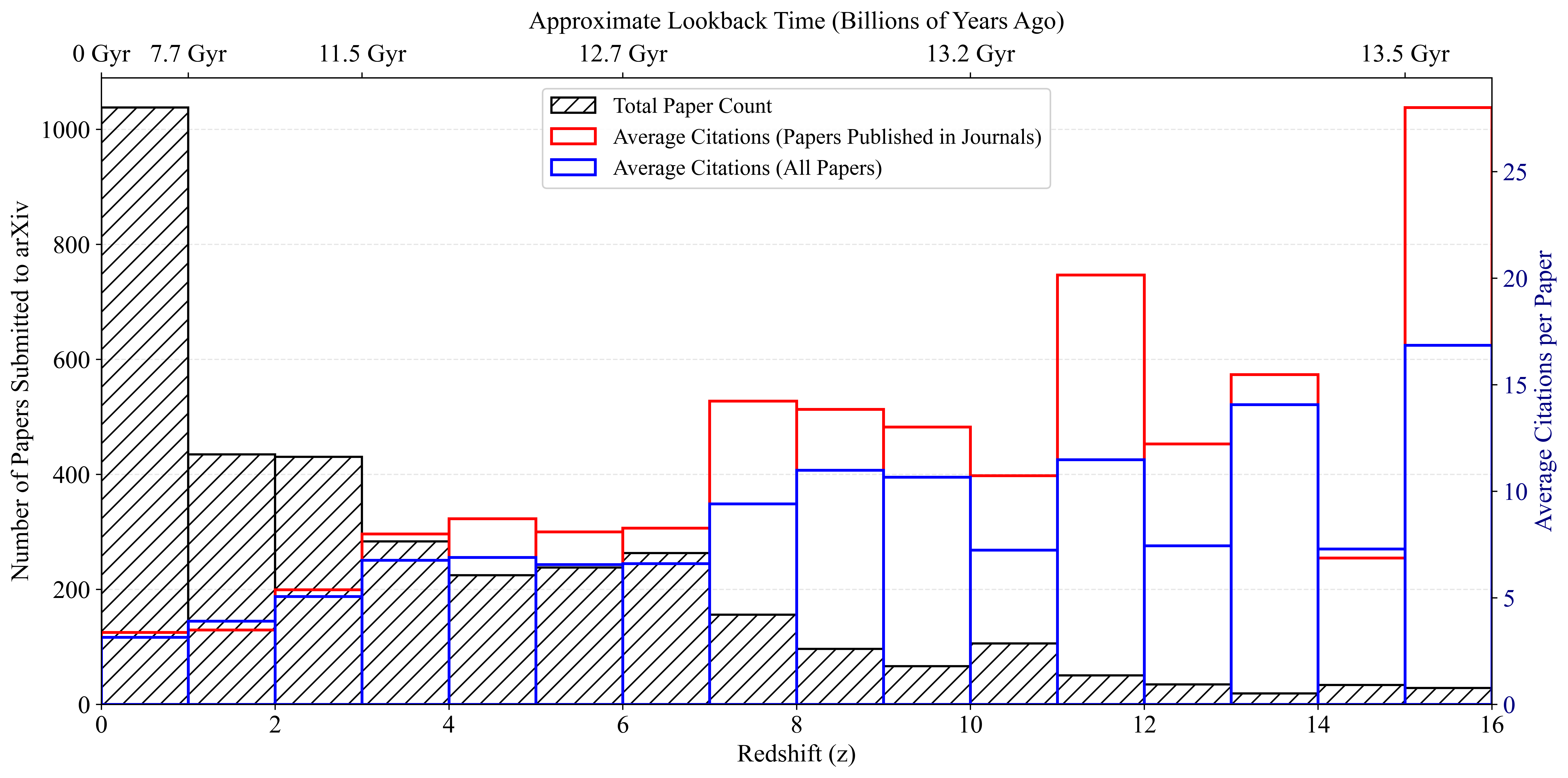}
    \caption{}
    \label{fig:redshift}

    \end{subfigure}
 \begin{subfigure}{\textwidth}
	\includegraphics[width=\linewidth]{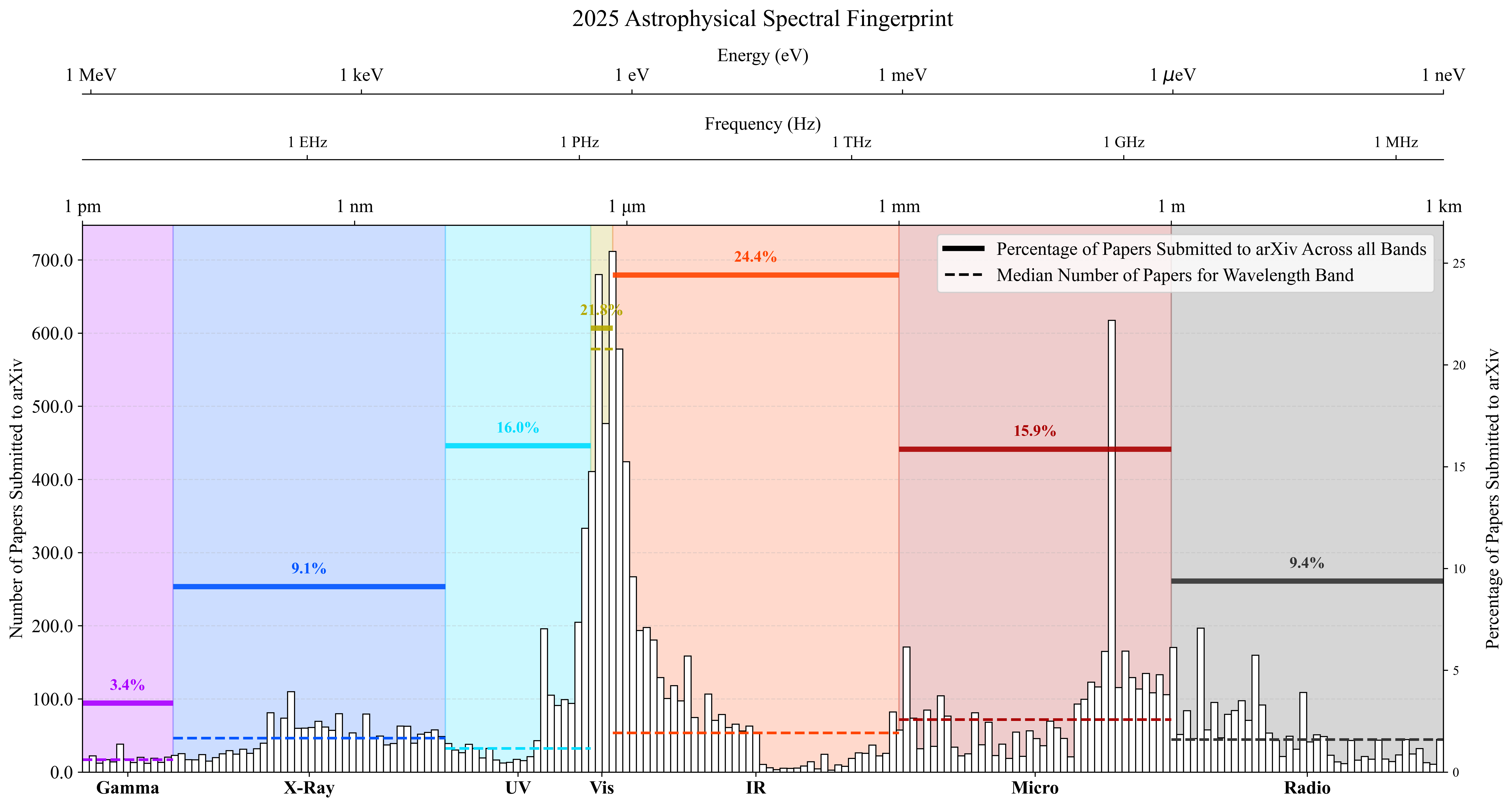}
    \caption{}
    \label{fig:sed}

       \end{subfigure}
       \caption{Redshift distribution and 2025 Astrophysics Spectral Fingerprint. a, Distribution of reported redshifts in papers and citations as a function of redshift. The hatched histograms represent the distribution of redshifts reported in paper titles and abstracts. The colored histograms show the distribution of average citations per paper for each redshift bin, considering all papers on arXiv (blue) or only papers published in journals (red). If there is a redshift range instead of a specific number, for example, z = 0-5, then the contributions are equally spread between the bins spanning 0-5. b, Distribution of wavelengths, frequencies and energies (accounting for telescope observations and spectral lines) across the electromagnetic spectrum mentioned in paper titles and abstracts. Colored regions separate the conventional spectral bands, with solid horizontal lines indicating the relative fraction of mentions within that region. Peaks can be seen at specific wavelengths that correspond to popular emission/absorption lines.}
       \label{fig:combo}
\end{figure*}

Figure \ref{fig:redshift} shows the distribution of redshifts that the astrophysics community is interested in. Perhaps as expected, there is an abundance of mentions for a redshift of 0 with a decreasing trend towards higher redshifts. There is a temporary reversal in this trend at a redshift of 6, presumably due to interest in the epoch of reionization. Motivated by the redshift distribution, we also investigated the distribution of research across the entire electromagnetic spectrum. We refer to this as the 2025 Astrophysics Spectral Fingerprint, shown in Figure \ref{fig:sed}. The favourite region is the infrared band, with around one quarter of the papers from this year studying the low-infrared region in particular. This is the effect of the dominance of the James Webb Space Telescope in research this year. Another interesting observation is that the entire optical band is well saturated; presumably, as this is the range the human eye is sensitive to, and following the historical beginnings of astronomy with observations in this waveband. In general, Figure \ref{fig:combo} highlights the preferences and areas of interest in the redshift and spectrum for astrophysics research in 2025; as mentioned earlier, these patterns help inform the course of past, current, and future research.

\begin{figure*}[hbt!]
    \centering
    
    \begin{subfigure}{\textwidth}
        \includegraphics[width=\textwidth]{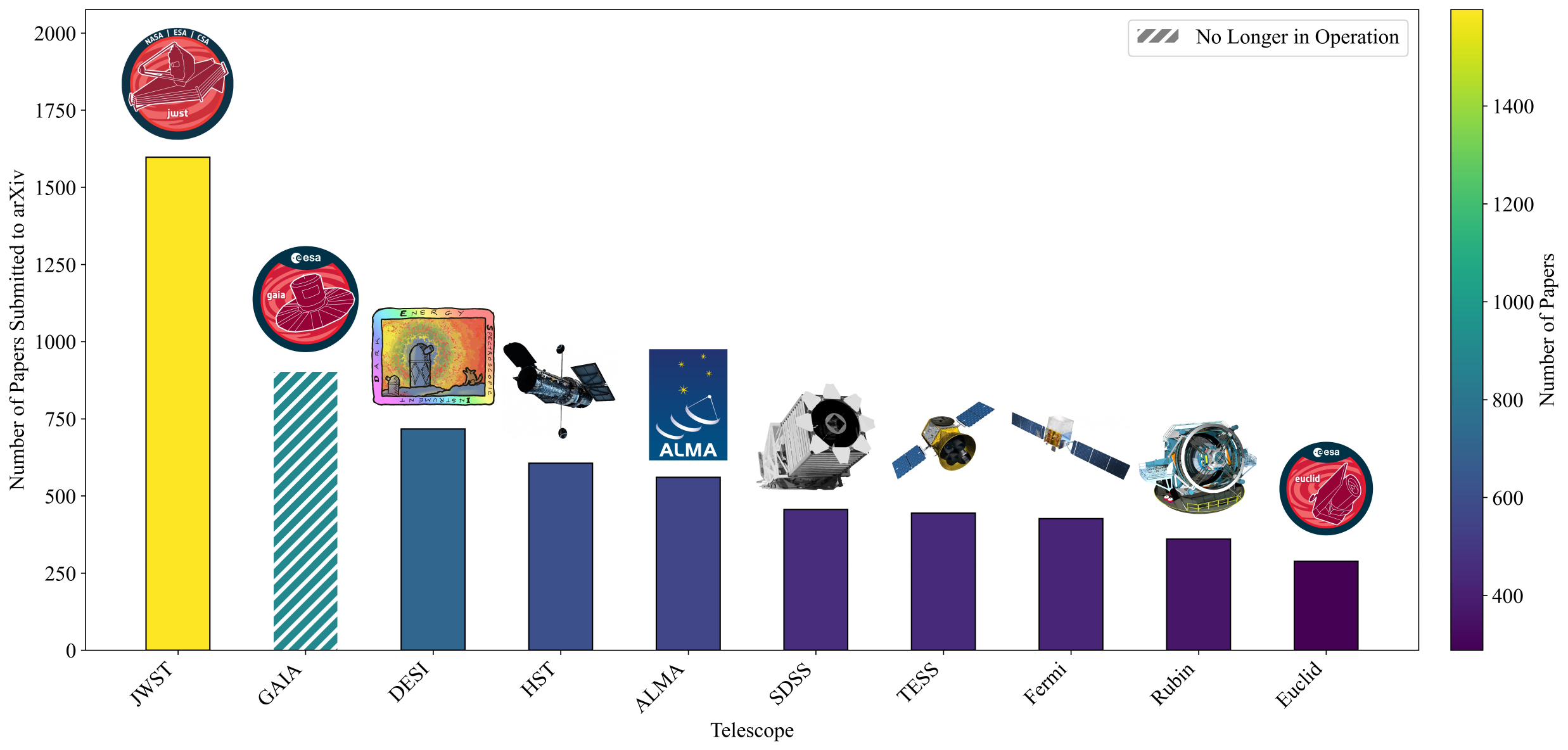}
        \caption{Top 10 telescopes in 2025}
        \label{fig:telescopes}
    \end{subfigure}
    
    \vspace{1.5em} 
    
    \begin{subfigure}{\textwidth}
        \includegraphics[ width=\textwidth]{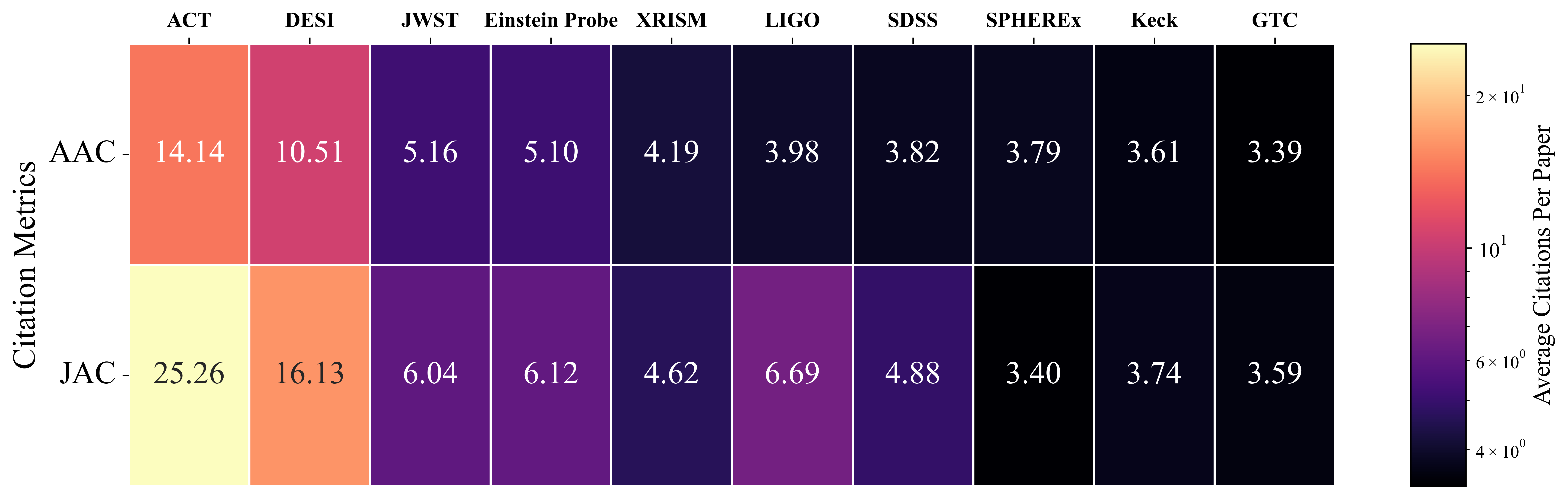}
        \caption{Top 10 telescopes in 2025 for average citations per paper}
        \label{fig:telescope_heatmap_citations}
    \end{subfigure}
    
    \caption{Distribution of the most mentioned telescopes and their respective citation metrics. a, Distribution of papers that refer to a telescope in their title and abstract (represented both as the relative height of the bars and as a colour map for readers with different visual preferences for interpreting data). GAIA is the only telescope with a hatched bar, as it is not operational anymore. b, Distribution of telescopes as a function of average citations per paper when considering all article citations (top panel) or only published journal article citations (bottom panel).}
    \label{fig:telescopes_combined}
\end{figure*}

\subsection{Telescopes, Instruments and Observations} \label{sec:telescopes}
In this section, we present some statistics on which telescopes and instruments we relied on the most during 2025 as well as the most studied astronomical objects. This analysis, as is the case with most of the sections in this paper, is based on matching names from an extensive list of 60 past, present and future telescopes to those appearing in the title and abstract of a paper. We ensure to account for telescope instrument names as well as short forms, to make the string matching catch all possible appearances of telescopes. For papers that mention multiple telescopes, we count all telescopes equally. Unsurprisingly, the James Webb Space Telescope was the most mentioned telescope this year by a decent margin, as is seen in Figure \ref{fig:telescopes}. The second and third most quoted are GAIA and DESI, respectively. We also see the Vera C Rubin Observatory in the top 10, despite being a more recent telescope compared to the other relatively older telescopes in the top 10. GAIA is the only telescope that has stopped operating, but is still highly used, considering how much data is contained within the archive and the fact that we still await GAIA DR4. The full bar plot with all the telescopes we searched for and their respective counts is provided in the Appendix for curious readers. 

\noindent For each telescope, we calculate four different citation indices (defined in \ref{methods}) which provide insight into how impactful the data from the telescope has been. For simplicity, in Figure \ref{fig:telescope_heatmap_citations} we show two indices that represent the average number of citations for: (i) all papers up- loaded to arXiv (AAC); and (ii) all papers published in journals (JAC). By a significant margin, the ACT and DESI are leading over other telescopes, presumably due to their contributions to cosmology this year. Encompassing broader research topics, JWST and the Einstein Probe lead the charge. What is particularly noteworthy is that despite not being among the 10 most used telescopes in 2025, the Einstein Probe is on par with JWST in terms of citations. This indicates that while there are fewer papers that use data from the Einstein Probe, these papers are usually highly cited. One might think that the Einstein Probe does well because of the fact that JWST is used in far more papers leading to the average being skewed down by many low citations papers. However, both JWST and the Einstein Probe have a similar ratio of low citation papers to total papers that mention the telescope meaning that both are skewed equally. This goes to show how impressive the Einstein Probe really is. The opposite trend occurs for the top 10 most mentioned telescopes which are not the most highly cited due to the fact that there are a good number of low-citation papers for the more-used telescopes, thereby skewing the indices to lower values. The complete list of citation indices for all telescopes are provided as a heat-map in the Appendix in Figure \ref{fig:telescope_heatmap_big}. 

\begin{figure*}[t]
	\includegraphics[width=\textwidth]{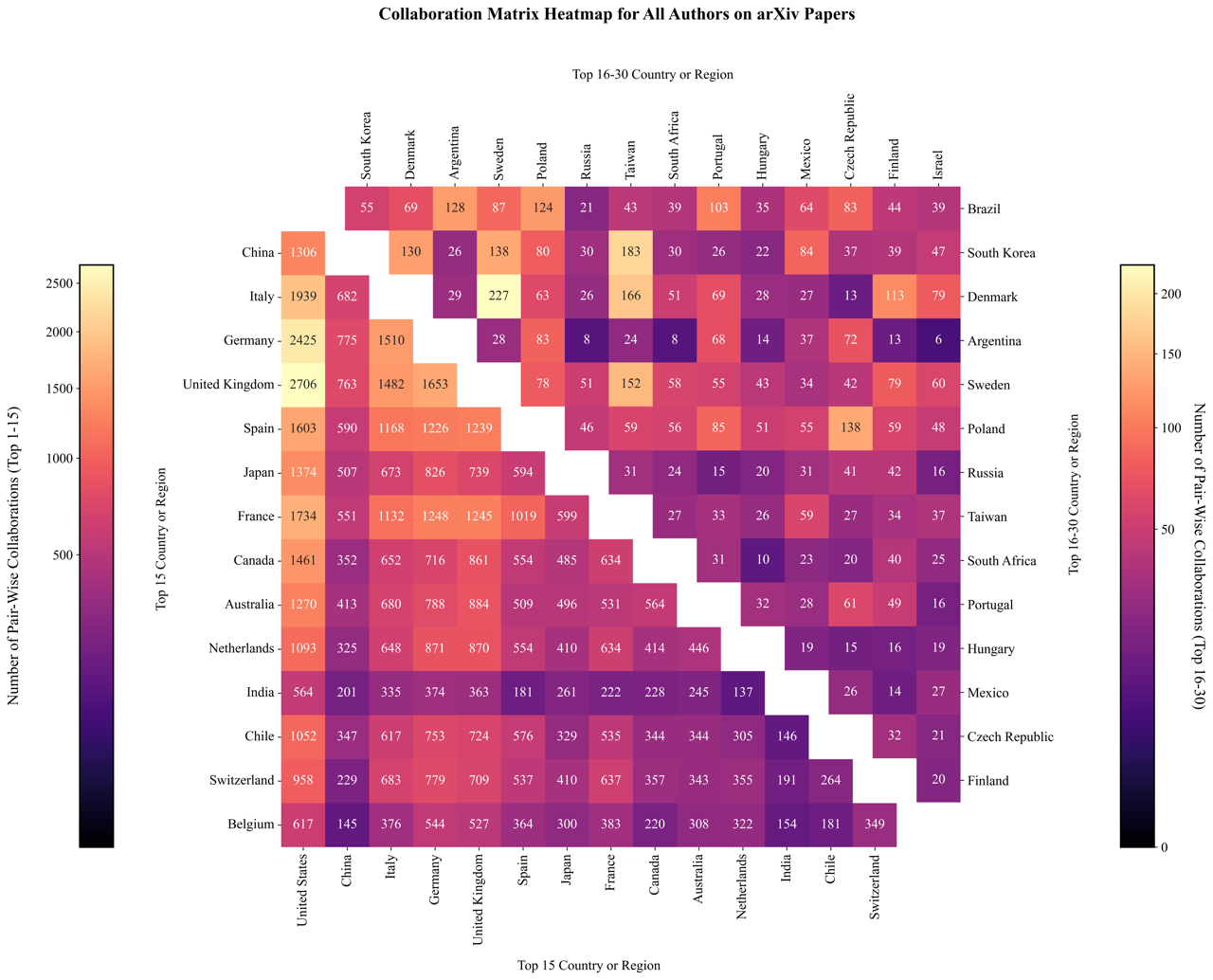}
    \caption{Heat map for top pair-wise collaborations. The colour scale and numbers indicate the number of pair-wise collaborations between countries and regions for the top 15 (bottom triangle) and top 30 (upper triangle).}
    \label{fig:country_heatmap}
\end{figure*}

\noindent Something interesting we can do is search for specific astronomical objects given a certain nomenclature. For example, LIGO events or gravitational wave (GW) events are usually quoted as `GWXXXXXX' in papers. This gives us a template to search for: the capital letters 'GW' followed by a set of six numbers. This also allows us to test for another trend in our data. We see that LIGO is not on the list of the top 10 most quoted telescopes. We find this hard to believe, considering how many papers the LIGO collaboration publishes. We believe the reason for this is the fact that our method relies on string matching. However, LIGO papers or those studying GW events may not mention 'LIGO' or an associated word in the title or abstract, as this is implied by the nature of the data. Since there is a standard format for reporting GW events as `GWXXXXXX', we have a way to test if LIGO is sincerely under-represented in the top telescopes by our string matching code or if this is a true statistic by searching for this specific format of object. With this method, we find 193 papers that studied GW events in 2025, most, if not all of which are from LIGO or related gravitational wave detectors. This proves our hypothesis that LIGO should indeed be much higher up on the list. We can apply this same kind of event format searching for other types of events like Gamma Ray Bursts (GRB) and Fast Radio Bursts (FRB). To compare the three, we calculate the number of unique GW, GRB and FRB events or sources that contribute to 80\% of the total sample of the respective events. We find this measure to be 6, 41 and 23 for GW, GRB and FRB events and sources, respectively, meaning that the GW community focuses the most on specific events, then followed by the FRB community and finally by the GRB community. In the same way we also search for Pulsars, Supernovae, Supernova Types, Exoplanets, molecular species as well as Messier and NGC Catalogue Objects. The detailed list of the top 5 events and sources we find using this method along with some other statistics are given in Tables \ref{tab:event_table_1} and \ref{tab:event_table_2} and \ref{tab:event_table_3}.

\noindent We acknowledge that this method can have errors and is likely to have some miscounting. However, we have done random checks as mentioned in Section \ref{methods} for each of the objects we search for and find near perfect matches for all of them. We only find a few miscounts when searching for Messier Catalogue objects owing to the fact that the string matching is very simplistic for such objects (since the search is for objects starting with a capital M or the word Messier and followed by a number), making the searching sensitive to errors. For example, we find a paper about testing Apple M1 chips counted as a Messier 1 reference. We do not make an effort to fix this, considering how few and far apart these errors are and the fact that they do not significantly change any of the conclusions. We would like to note that the more complex searches like those for GWs, GRBs, FRBs, exoplanets and SN are less sensitive to those kinds of errors and are extremely likely to have no errors at all. This is supported by the fact that we encountered no errors during our random checks for these objects. This is currently the best method we have for searching for these objects. Although this method already works very well, we do plan to experiment with other, possibly cleaner, methods in the future. 

\begin{table}
\centering
\renewcommand{\arraystretch}{1.2}
\resizebox{\columnwidth}{!}{%
\begin{tabular}{|c|c|c|}
\hline
\textbf{Category} & \textbf{Name} & \textbf{Count} \\
\hline

 & GW170817 & 76 \\
 & GW231123 & 31 \\
 GW Events  & GW190521 & 14 \\
 & GW190814 & 14 \\
 & GW230529 & 13 \\
\hline

 & GRB221009A & 39 \\
 & GRB230307A & 16 \\
 GRB Events & GRB211211A & 16 \\
 & GRB170817A & 14 \\
 & GRB250702B & 9 \\
\hline

 & FRB20240114A & 13 \\
 & FRB20201124A & 12 \\
 FRB Sources & FRB20121102A & 11 \\
 & FRB20190520B & 9 \\
 & FRB20180916B & 5 \\
\hline

 & PSRJ0740+6620 & 12 \\
 & PSRJ0030+0451 & 7 \\
 Pulsars & PSRJ0437-4715 & 6 \\
 & PSRJ1023+0038 & 5 \\
 & PSRJ0614 & 5 \\
\hline

 & SN1987A & 25 \\
 & SN2023i & 25 \\
 SN Events & SN2024g & 15 \\
 & SN2024a & 11 \\
 & SN2009i & 8 \\
\hline
\end{tabular}}
\caption{Top 5 Most Mentioned Events/Sources by Category (Part 1)}
\label{tab:event_table_1}
\end{table}

\begin{table}
\centering
\renewcommand{\arraystretch}{1.2}
\resizebox{\columnwidth}{!}{%
\begin{tabular}{|c|c|c|}
\hline
\textbf{Category} & \textbf{Name} & \textbf{Count} \\
\hline

 & Type Ia & 323 \\
 & Type II & 66 \\
 SN Type & Type IIb & 27 \\
 & Type Ic & 19 \\
 & Type IIn & 16 \\
\hline

 & K2-18b & 32 \\
 & WASP-39b & 23 \\
 Exoplanets & WASP-121b & 16 \\
 & TOI-270d & 14 \\
 & WASP-107b & 12 \\
\hline

 & $CO$ & 622 \\
 & $H_{2}$ & 365 \\
 Molecular Species & $HI$ & 335 \\
 & $H_{2}O$ & 238 \\
 & $CO_{2}$ & 207 \\
\hline

 & M31 & 119 \\
 & M87 & 63 \\
 Messier Objects  & M4 & 43 \\
 & M33 & 33 \\
 & M1 & 33 \\
\hline

 & NGC1068 & 29 \\
 & NGC628 & 16 \\
 NGC Objects  & NGC253 & 14 \\
 & NGC4258 & 13 \\
 & NGC1275 & 9 \\
\hline
\end{tabular}}
\caption{Top 5 Most Mentioned Events/Sources by Category (Part 2)}
\label{tab:event_table_2}
\end{table}

\begin{table*}
\centering
\renewcommand{\arraystretch}{1.2}
\small
\resizebox{\textwidth}{!}{%
\begin{tabular}{|l|c|c|c|c|}
\hline
\textbf{Category} & \textbf{Total Papers} & \textbf{Number of Unique Events} & \textbf{Top 5 \%} & \textbf{Unique Events that make up 80\% of Sample} \\
\hline
GW & 193 & 23 & 77\% & 6 \\
GRB & 217 & 84 & 43\% & 41 \\
FRB & 104 & 43 & 48\% & 23 \\
Pulsar & 253 & 188 & 14\% & 138 \\
SN & 311 & 146 & 27\% & 84 \\
SN Type & 487 & 10 & 93\% & 3 \\
Exoplanet & 715 & 512 & 1.4\% & 369 \\
Molecular Species & 2389 & 24 & 74\% & 7 \\
Messier & 589 & 58 & 49\% & 15 \\
NGC & 790 & 406 & 10\% & 248 \\
\hline
\end{tabular}}
\caption{Astronomical Event Statistics: The fourth column shows the percentage of the total number of papers in a particular category that come from the top 5 most mentioned unique events. The fifth column on the other hand shows the number of unique events that make up 80\% of the papers. Both columns show how different areas of the field study data.}
\label{tab:event_table_3}
\end{table*}

\subsection{Journal Costs} \label{sec:journal_costs}

\noindent We think this section is of great importance to the astrophysics community, and the scientific community in general. Science is underpinned by communication and collaboration. Systems that disrupt this are naturally abandoned for better ones given sufficient time. We provide a simple cost analysis for the papers published this year to put into perspective how expensive (and lucrative) publishing can be (brace yourselves!).

\noindent Throughout the years, there has been a gradual increase in the publishing costs of journal articles. With more journals shifting to gold open-access where authors need to pay an Article Processing Charge (APC) for readers to view the article for free, there has been mounting pressure on authors and their grant funding to keep up with publication costs. For example, typical APCs are in the range of thousands of USD; starting at about \$1400 USD and can go up to a staggering $\sim$ \$13,000 USD. Especially for readers who are not in the astrophysics or broader scientific community, we emphasise a few key points that one may find shocking: 

\begin{itemize}

\item Science (for the most part) is funded by the tax-paying general public. Despite this, authors are typically required to pay exorbitant amounts of money for publication (mostly online only), and more so for open-access publication. In addition, readers are also generally expected to pay a subscription cost to the journals, despite most of these readers also being the tax-paying general public.

\item The above is exacerbated by the fact that the reviewers in peer-review journals are not paid - they do it purely as charity and out of a sense of responsibility to the community. One would think such free labour is something that would not exist in this day and age simultaneously with large APCs. The same authors who pay to publish must also review articles for free, and even pay to read other articles.

\item Almost all primary journals are digital only - there are no hardcopy prints of the publications. It is hard to imagine how it is justified to charge thousands of USD for a single article, especially when there are no cost breakdowns provided and a lack of transparency. This is very surprising considering that current peer-review journals would not exist without the main ingredient: free reviewers.

\end{itemize}

\noindent Despite this inane system, scientists have been forced to adhere to it. The arXiv is therefore an important repository, especially for students, who, when entering the field, need to read a lot of papers and might not be part of an institution that has an open-access agreement with these journals. With mounting pressure on APCs, however, an increasing number of scientists are moving to fully open-access peer-review journals with no publication costs or APCs, termed \textit{diamond open access journals}. A leading and up-and-coming example of such a journal is the Open Journal of Astrophysics.

\noindent We estimate the total amount of money spent in publishing to paid journals, assuming every paper is published under Gold Open Access. Our calculation takes into account the publishing fees for different journals, the cost per page where applicable, as well as discounted rates for authors from certain `member' countries for specific journals. We remind the reader that even if a publication cost has discounted rates, someone is paying for it. It may not be the authors, but certainly the tax-paying general public - in the end, the profits still remain the same. In total, we estimate the community spent a staggering 17 million USD on publishing fees this year. To allow the reader to visualize this sum, the cost of the Zwicky Transient Facility was $\sim$ 24 million USD. Counting only the papers that were published and ignoring the zero cost of open source journals, this rounds out to an absolutely stunning 2,400 USD per publication. If every astrophysics paper published on the arXiv in 2025 were to be published under the same standards (average cost from the previous calculation applied to every paper), that would mean a total cost of 45 million USD on publishing. These numbers are similar to those obtained in \cite{Coles_2025a}. Astrophysics is, according to our calculations, a multi-million dollar business, but for whom? Certainly not for the people who make it possible. Definitely not for the scientists and not for the general public. These numbers are lower limits, assuming that authors do take full advantage of discounts, which is not necessarily true. In some cases, the institutions bear the cost. To drive the point home, we find the average number of citations per dollar is 0.002 for all citations and 0.0014 excluding self-citations, meaning the cost per citation is between 500 and 700 USD per citation.

\begin{figure}
	\includegraphics[width=\columnwidth]{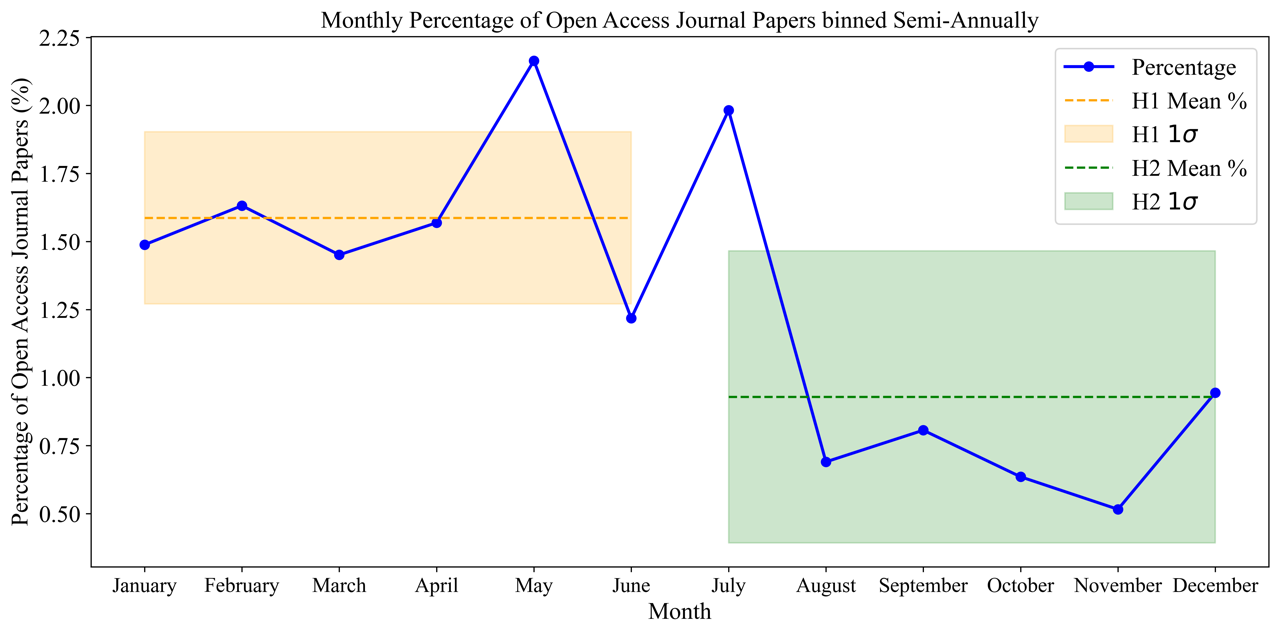}
    \caption{Yearly Track of the Percentage of Open Access Journal Papers. The two bins correspond to the two halves of the year with the dotted line as the mean and shaded regions as the 1$\sigma$ regions. The drop in the second half is not statistically significant and is due to long journal processing times.}
    \label{fig:journal_costs}
\end{figure}

\noindent We estimate that this year, of the 10140 papers that were published in journals, only 231 or 2.28\% of the published papers or 1.24\% of all papers, were published in open access journals. We define open-access journals as those that are free to read and free to publish in, i.e. diamond open-access journals. Figure \ref{fig:journal_costs} tracks the percentage of open journal papers published per month over the year, relative to all the papers uploaded to the arXiv. The plot also includes semi-annual bins for the first and second half of the year, with mean and standard deviation represented. We expected to see an increase in the number of papers published in open access journals in the second half of the year when some journal prices increased. However, we see the opposite trend, though it is not statistically significant. The drop in the second half is probably because most of those papers have been submitted to journals and are still being reviewed or edited before final publication in the coming months. This argument is supported by Figure \ref{fig:journals_monthly_pub} in Section \ref{sec:journals}. This tells us that the increase in journal costs has not yet influenced the community enough to enter a paradigm shift to publishing in open access journals.

\subsection{Location Statistics and Collaboration} \label{sec:location_and_collab}
\begin{figure}
	\includegraphics[width=\columnwidth]{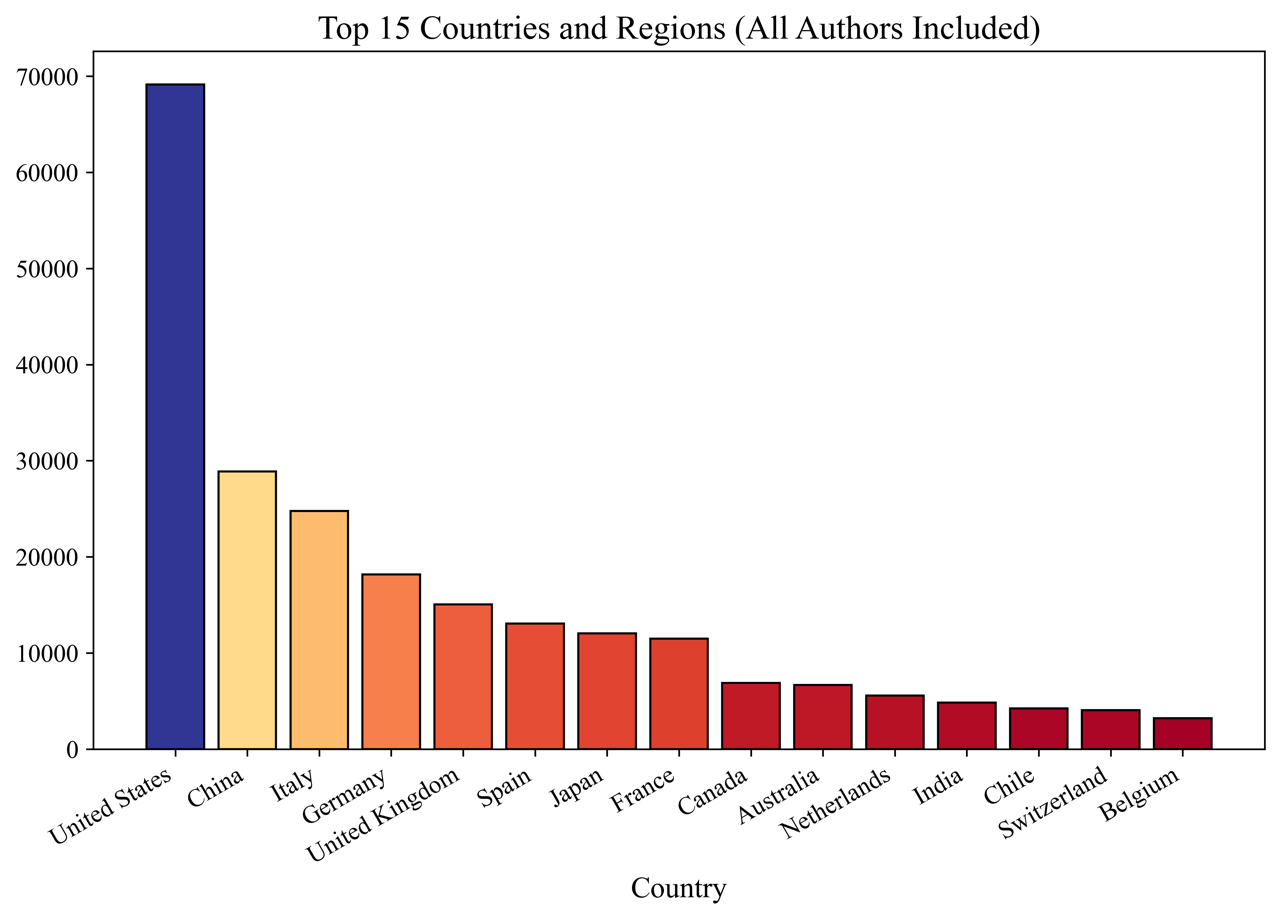}
    \caption{Top 15 Countries or Regions that submitted papers to the arXiv with all authors included. Multiple authors from the same country on the same paper are counted separately.}
    \label{fig:figure7}
\end{figure}

\begin{figure}
	\includegraphics[width=\columnwidth]{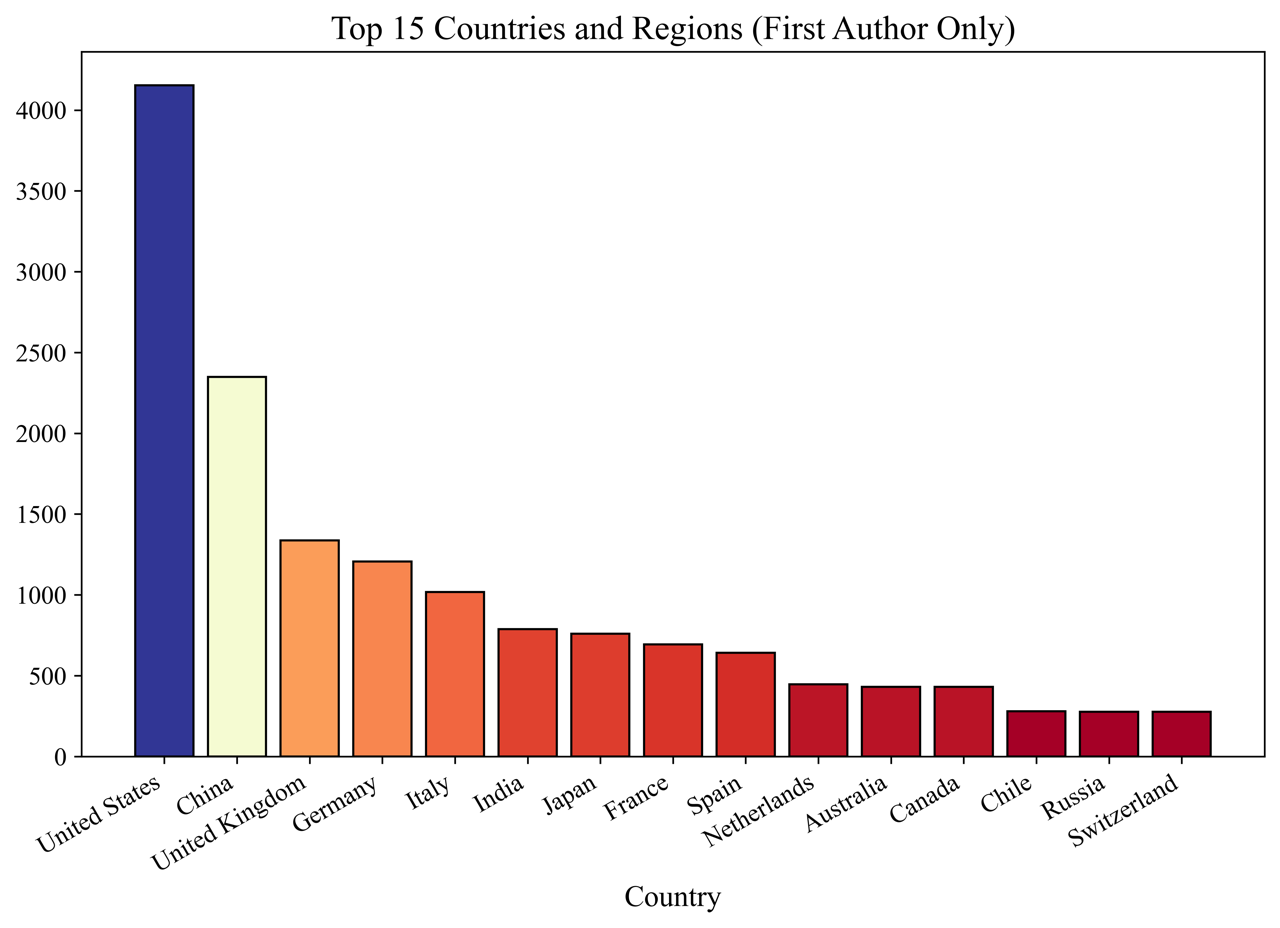}
    \caption{Top 15 Countries or Regions that submitted papers to the arXiv counting only first author papers. First authors with multiple affiliations from the same or different country are counted separately.}
    \label{fig:figure8}
\end{figure}

This section aims to provide some geolocation statistics about the authors as well as the different collaborations and research groups across the world. For this section, we use location data obtained from two different methods as outlined in Section \ref{subsec:data_collection}- using the NASA ADS API and from the LaTeX file submitted to the arXiv. For this section, we are only interested in the country or region name and not the full affiliation, so we rely heavily on finding country or region names in the LaTeX files to fill in location information. As we have done with other methods in this paper, we conduct some random checks and find this LaTeX method to be reliable most of the time, though there are some inconsistencies. We emphasise that these inconsistencies are few and far apart and do not influence any overarching trends we find. Combining these two methods, we are able to get the country or region information for all authors on 97\% of the papers in our dataset. From this, we find that in 2025, the authors who published papers were spread across 129 different countries and regions. Authors from the United States and China represent 26\% and 11\% of all authors publishing across the world. The top 15 countries and regions in terms of author affiliation are given in Figure \ref{fig:figure7}. Considering only first authors, they are spread across 105 different countries and regions, with those from the United States and China representing 23\% and 13\% of the whole sample of first authors, respectively. The top 15 countries and regions by only the first author are given in Figure \ref{fig:figure8}. 

Next, we analyse some basic collaboration statistics, starting with the most pair-wise country or region collaborations. For each paper, we keep count of the number of country or region pairs that appear. For example, a paper with the United States, China and the United Kingdom will have three collaboration pairs- the United States and China, the United States and the United Kingdom, and China and the United Kingdom. For repeated countries or regions we only consider the number of unique country pairs as this gives us the number of papers where any particular country or region pair collaborated rather than individual author pairings. The top 10 country or region collaborations pairs, along with the number of papers with these collaboration pairs, are as follows:

\begin{itemize}
    \item United Kingdom - United States: 2706
    \item Germany - United States: 2425
    \item Italy - United States: 1939
    \item France - United States: 1734
    \item Germany - United Kingdom: 1643
    \item Spain - United States: 1603
    \item Germany - Italy: 1510
    \item Italy - United Kingdom: 1482
    \item Canada - United States: 1461
    \item China - United States: 1306
\end{itemize}

For a more detailed overview, the heat map in Figure \ref{fig:country_heatmap} shows the number of collaborations between the top 15 and top 30 most collaborative countries.

In 2025, 1330 papers were published with a single author. Excluding these, only 730 papers were published with multiple authors, each being from a unique country that another author is not from. On average, every paper published in 2025 had 10 authors. Generally, 1.67 authors were from the same country or region as another author on the paper. To better understand the diversity of collaboration, we provide statistics relating to the collaboration indices we defined in Section \ref{subsec:collab_indices}.

Before we begin the next part, we provide a quick, one-line recap of those collaboration indices:
\begin{itemize}
    \item Local Collaborative Index (LCI) - Number of collaborators from the same country or region as the first author
    \item Local Collaborative Ratio (LCR) - Ratio of LCR to the total number of collaborators on a paper
    \item Global Collaborative Index (GCI) - Number of collaborators from a country or region different from that of the first author 
    \item Non-repeated Global Collaborative Index (NGCI) - Number of unique countries or regions involved in a paper other than the first author's
\end{itemize}

The median LCI taken over the entire dataset is 4.00, while the average is 6.48, skewed most likely by the papers from large collaborations or teams. When compared with the average number of authors on a paper (10), this tells us that most papers are published with around 60\% or almost a two-thirds majority of collaborators from the same country or region as the first author. The median LCR for all 2025 papers is 0.86, which corroborates the previous statement that research is typically very domestic, and the average LCR of 1.03 indicates a very small amount of international collaboration. We note here that the LCR can have a value greater than 1 since some authors can have multiple affiliations. The median GCI is 1.00, meaning that the typical paper only 1 foreign author or collaborator, where `foreign' here is used merely to indicate a difference from the first author's country. The average GCI is 6.64, skewed again, probably owing to the big collaborations which have a large number of authors from a country other than the first author's country. The NGCI median value of 1.00 means that papers usually involve only one foreign country or authors from a country other than the first author's country. In a sense, research usually takes place between two countries or regions. The average NGCI is 1.91, demonstrating that even in the large collaborations, which skew the statistic, these large teams will have authors from only around 2 countries or regions. 

The median statistic essentially represents the typical paper, which comprises a small number of collaborators, most of whom are from the same region as the first author, with at least 1 author from a foreign country. These papers have a bilateral collaboration. The average statistic for the indices is distorted by the large teams or mega-collaborations, which have a large number of authors on a single paper. We note that the average GCI is slightly higher than the average LCI, suggesting that in these mega collaborations, the foreign authors dominate and outnumber the local authors, contrary to what is seen for the smaller collaborations.

Since we mention these large collaborations, we report some number statistics to give some context to the frequency of such big-team-papers in 2025. There were 1093 papers from large teams (defined as having more than 30 authors) contributing to 5.86\% of the overall dataset. If we consider a less stringent cut-off of 20 authors, this number almost doubles to 1979 or 10.61\% of the dataset. On the opposite end, there are 10001 papers (53.60\%) that have 5 or less authors, clearly pointing to the dominance of small teams in astrophysical research.

We further utilise these collaboration indices to understand how collaboration trends differ by geolocation. The top 10 most locally collaborative countries and regions by number of authors (highest LCI) are:
\begin{itemize}
    \item Mauritania: 127.00
    \item China: 73.15
    \item Morocco: 70.81
    \item Thailand: 68.52
    \item Algeria: 58.94
    \item Georgia: 48.22
    \item Burkina Faso: 47.19
    \item Belgium: 43.59
    \item Armenia: 40.84
    \item Germany: 38.14
\end{itemize}

The top 10 most locally collaborative countries and regions by ratio of local authors to total authors (highest LCR) are:
\begin{itemize}
\item China: 35.82
\item Thailand: 33.73
\item Armenia: 18.77
\item Hong Kong: 15.31
\item Russia: 14.24
\item Mayotte: 13.95
\item Morocco: 11.85
\item Algeria: 11.64
\item Romania: 7.62
\item Georgia: 6.67
\end{itemize}

The top 10 most internationally collaborative countries and regions (highest GCI) are:
\begin{itemize}
\item Mali: 1078.33
\item Burkina Faso: 904.00
\item Bosnia and Herzegovina: 477.00
\item Albania: 477.00
\item Taiwan: 364.37
\item Georgia: 351.41
\item Italy: 342.19
\item Qatar: 331.26
\item Belgium: 315.34
\item Netherlands: 312.56
\end{itemize}

The top 10 most internationally diverse collaborative countries and regions (highest NGCI) are:
\begin{itemize}
\item Bosnia and Herzegovina: 57.00
\item Albania: 57.00
\item Qatar: 34.48
\item Mauritania: 19.00
\item Philippines: 17.39
\item Mayotte: 17.30
\item Mali: 16.42
\item Romania: 16.31
\item Burkina Faso: 15.62
\item Malta: 15.20
\end{itemize}

\begin{figure*}
    \centering
    \begin{subfigure}{\columnwidth}
        \centering
        \includegraphics[width=\columnwidth]{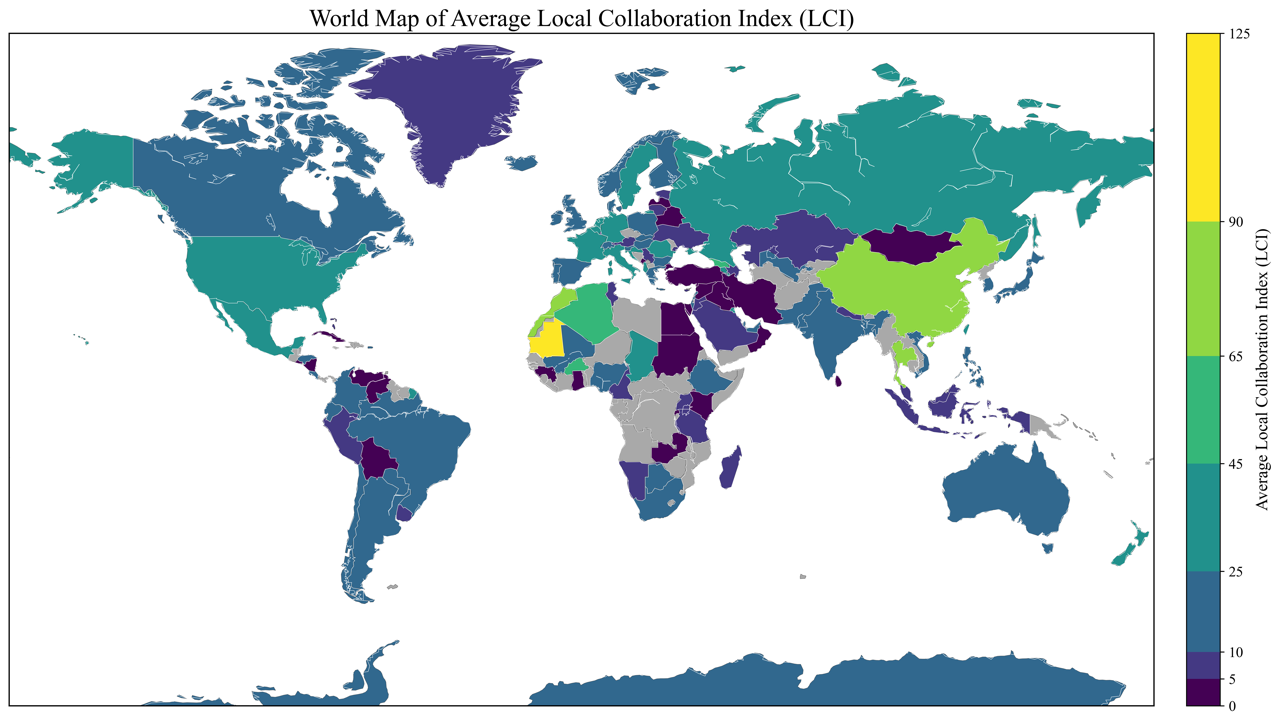}
        \caption{LCI Global Distribution}
        \label{fig:lci_map}
    \end{subfigure}
    \hfill 
    \begin{subfigure}{\columnwidth}
        \centering
        \includegraphics[width=\columnwidth]{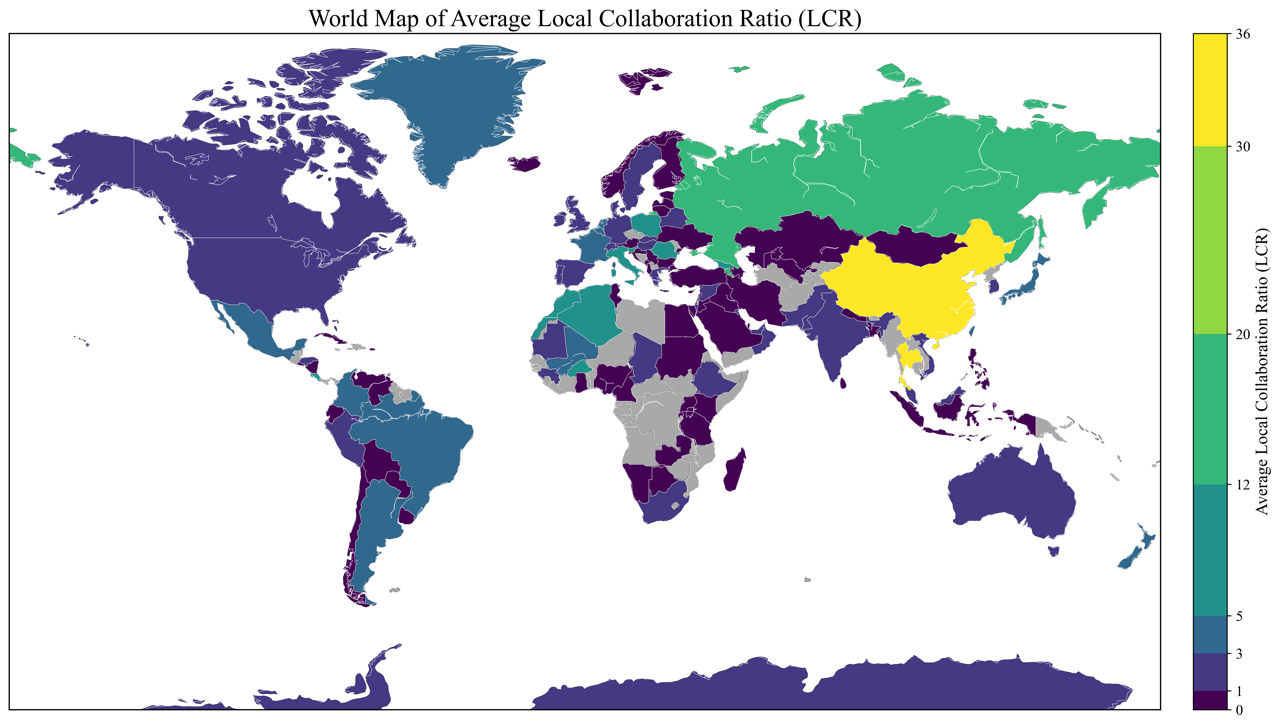}
        \caption{LCR Global Distribution}
        \label{fig:lcr_map}
    \end{subfigure}
    \hfill
    \centering
    \begin{subfigure}{\columnwidth}
        \centering
        \includegraphics[width=\columnwidth]{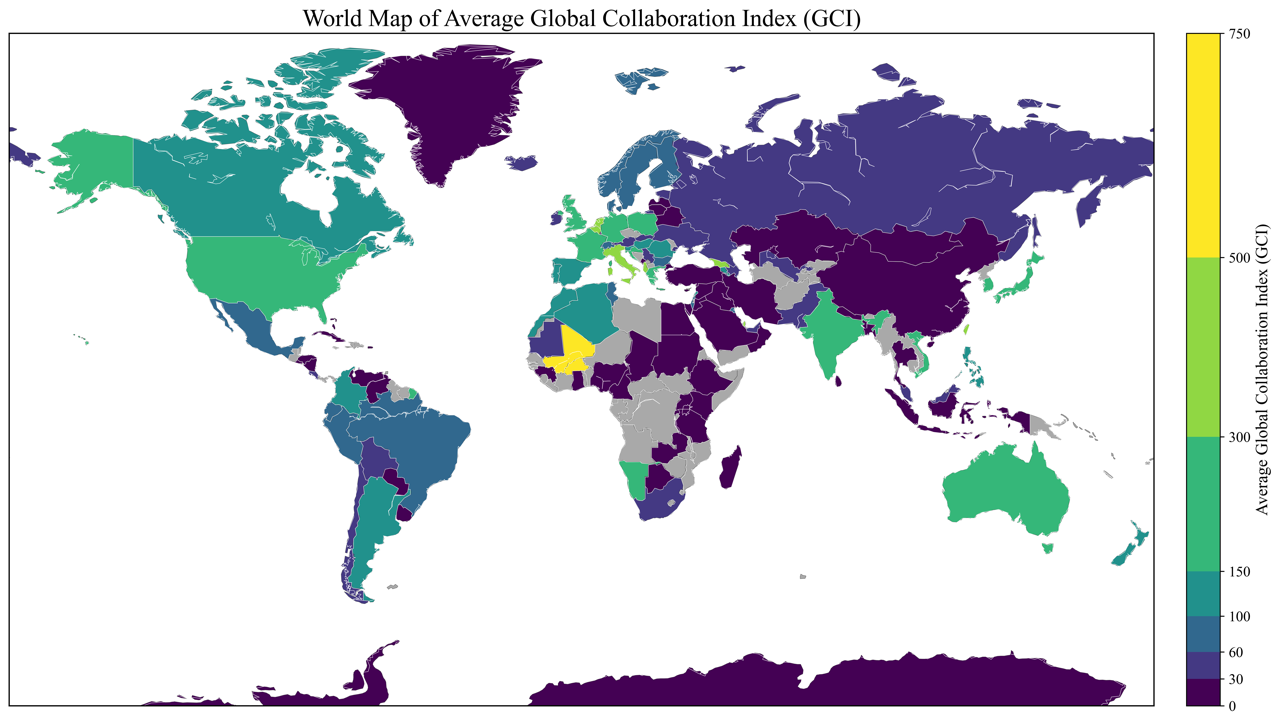}
        \caption{GCI Global Distribution}
        \label{fig:gci_map}
    \end{subfigure}
    \hfill
    \begin{subfigure}{\columnwidth}
        \centering
        \includegraphics[width=\columnwidth]{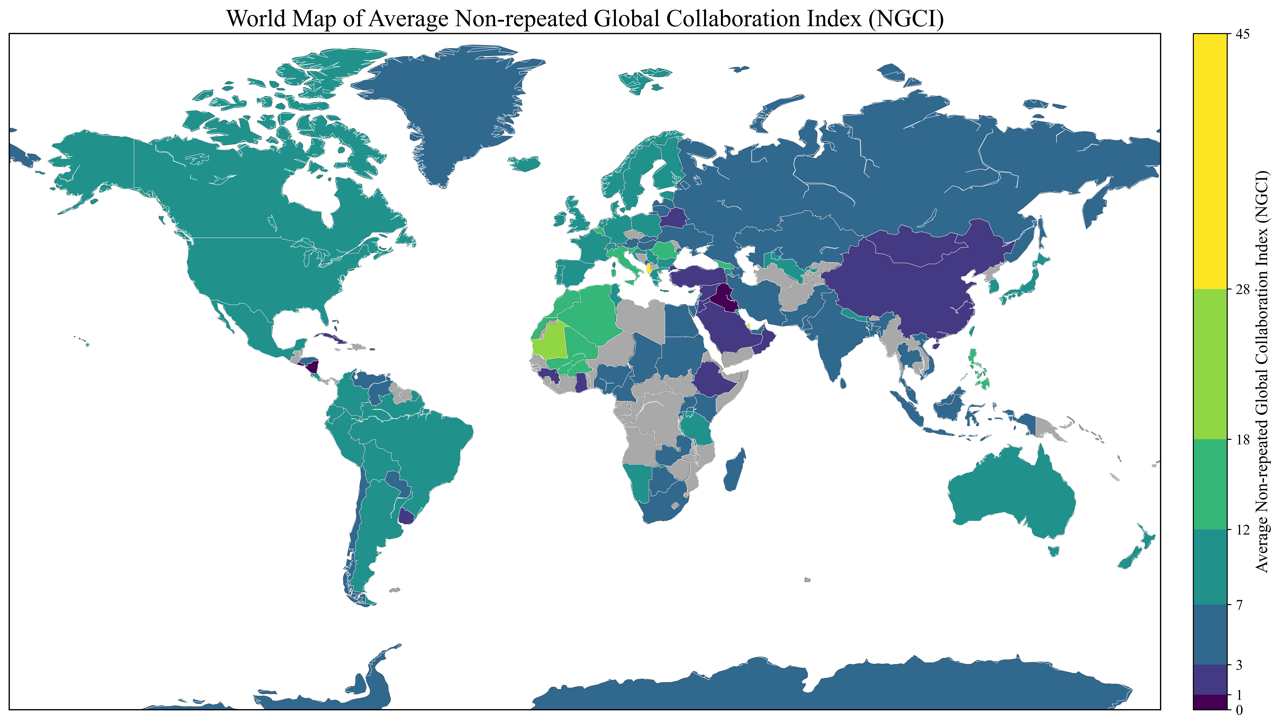}
        \caption{NGCI Global Distribution}
        \label{fig:ngci_map}
    \end{subfigure}
    
    \caption{Global Distribution of Collaboration Indices}
    \label{fig:figure_colab_indices}
\end{figure*}

From the above lists, we see that many of the countries with the most submitted papers (with the exception of China) do not appear. One way to understand this is that those countries or regions, due to their large volume of papers, fall in between being highly locally collaborative and highly globally collaborative. For the high index countries or regions in the four lists above that seem atypical, some dominate due to small number statistics. For example, we find some fairly high statistics for smaller countries (in terms of astrophysics research funding and output) like Burkina Faso and Mali. We interpret this as authors from these countries being present on some large collaboration teams and being attributed the high collaborative indices that are not diluted down (as happens with the more published countries) owing to lesser number of overall publications. We checked for this and found an author from Burkina Faso on several very large collaboration papers (specifically a LIGO collaboration paper on a binary black hole merger), thus confirming our hypothesis. Despite these outliers, we find these statistics very informative and hence choose to report them as is. For a more general view of the distribution of collaboration indices across the world, we show maps with the four indices in Figures \ref{fig:lci_map}, \ref{fig:lcr_map}, \ref{fig:gci_map}, \ref{fig:ngci_map}. 

\begin{figure*}
	\includegraphics[width=\linewidth]{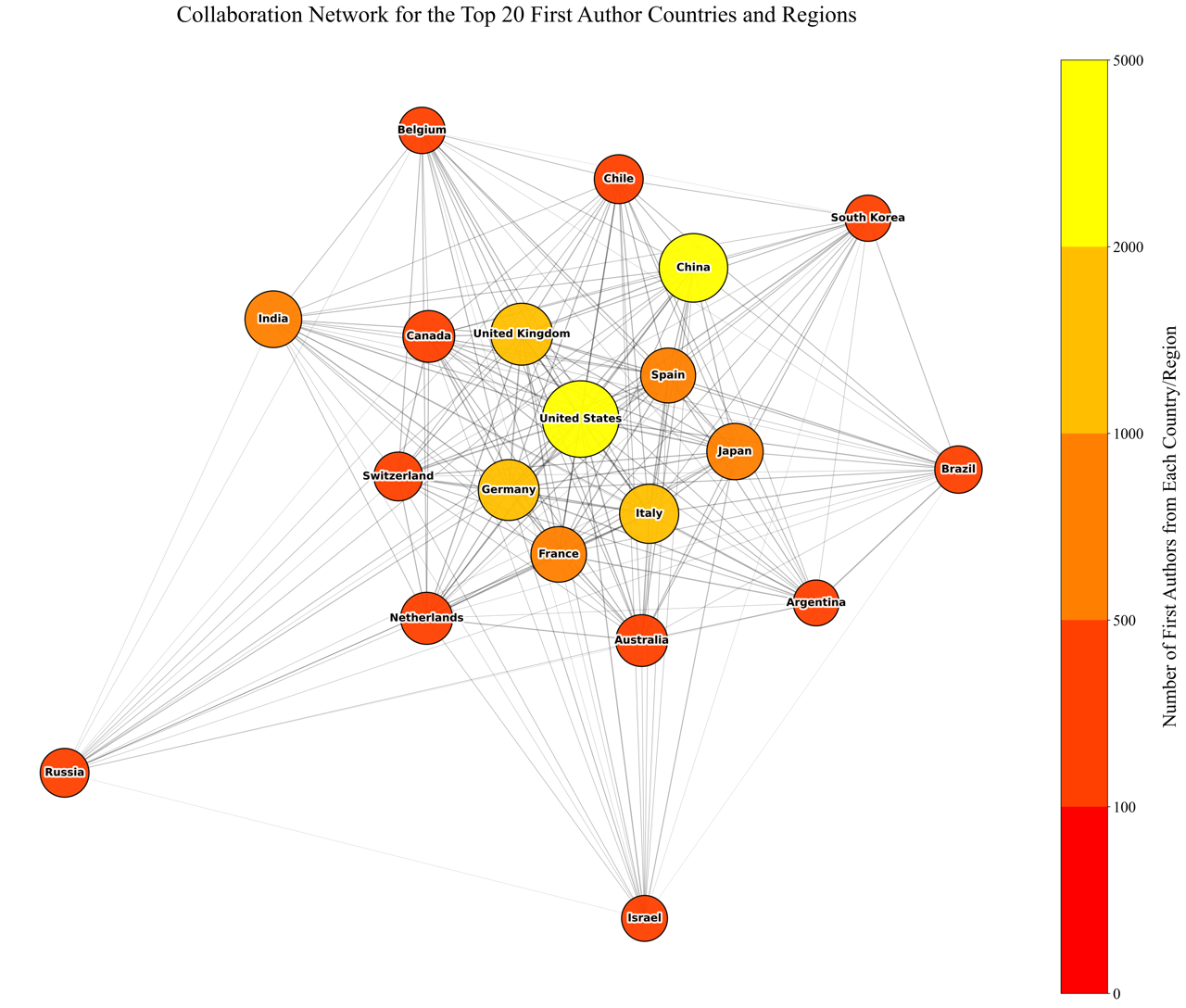}
    \caption{Global Network of Collaboration (Top 20 First Author Countries or Regions)}
    \label{fig:collab_network}
\end{figure*}

One plot we thought would be interesting to show is the global collaboration network. To do so, we create a graph with each node as one of the top 20 first author countries or regions. The nodes are interconnected by the authors that are affiliated with regions other than the first author's region. Figure \ref{fig:collab_network} shows this graph.

Now we can look at some more typical metrics like primary subject which we discussed earlier in the paper through the lens of collaboration indices. The full range of statistics for each primary subject is given in Table \ref{tab:colb_primary_sub}. While we cannot comment on every trend in the table, we point out some prominent ones. We find that across all indices the variation in the primary subject collaboration statistics is within $1\sigma$ of the mean for the whole sample. Despite this, we would like to discuss these differences as we believe the discussion offers some insights into the meaning of the collaboration indices.

The most internationally collaborative primary subject (highest average GCI) is Astrophysics, most likely due to the fact that the big survey teams and mega-collaborations publish under this subject. This is supported by the fact that it is also the subject with the largest number of authors per paper by some margin. The most locally collaborative subject is High Energy Astrophysical Phenomena (the highest average LCI). Our best guess for this is that the research in this field involves using local instrumentation and data, which is naturally studied more by local groups. The least globally collaborative subject is Solar and Stellar Astrophysics, with the lowest GCI and the lowest number of authors per paper, indicating that the teams involved in solar physics tend to be smaller, localised teams. There are also some interesting trends highlighted by the LCR. As mentioned before, an LCR $\gt$ 1 is possible if authors have multiple affiliations. High Energy Astrophysics has both a high LCI and a high LCR ($\gt$ 1). This tells us that the researchers working under this subject are locally well networked, as they have a large number of local collaborators (high LCI) and are affiliated with a large number of local institutions (high LCR). Papers written under Astrophysics of Galaxies tend to be the most international, as they have the lowest LCR or the highest ratio of international to local collaborators. This is easily explainable considering that research relating to galaxies typically involves data from multiple surveys or resources from a large number of different groups across the globe. Finally, the Cosmology and Nongalactic Astrophysics and Solar and Stellar Astrophysics groups are very similar across most indices, especially LCR, meaning the groups work in similar ratios of local to international collaborators despite studying on opposite ends of the astrophysical distance scale.

\begin{table*}
\resizebox{\textwidth}{!}{%
\begin{tabular}{lccccc}
\hline
\textbf{Primary Subject}                              & \textbf{LCI}  & \textbf{LCR}  & \textbf{GCI}   & \textbf{NGCI} & \textbf{Avg Authors$^{\dagger}$} \\
\hline
Astrophysics of Galaxies                     & 6.60 & 0.82 & 6.50  & 2.43 & 11.33                 \\
High Energy Astrophysical Phenomena          & 7.84 & 1.51 & 10.93 & 2.06 & 10.05                 \\
Instrumentation and Methods for Astrophysics & 7.52 & 1.14 & 5.64  & 1.47 & 10.99                 \\
Solar and Stellar Astrophysics               & 5.56 & 0.90 & 3.69  & 1.67 & 7.55                  \\
Cosmology and Nongalactic Astrophysics       & 4.61 & 0.90 & 5.02  & 1.51 & 9.36                  \\
Earth and Planetary Astrophysics             & 6.59 & 0.85 & 6.55  & 2.06 & 10.65                 \\
Astrophysics                                 & 7.04 & 1.20 & 17.86 & 3.35 & 20.91                \\
\hline
\end{tabular}}
\caption{Collaboration Indices across Primary Subjects ($\dagger$ - per Paper)}
\label{tab:colb_primary_sub}
\end{table*}

Keywords are another interesting metric we can compare collaborative indices against. We dedicate a separate section to keyword statistics relative to other metrics later on (section \ref{sec:keywords}), but we give a brief discussion here. We find that the top 30 most-used keywords all have collaborative indices within one standard deviation of the mean of the whole sample and hence show no significant difference. This conclusion remains valid irrespective of if we include self-assigned keywords (for more details, refer to section \ref{sec:keywords}) or not. Even when considering subfields, another metric we talk about later in sub-section \ref{sec:subfields}, this conclusion remains the same with the exception of the sub-field `Multi-Messenger Astronomy', which has an LCI of 27 (1.4$\sigma$ from the mean), indicating that `Multi-Messenger Astronomy' papers tend to have a large number of local collaborators. 

Next, looking towards journal metrics, we can check the collaboration indices for papers that are published versus those that are not and also compare across papers published in different journals. When comparing journal versus non-journal published papers, we find the mean across the two samples to be within 1$\sigma$ of the mean of the entire sample for all four indices. We also calculate the indices for all papers from each of the top 10 most published journals in our dataset and find no variation across journals that exceeds 1$\sigma$ of the mean for any of the four indices. 

We also do a comparison with citation metrics and find no significant correlation between any of the collaboration indices and citation metrics (defined in \ref{methods}). We also find no variation in collaboration indices outside 1$\sigma$ of the mean when comparing these indices between papers published in journals versus those that are not published. This does not change when the sample includes papers that have been submitted for publication but have not yet been accepted for publication. 

From the telescopes associated with each paper, we calculate the collaboration indices for each telescope. All telescopes have their collaboration indices within or barely exceeding one standard deviation but are all within two times the standard deviation. Hence, there is no significant difference in the groups or collaborations that work on different telescopes. 

From all of this, we can conclude that the type of collaboration does not change with the broad category, specific subject matter or citation score of a paper. Generally, all telescopes, subjects and subfields are being equally worked on by both small and large, local and global collaborators and groups without any bias. A simple and elegant way to demonstrate this is to check if any of these metrics (primary subject, keywords, etc) differ by the total number of authors per paper. We find that across all metrics, the average number of authors per paper is within one standard deviation of the mean, confirming that across all these metrics, there are both mega-collaborations and small, focused teams working. We welcome this conclusion, considering it implies that the split between big groups and small teams working on any given area of the field is consistent across the field, though all areas may not have the same number of people working in them. This means all subjects and sub-fields within the field are being equally investigated by individuals and small groups as well as large multinational teams. Irrespective of if an individual prefers working in a smaller group or as part of a large team, no sub-field is out of reach. We just need to find the right people.

Following this, one piece of information we thought would be helpful, was to give readers an idea about the best places to do a specific kind of research or where in the world are the highest concentrations of collaborators working on a specific area. To do this, we use the keywords and subfields and find the research specialisation for all 129 unique countries that astrophysics authors come from. Here, it is important to keep in mind that these are recommendations based solely on number statistics and do not reflect the quality of the research output. The full list is provided in the Appendix. 

\subsection{Institutions} \label{sec:institution}
Given that we have affiliations for a large number of authors, we can report institution-related number statistics. Before we delve into them, we would like to reiterate that the point of this paper is not to encourage unhealthy competition or put down institutions, but rather to highlight achievement and the positives of institutions that promote Astrophysical research. We made a conscious decision not to analyse citation scores of institutions, though the data is available for us to do so. This section is only dedicated to showing the community what institutions are contributing the most positively and should not be interpreted in a negative way. We have purposefully avoided presenting metrics that we believe have a negative tone or implication. 

Owing to this reasoning, the only statistic we provide in this section is the list of institutes that produced the most number of papers in 2025. We count the contributions of multiple authors from the same institute on the same paper as separate contributions for the institute. The top three institutes this year were the Chinese Academy of Sciences, the University of California and the Harvard \& Smithsonian Centre for Astrophysics. The full list of the top 20 institutes and their metrics is given in table \ref{tab:table_institutes}. 

\begin{table*}
\resizebox{\textwidth}{!}{
\begin{tabular}{llccccclc}
\hline
\textbf{Institute} & \textbf{Top Primary Subject} & \multicolumn{1}{c}{\textbf{Count}} & \multicolumn{1}{c}{\textbf{LCI}} & \multicolumn{1}{c}{\textbf{LCR}} & \multicolumn{1}{c}{\textbf{GCI}} & \multicolumn{1}{c}{\textbf{NGCI}} & \multicolumn{1}{c}{\textbf{Top Keyword}} & \multicolumn{1}{c}{\textbf{Top Telescope}} \\
\hline
\hline
Chinese Academy of Sciences & High Energy Astrophysical Phenomena & 7837 & 67.38  & 26.60  & 15.40  & 2.68 & Galaxy Clusters & FAST \\
University of California & Astrophysics of Galaxies & 3209 & 22.89  & 1.19   & 56.83  & 4.88 & Supernovae & JWST \\
Harvard \& Smithsonian Center for Astrophysics & Astrophysics of Galaxies & 2848 & 26.59  & 4.79   & 58.74  & 6.06   & Accretion & JWST \\
California Institute of Technology & High Energy Astrophysical Phenomena & 2799 & 32.30  & 2.14   & 262.87 & 6.93 & Exoplanets & JWST \\
University of Chinese Academy of Sciences & High Energy Astrophysical Phenomena & 2113 & 55.75  & 21.32  & 11.15  & 1.99 & Surveys & GAIA \\
University of Tokyo & High Energy Astrophysical Phenomena & 2097 & 29.52  & 1.84   & 247.45 & 8.58 & High-redshift Galaxies & JWST \\
University of Science and Technology of China & High Energy Astrophysical Phenomena & 1705 & 48.11  & 17.43  & 14.03  & 1.84 & Active Galactic Nuclei & JWST \\
Space Telescope Science Institute & Astrophysics of Galaxies & 1676 & 16.11  & 0.71   & 15.60  & 4.61 & High-redshift Galaxies & JWST \\
University of Arizona & Astrophysics of Galaxies & 1604 & 16.66  & 0.95   & 47.34  & 4.98 & High-redshift Galaxies & JWST \\
University of Cambridge & Astrophysics of Galaxies & 1407 & 12.71  & 0.85   & 51.61  & 5.35 & Reionization & JWST \\
University of Chicago & Cosmology and Nongalactic Astrophysics & 1255 & 30.26  & 1.18   & 72.62  & 6.10 & Surveys & JWST \\
NASA Goddard Space Flight Center & High Energy Astrophysical Phenomena & 1182 & 17.00  & 1.01   & 38.64  & 4.20 & Micro-lensing & JWST \\
Leiden University & Astrophysics of Galaxies & 1140 & 8.33   & 0.84   & 23.94  & 5.39 & Protoplanetary Disks & JWST \\
Peking University & Astrophysics of Galaxies & 1061 & 21.49  & 4.94   & 26.01  & 3.47 & Active Galactic Nuclei & JWST \\
National Astronomical Observatory of Japan & Astrophysics of Galaxies & 1049 & 20.50  & 1.49   & 137.57 & 6.11 & Galaxy Evolution & ALMA \\
University of Maryland & High Energy Astrophysical Phenomena & 1043 & 31.45  & 2.30   & 96.79  & 6.24 & Neutrino Astronomy & XRISM \\
Ohio State University & Cosmology and Nongalactic Astrophysics & 1038 & 27.52  & 0.81   & 54.05  & 6.49 & Galactic Winds & DESI \\
University of Oxford & Astrophysics of Galaxies & 983 & 11.04  & 0.63   & 22.48  & 5.27 & Accretion & JWST \\
Nanjing University & High Energy Astrophysical Phenomena & 978 & 46.02  & 18.59  & 10.04  & 2.62 & Supernova Remnants & Fermi \\
Pennsylvania State University & Astrophysics of Galaxies & 913 & 34.85  & 1.94   & 214.25 & 7.23 & High-redshift Galaxies & JWST \\  
\hline
\end{tabular}
}
\caption{Statistics of the Top 20 Institutes that contributed the most number of papers in 2025}
\label{tab:table_institutes}
\end{table*}

\subsection{Keywords} \label{sec:keywords}
In this section, we will analyse keywords specified for each paper. In our dataset, 12753 papers or 68\% of the papers already have keywords assigned by the author's during submission. For the papers that already contain keywords, we simply extract them as is. As a reminder, for papers without keywords, we assign three keywords based on matching words in the title and abstract to a predefined list of approximately 1000 of the most frequently used keywords from the already assigned papers. The three most frequently matched words are taken as keywords. While conducting our analysis, we realised that a few trends change when we include our self-assigned keywords. For future reference, the keywords assigned by our matching technique will be referred to as self-assigned keywords. We examine the trends in keywords both before and after the addition of self-assigned keywords. 
\\\\
We find that `accretion' is the most used keyword of the year by some margin, both with and without the self-assigned keywords. In second place is `galaxy evolution' followed by `active galactic nuclei', even when we add in our own keywords re-emphasising the community's focus on galaxies. `Exoplanets', `star formation' and `cosmology' are also common highly used keywords. We see that black holes are also highly studied by the community owing to the fact that both `supermassive black holes’ and `black holes’ are among the top keywords.

\begin{figure*}
    \centering
    \begin{subfigure}{\columnwidth}
        \centering
        \includegraphics[width=\columnwidth]{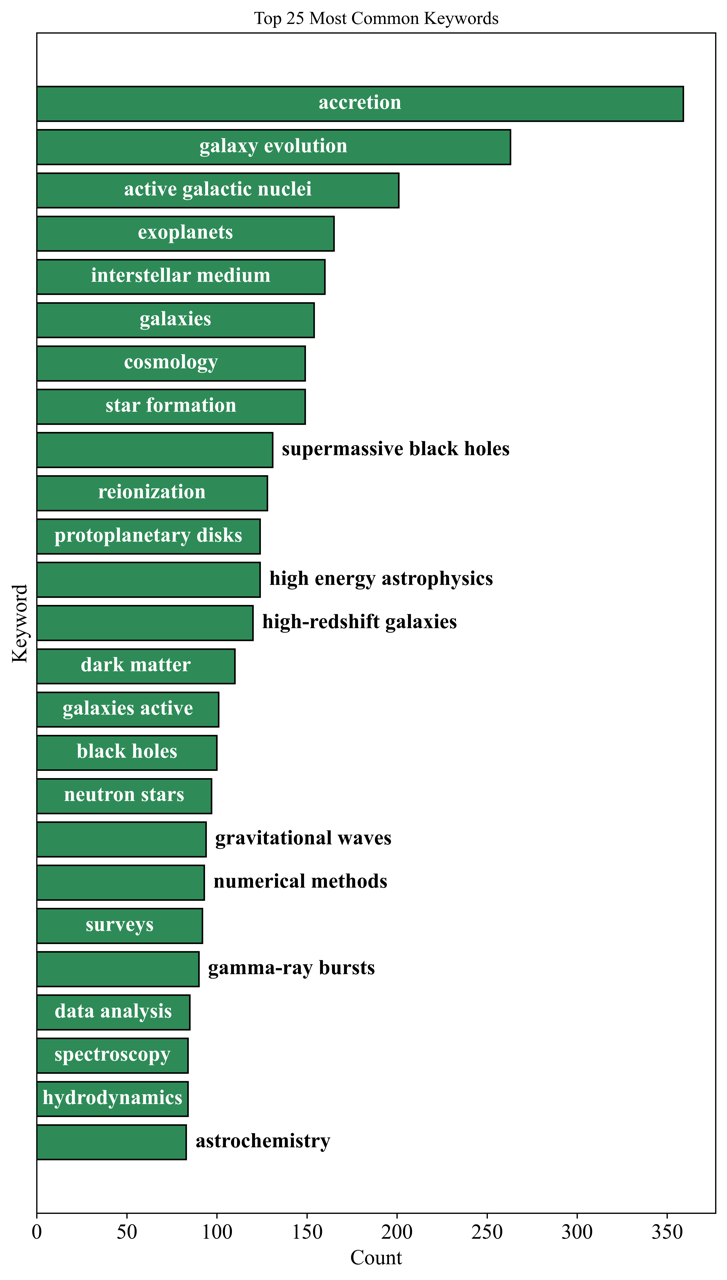}
        \caption{Without self-assigned keywords}
        \label{fig:keywords_a}
    \end{subfigure}
    \hfill 
    \begin{subfigure}{\columnwidth}
        \centering
        \includegraphics[width=\columnwidth]{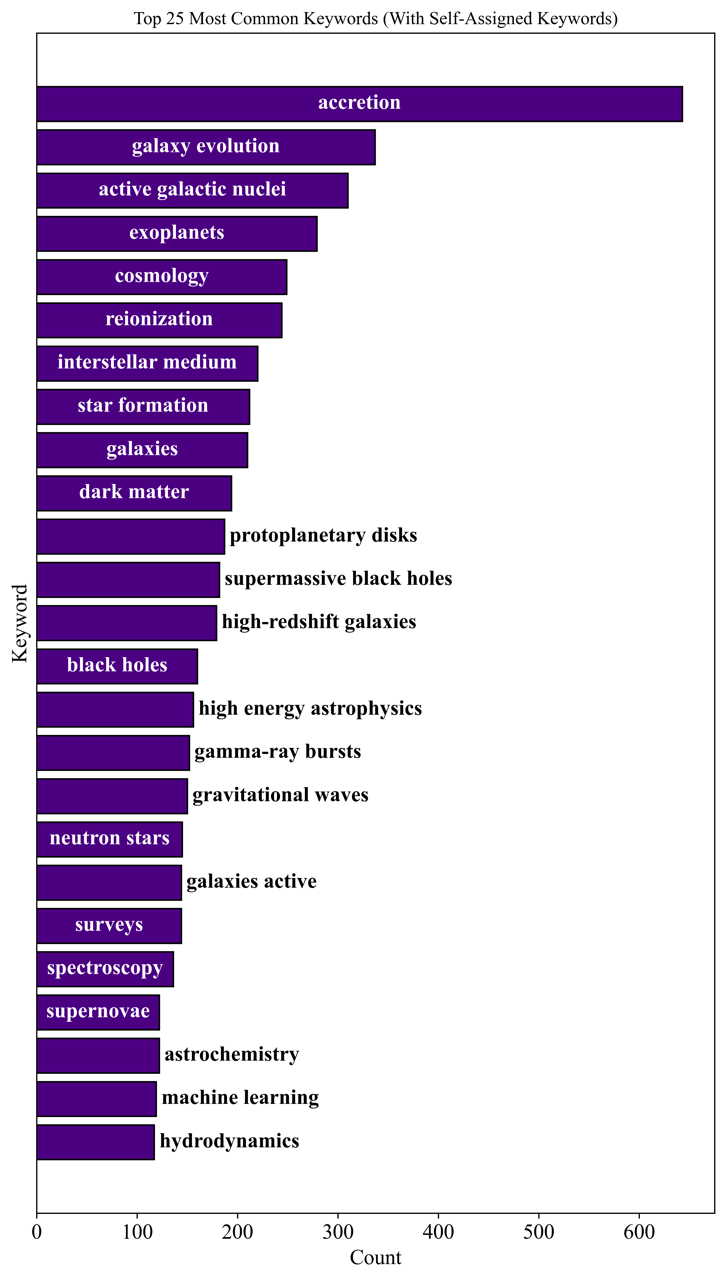}
        \caption{With self-assigned keywords}
        \label{fig:keywords_b}
    \end{subfigure}
    
    \caption{Most Common Keywords for 2025}
    \label{fig:figure2_keywords}
\end{figure*}

As we have done with other metrics, we can see if there are any trends in the keywords when compared with the citations. This is done by counting the average citation indices per paper for all papers where a given keyword is mentioned. We find that `high-redshift galaxies' is the keyword with the most citations per paper (AAC) at 7.08, followed by `reionization' at 4.83 citations per paper. The full list of values is reported in Table \ref{tab:table3}. Figures \ref{fig:keywords_a} and \ref{fig:keywords_b} show the top 25 most commonly used keywords both with and without the self-assigned keywords.

\begin{table}
\centering
\resizebox{\columnwidth}{!}{%
\begin{tabular}{lcccc}
\hline
\textbf{Keyword}                  & \textbf{AAC}  & \textbf{EAAC} & \textbf{JAC} & \textbf{EJAC} \\
\hline
\hline
High-redshift galaxies   & 7.08 & 4.55 & 8.87 & 6.05 \\
Reionization             & 4.83 & 3.03 & 6.92 & 4.36 \\
Supermassive Black Holes & 3.78 & 2.48 & 4.19 & 2.78 \\
Galaxy Evolution         & 3.59 & 2.25 & 3.84 & 2.48 \\
Black Holes              & 3.53 & 2.39 & 3.86 & 2.60 \\
Cosmology                & 3.08 & 2.10 & 4.17 & 2.88 \\
Protoplanetary Disks     & 3.02 & 1.48 & 2.52 & 1.44 \\
Galaxies                 & 2.98 & 1.89 & 3.30 & 2.15 \\
Surveys                  & 2.96 & 1.76 & 3.55 & 2.20 \\
Galaxies Active          & 2.77 & 1.70 & 2.84 & 1.73 \\
Numerical Methods        & 2.40 & 1.51 & 2.92 & 1.90 \\
Star Formation           & 2.40 & 1.60 & 2.60 & 1.78 \\
Gravitational Waves      & 2.39 & 1.72 & 2.75 & 1.97 \\
Spectroscopy             & 2.38 & 1.39 & 2.64 & 1.67 \\
Gamma-ray Bursts         & 2.31 & 1.54 & 2.55 & 1.63 \\
Neutron Stars            & 2.28 & 1.75 & 2.34 & 1.74 \\
Astrochemistry           & 2.22 & 1.50 & 2.16 & 1.47 \\
Dark Matter              & 2.20 & 1.42 & 3.00 & 1.97 \\
Exoplanets               & 2.14 & 1.51 & 2.75 & 1.96 \\
Accretion                & 2.08 & 1.43 & 2.24 & 1.50 \\
Interstellar Medium      & 2.01 & 1.30 & 2.09 & 1.36 \\
Hydrodynamics            & 1.96 & 1.37 & 2.27 & 1.60 \\
Data Analysis            & 1.88 & 1.21 & 1.95 & 1.30 \\
Active Galactic Nuclei   & 1.84 & 1.21 & 1.97 & 1.26 \\
High Energy Astrophysics & 1.84 & 1.42 & 2.36 & 1.75 \\
\hline
\end{tabular}}
\caption{Citation Indices for all Papers containing a certain keyword (Top 25 Keywords (excluding assigned-keywords))}
\label{tab:table3}
\end{table}

\subsection{Sub-Fields and Research Areas} \label{sec:subfields}
\begin{figure}[t]
	\includegraphics[width=\columnwidth,height=13cm]{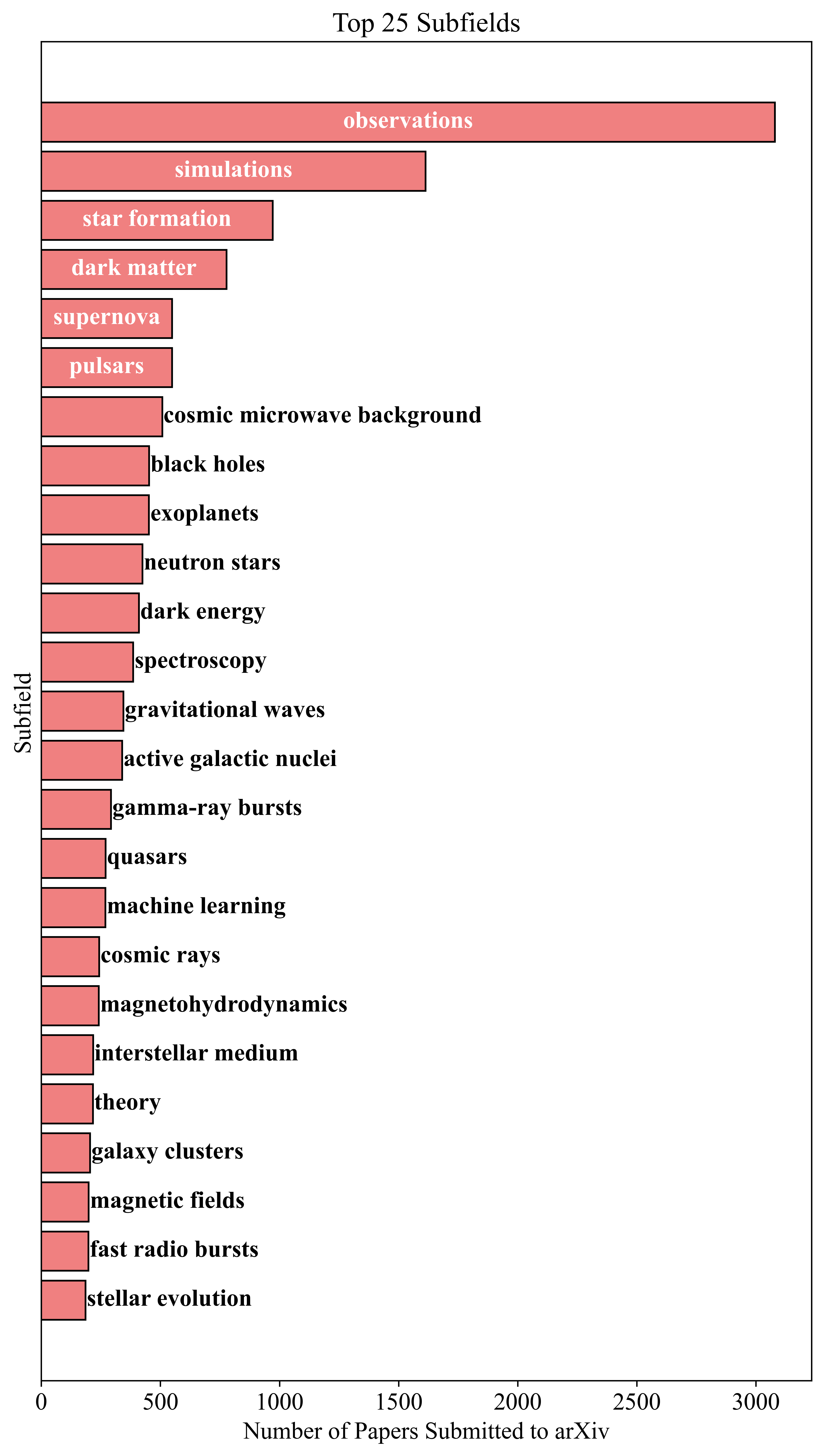}
    \caption{Top 25 sub-fields and research areas. Distribution of sub-fields and research areas based on paper titles and abstracts.}
    \label{fig:subfields}
\end{figure}

Although other sections have covered certain aspects of the type of research the community has engaged in this year, we dedicate this sub-section to discussing trends in relation to more specific sub-fields and research areas. Just as we have done before, we create a list of about 100 sub-fields and apply string matching to the words in the title and abstract. Each paper is assigned one subfield based on the most frequently occurring word from our list. Since sub-fields are based entirely on string matching, they more so represent what the community is talking about or rather writing about in their papers. The title and abstract are the first thing one reads when opening a new paper. Hence, authors should only include what they deem to be the most relevant details or the most precise description of their work in the title and abstract. Thus, we feel this sub-field analysis based only on string matching is justified. The top 10 fields for the year are shown in Figure \ref{fig:subfields}. `Observations' was the largest research area, followed by `Simulations', `Star Formation' and `Dark Matter'. While this is not strong enough evidence to conclude that the community relies more on observations than simulations for their work it, does give us some sense of how much of current analysis uses simulated data. Further down the list, we see supernova, pulsars and exoplanets. We purposefully avoided using 'accretion' as a sub-field and try to target the accretion-related events or astronomical objects, which is why we see no mention of 'accretion' but instead of 'black holes' in the top fields.

We also test for significant deviation in the number of pages, tables and figures across different subfields by checking if the mean number of pages per paper is within 1$\sigma$ of the overall mean. We find no significant deviation between the number of pages, tables or figures across sub-fields. While the papers associated with the sub-field `Cosmic Reionization' have pages outside one standard deviation, it is only marginal ($1.1 \sigma$ deviation), and hence we do not consider it significant.
\\\\
Following this, we also analyse the citation metrics for the different research areas. `Scalar-Tensor Gravity' has the most citations per paper across all citation indices. This is most likely due to the fact that there are a small number of highly cited (excluding self-citations) papers. Other highly cited fields include
`Dark Energy' and `Cosmic Microwave Background'. The most cited fields are consistent across all citation indices, though there is some shuffling. Table \ref{Tab:table5} shows the top 10 most highly cited subfields.

\begin{table}
\centering
\resizebox{\columnwidth}{!}{%
\begin{tabular}{lcccc}
\hline
\textbf{Subfield}                    & \textbf{AAC}  & \textbf{EAAC} & \textbf{JAC}   & \textbf{EJAC}  \\
\hline
\hline
Scalar-Tensor Gravity       & 11.5 & 11.5 & 23.0  & 23    \\
Dark Energy                 & 8.59 & 7.25 & 9.63  & 8.07  \\
Cosmic Microwave Background & 7.89 & 6.70 & 15.57 & 13.29 \\
Cosmic Reionization         & 6.82 & 4.64 & 4.0   & 2.25  \\
Inflation                   & 5.44 & 4.85 & 8.67  & 7.70  \\
Early Universe              & 4.61 & 3.25 & 6.61  & 4.97  \\
Epoch of Reionization       & 4.11 & 2.67 & 7.07  & 4.48  \\
Hubble Tension              & 4.05 & 3.02 & 4.44  & 2.33   \\
Black Holes                 & 3.83 & 2.82 & 4.17  & 3.06  \\
Galaxy Formation            & 3.58 & 2.04 & 3.37  & 2.10  \\
\hline
\end{tabular}}
\caption{Top 10 Subfields by Citation Indices}
\label{Tab:table5}
\end{table}

\subsection{Journals} \label{sec:journals}
\begin{figure}
	\includegraphics[width=\columnwidth]{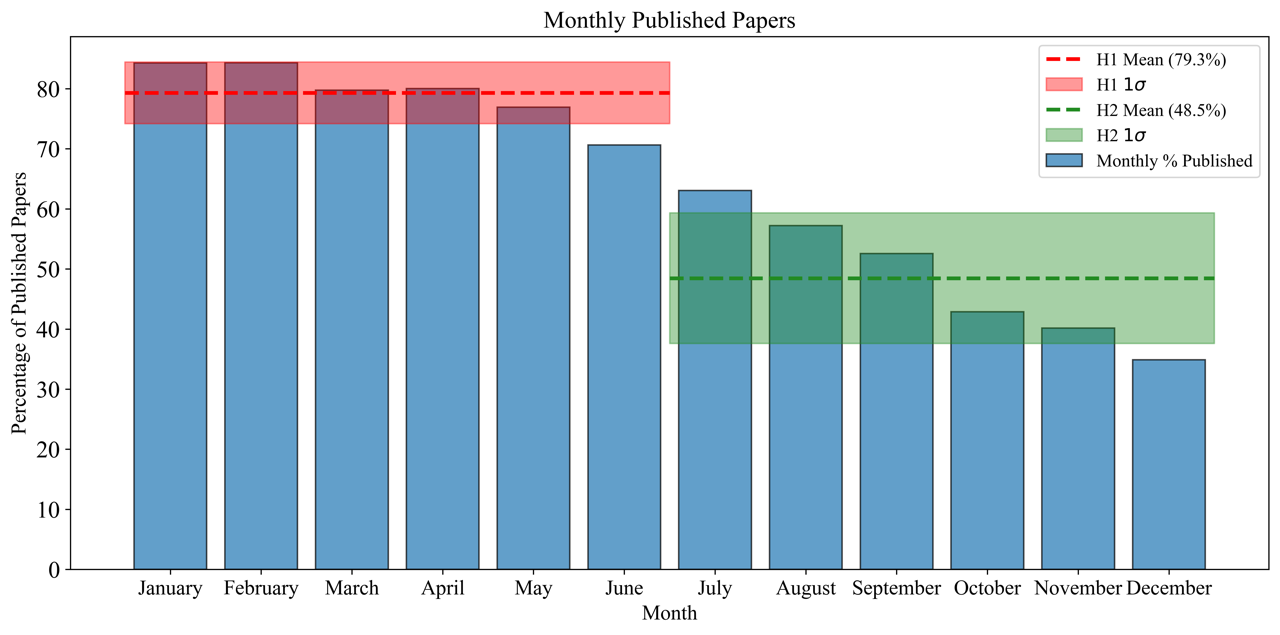}
    \caption{Monthly Track of the Percentage of Journal Published Papers. The two bins correspond to the two halves of the year with the dotted line as the mean and shaded regions as the 1$\sigma$ regions. The drop in the second half is not statistically significant and is due to long journal processing times.}
    \label{fig:journals_monthly_pub}
\end{figure}

This sub-section is dedicated to exploring any statistics that differ across journals. We report a pleasantly surprising statistic - 62\% of the papers released on the arXiv this year were published in a journal (as of December 2025). We estimate that there are about 608 arXiv papers from 2025 that have been submitted to journals and are currently in review. If we assume all papers that mention submission in the comments metadata will eventually be published in some journal, this increases the proportion of arXiv papers that end up being published to 65.73\%. This is almost a two-thirds majority, an amazing achievement for the field. It means that a majority of the research work that is being done in the field is good enough to be published in some journal. It implies a certain quality to the output that the field generates, and we should commend ourselves for the same. Figure \ref{fig:journals_monthly_pub} shows how the percentage of papers published per month varies over the year. In the first half of the year, we see that the values are fairly consistent. In fact, they are so consistent that both January and February have exactly the same percentage of published papers. We checked to make sure this was not a fault in our data and found no obvious inconsistencies. During the first six months, on average, 79\% of the arXiv papers were published in some journal. As we go to the end of the year, there is almost a step-wise decrease in the percentage of published papers. Seeing how consistent the values were in the first half, we feel confident in concluding this is because of the fact that papers from the later half of the year are yet to be published. Thus, in reality, the percentage of papers uploaded to the arXiv that get published in a journal is closer to 80\% and not the earlier reported value of 66\%  when we allow a sufficient window for journal processing.

For journal specific analysis, we identified journal names by string matching from a list of top journals, just as we have done with many of the other metrics. We have tried to identify and include as many journals as possible from the sample, but it is possible some have been missed. We have also included some regional journals that we saw appear numerous times in the dataset, such as `Publications of the Astronomical Society of Japan', `Science China Physics, Mechanics \& Astronomy', `Revista Mexicana de Astronomía y Astrofísica', etc. The list has been made as inclusive as possible. Figure \ref{fig:journals_pie} shows the top 15 journals in which arXiv papers were published this year. `Astronomy and Astrophysics' and `The Astrophysical Journal' top the list, followed by `Monthly Notices of the Royal Astronomical Society'. Nature and Nature Astronomy do not feature on the list, most likely due to the long article processing and reviewing time. It is very unlikely for an article uploaded to the arXiv this year to be published in Nature within the same year. Additionally some Nature papers might simply need be release as a preprint on the arXiv.

\begin{figure}
	\includegraphics[width=\columnwidth]{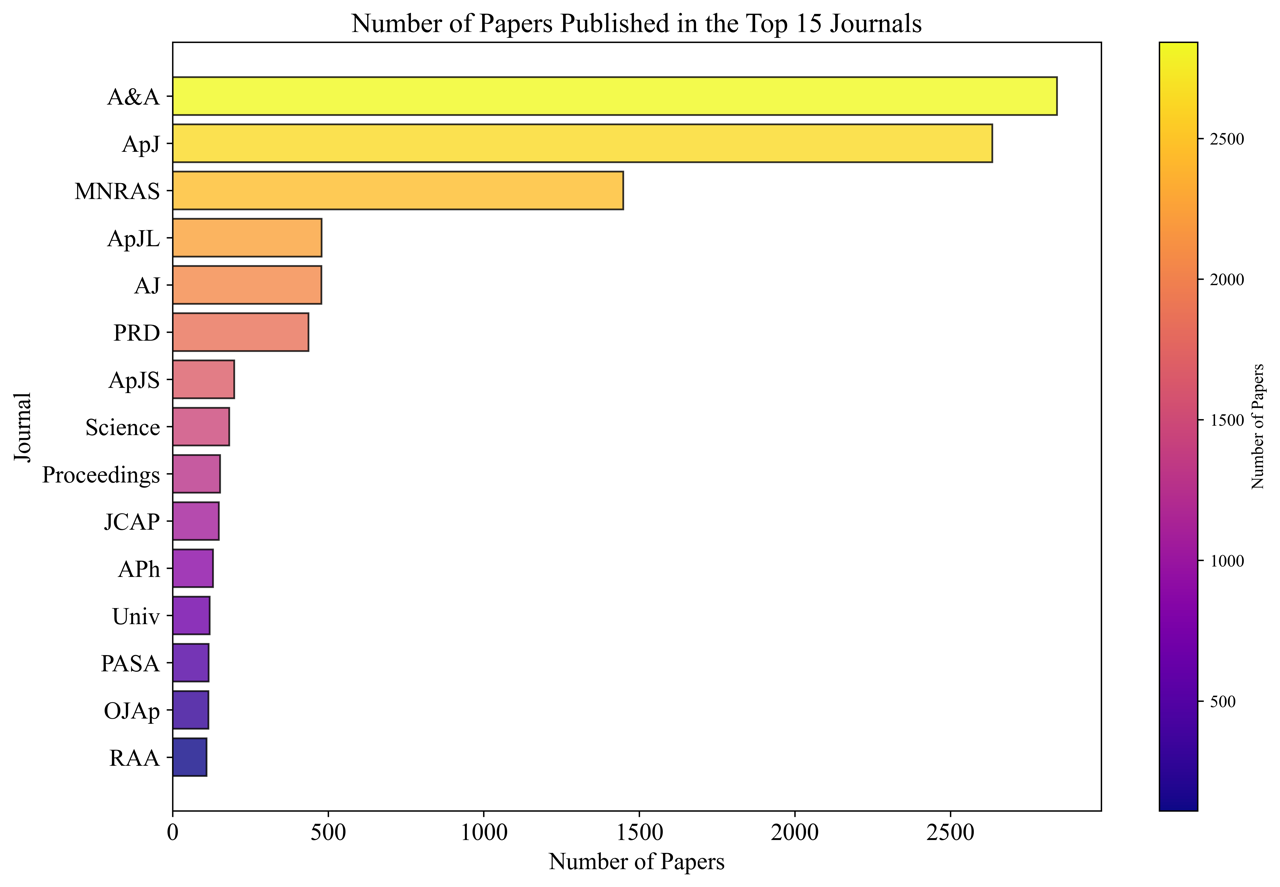}
    \caption{Top 15 Journals of the Year (by number of arXiv papers published)}
    \label{fig:top_15_journals}
\end{figure}

\begin{figure}
	\includegraphics[width=\columnwidth]{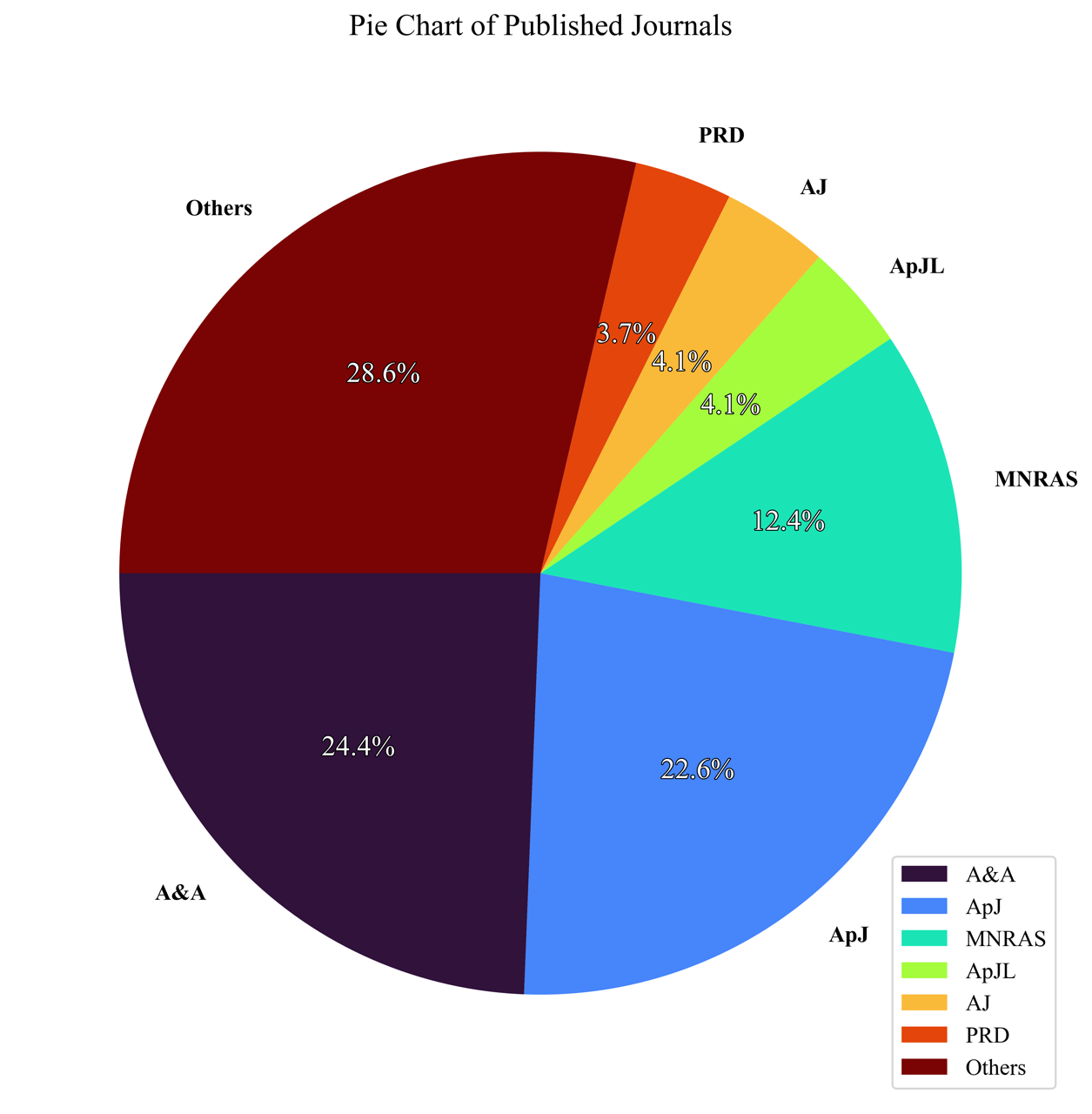}
    \caption{Pie Chart of the Top Journals into which arXiv articles were published}
    \label{fig:journals_pie}
\end{figure}

For the top 5 journals, we report some primary subject, citation and keyword statistics in Table \ref{tab:table_journals}. Since all the papers analysed here are journal papers, there is no difference between the AAC and JAC or the EAAC and the EJAC. Hence, we report the latter. We also conduced analysis of the title and abstract length across different journals and found both, for all journals to be within 1$\sigma$ of the overall mean. The differences in primary subjects and most used telescopes among journals are statistically significant. Conversely, for the case of keywords, the differences in the keywords are marginal and not statistically significant. \\

\begin{table*}
\resizebox{\textwidth}{!}{%
\begin{tabular}{llcccc}
\hline
\textbf{Journal}                                           & \textbf{Top Primary Subject}                    & \multicolumn{1}{l}{\textbf{JAC}} & \multicolumn{1}{l}{\textbf{EJAC}} & \multicolumn{1}{l}{\textbf{Top Keyword}} & \multicolumn{1}{l}{\textbf{Top Telescope}} \\
\hline
\hline
Astronomy and Astrophysics                        & Astrophysics of Galaxies               & 2.61                    & 1.82                     & accretion                       &  GAIA                                 \\
Astrophysical Journal                             & Astrophysics of Galaxies               & 2.18                    & 1.92                     & accretion                       &  JWST                                \\
Monthly Notices of the Royal Astronomical Society & Astrophysics of Galaxies               & 2.97                    & 2.03                     & accretion                       &  JWST                                 \\
Astrophysical Journal Letters                     & Earth and Planetary Astrophysics       & 4.58                    & 3.07                     & accretion                       &  JWST                                 \\
Astronomical Journal                              & Earth and Planetary Astrophysics       & 3.00                    & 1.26                     & exoplanets                      &  GAIA                                 \\
Others                                            & Cosmology and Nongalactic Astrophysics & 3.68                    & 2.76                     & exoplanets                      &  GAIA  
\\
\hline
\end{tabular}}
\caption{Statistics of the Top 5 Journals by arXiv Publication}
\label{tab:table_journals}
\end{table*}

\subsection{Pages, Tables and Figures} \label{sec:pages_tables_figures}
While this section might not be the most revealing, we chose to include it to give readers a better idea on the content in different papers and because these statistics might be interesting to some.

On average, the papers published this year had 18 pages, 2 tables and 9 figures. We can take this further and look at these same statistics through the lens of each paper's primary subject to understand if the papers of different subjects are written in a different manner. A simple KS test between the distribution of the pages, tables and figures of the entire sample of papers and the distribution of the respective primary subjects yields KS statistic values $\leq$ 0.1 with p-values = 0, showing us that there is no significant difference in how papers across sub-fields are written. 

We conducted a similar test with only the papers that were published in journals and found no significant difference between the distributions of pages, tables or figures and the overall distribution. Finally, we also conducted this test for papers across journals, specifically comparing the top three journals of the year - The Astrophysical Journal (ApJ), Astronomy and Astrophysics (A\&A) and Monthly Notices of the Royal Astronomical Society (MNRAS) and again found no significant difference between the content metric distributions of the arXiv papers published in these three journals. 

Given the citation information, we also checked for correlations with the number of pages, tables and figures and found no significant correlation even when self-citations were removed, indicating no preference for longer papers or papers with more tables or figures to be cited more.

All of this tells us that different sub-fields in the community might differ stylistically in terms of the papers they write; however, overall, we write the same. All of our papers are the same length and contain the same number of tables and figures. This result is somewhat synonymous with the result we got earlier in sub-section \ref{subsec:collab_indices}, that metrics do not differ by collaborations. We all seem to work in similar patterns and structures on very different aspects of the field. 

\begin{figure}
	\includegraphics[width=\columnwidth]{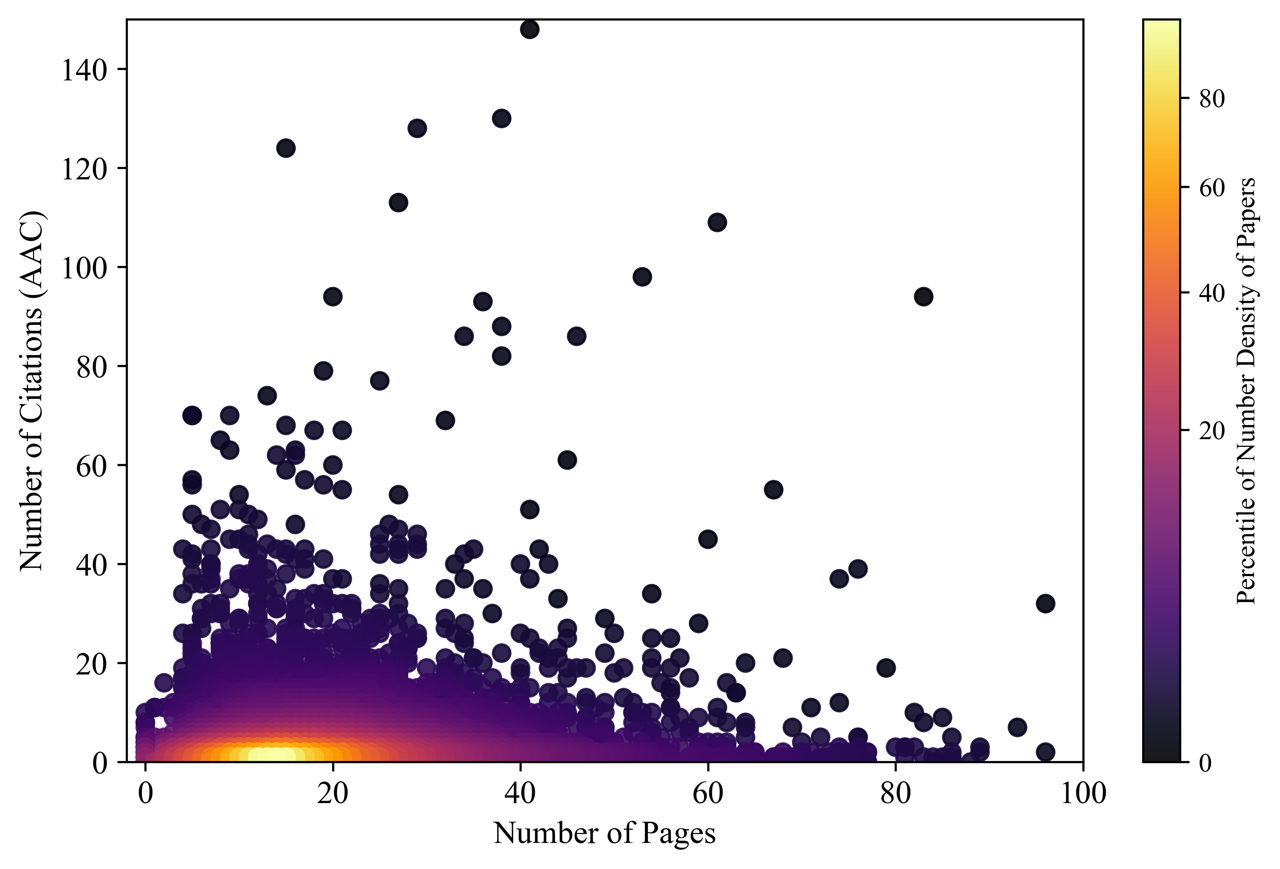}
    \caption{Citation distribution of the number of paper pages coloured by percentile of number density}
    \label{fig:pages}
\end{figure}

\begin{figure}
	\includegraphics[width=\columnwidth]{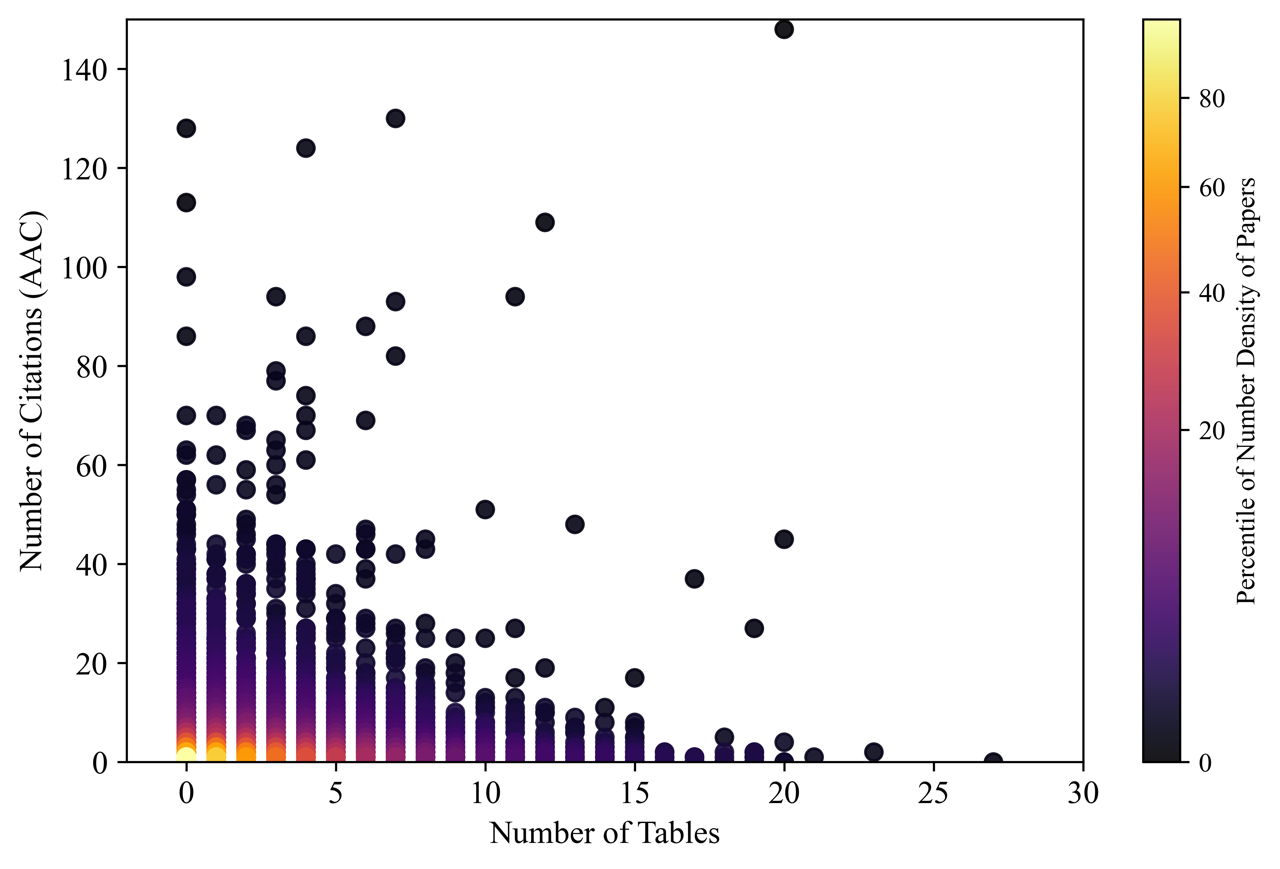}
    \caption{Citation distribution of the dumber of paper tables coloured by percentile of number density}
    \label{fig:tables}
\end{figure}

\begin{figure}
	\includegraphics[width=\columnwidth]{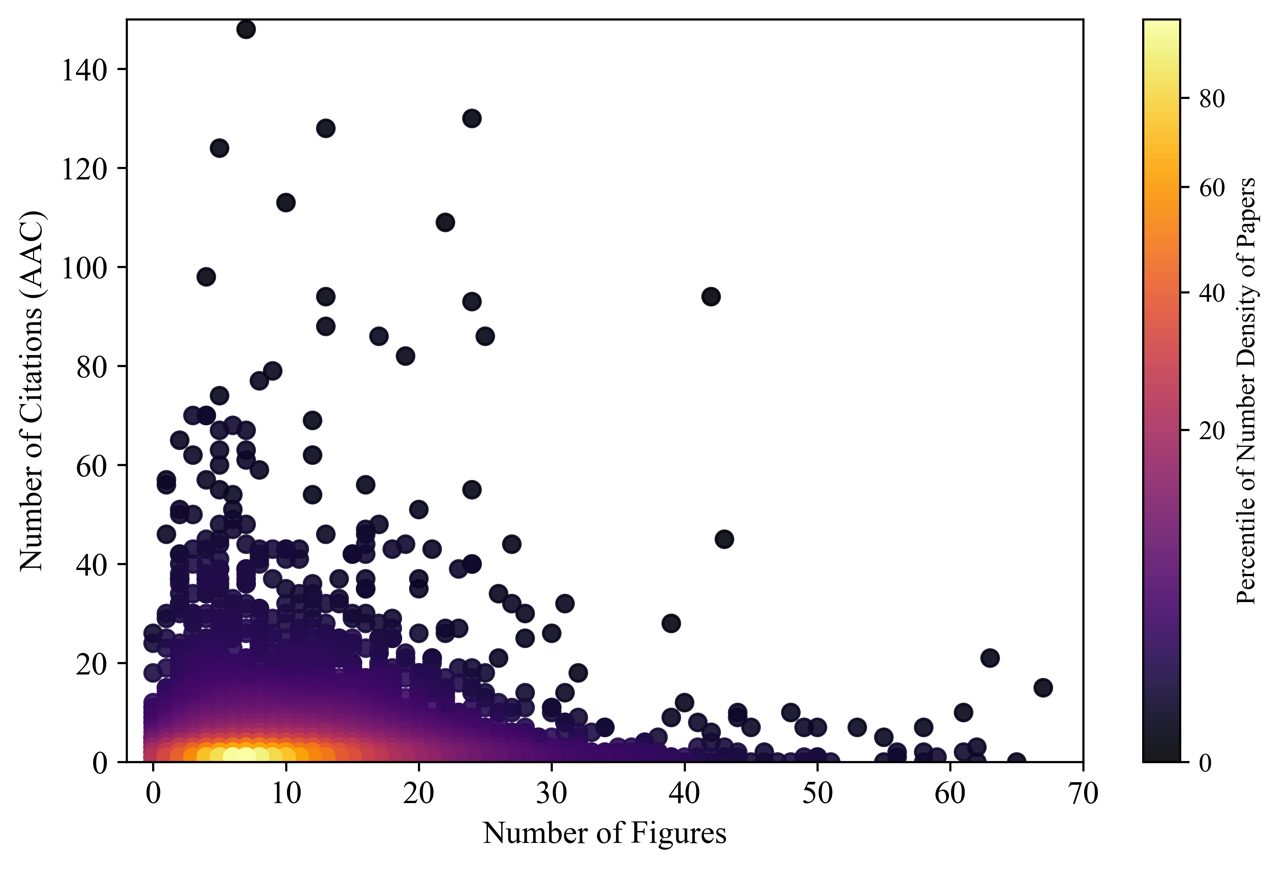}
    \caption{Citation distribution of the number of paper figures coloured by percentile of number density}
    \label{fig:figures}
\end{figure}

Figures \ref{fig:pages}, \ref{fig:tables}, and \ref{fig:figures} show the distributions of the pages, tables and figures for all papers in 2025 coloured by the local number density. We also have a few large outliers in our dataset for pages, tables and figures. However, we limit our plots to a reasonable range to prioritise visualising where the majority of papers lie. \\

\subsection{Titles and Abstracts} \label{sec:titles_and_abstract}
We include a small section on the specifics of titles and abstracts since much of the statistics from this section have already been included elsewhere. Two basic but nonetheless insightful metrics we report are the average length of titles at 13 words and the average length of the abstract at 211 words per paper. Across primary subjects, the word count for titles and abstracts is within 1$\sigma$ of the mean, demonstrating that there is no significant difference. 

We can also analyse if the papers associated with different telescopes tend to have longer or shorter titles or abstracts. While both the length of the title and abstract for all telescopes is within 1$\sigma$ of the mean of the overall sample, we would like to point out that papers associated with the Fermi telescope have both the longest titles and abstracts among all the telescopes. We also compare with the number of pages, tables and figures and find no strong correlation between any of these and the length of the title or abstract. 

\begin{figure}
	\includegraphics[width=\columnwidth]{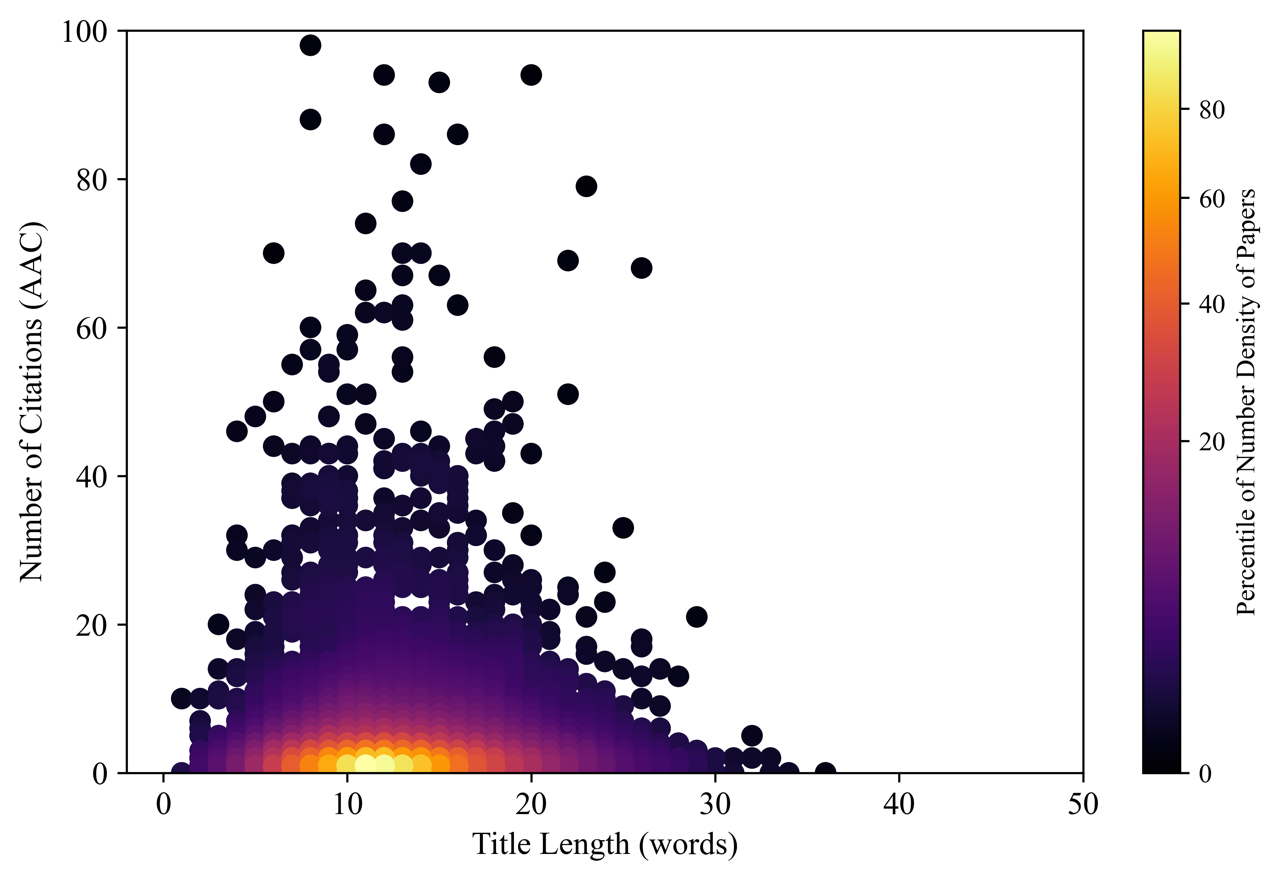}
    \caption{Citation distribution of the title length coloured by percentile of number density}
    \label{fig:title}
\end{figure}

\begin{figure}
	\includegraphics[width=\columnwidth]{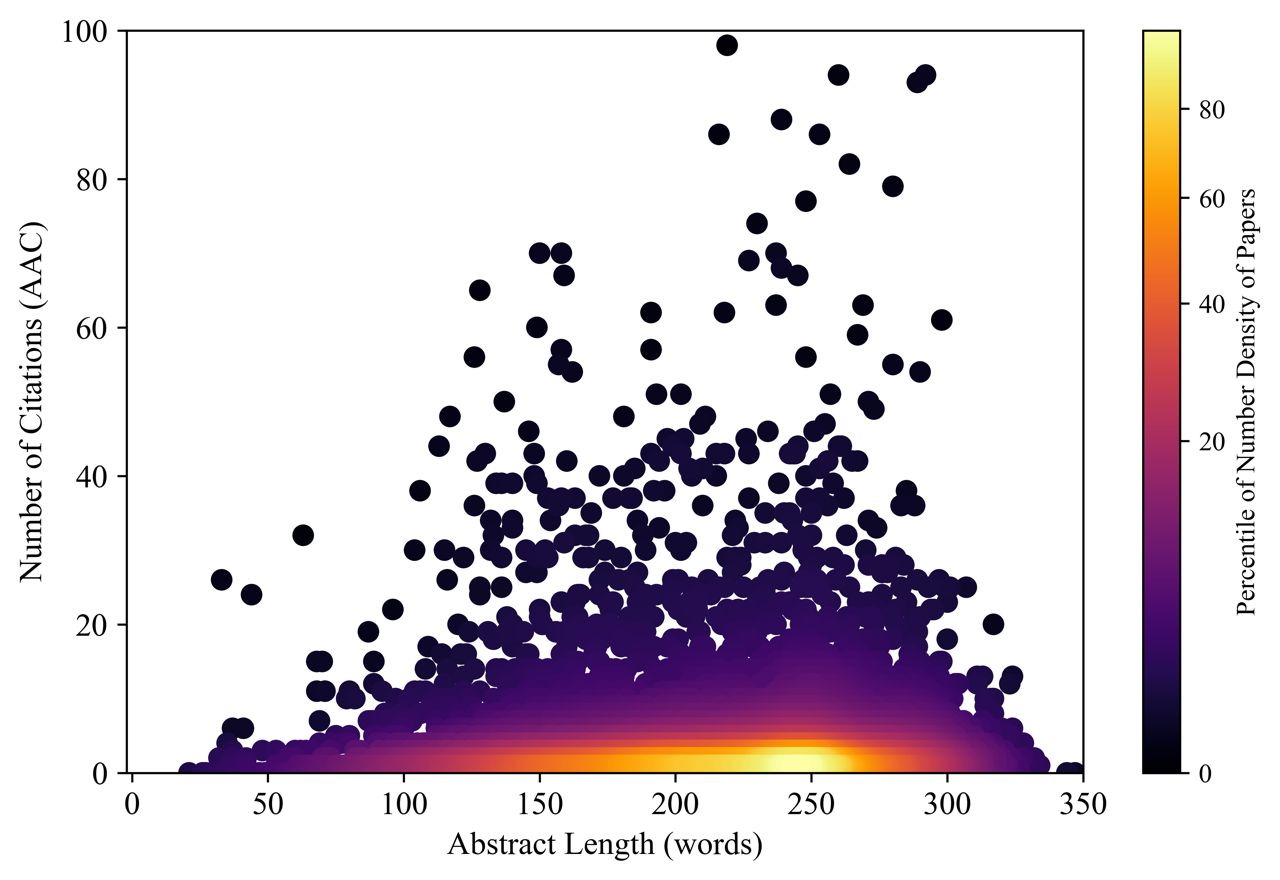}
    \caption{Citation distribution of the abstract length coloured by percentile of number density}
    \label{fig:abstract}
\end{figure}

Lastly, we look at if and how citations have any relation to the length of a title or abstract. Just as an example, we show the plot of citations versus title and abstract length coloured by the local density in Figures \ref{fig:title}, \ref{fig:abstract}. From the figure, there is no clear dependence of either of the length's on the number of citations. This does not change when switching to the other citation indices. Owing to how different journals have different rules for their titles and abstracts, this makes sense. Considering the nature of the data, we calculate the Spearman's rank correlation. We obtain values of $\rho = 0.051$ for title length and $0.121$ for abstract length (both calculated for AAC) with very small p-values ($p < 0.001$). This indicates a highly statistically significant but very weak positive monotonic relation between title and abstract length and the number of citations, allowing us to conclude that title and abstract length are not major determiners of the number of citations. Changing the citation index does not change the conclusion of this result. 

\section{Honours and Medals} \label{sec:honors}
In honour of the Spotify Wrapped (the inspiration for this paper) awarding listeners with listening achievements at the end of the year, we would like to hand out some unconventional awards to the community in general and certain papers and authors. This also serves as a place for us to report some fascinating statistics we picked up while conducting the analysis for this paper that could not fit in anywhere else.

\begin{itemize}
    \item Longest Paper: Statistical Machine Learning for Astronomy -- A Textbook
    \\
    Pages: 677
    \item Longest Paper (excluding books): The TESS Grand Unified Hot Jupiter Survey. III. Thirty More Giant Planets
    \\
    Pages: 96 
    \item Paper with the most Figures: Thermodynamic Origin of the Tully-Fisher Relation in Dark Matter Dominated Galaxies: A Theoretical-Empirical Derivation
    \\
    Figures: 609
    \item Paper with the most Tables: Thermodynamic Origin of the Tully-Fisher Relation in Dark Matter Dominated Galaxies: A Theoretical-Empirical Derivation
    \\
    Tables: 953
    \item Community's Favourite Redshift (Most quoted redshift-related phrase): z = 0
    \\ No of times: 270 (directly quoted in paper abstract/title)
    \item Community's Favourite Redshift Range (Interquartile Range): 0.76 $<$ z $<$ 6.00
    \item Most Published Author: J. Carretero
    \\
    Number of Papers Published: 138
    \item Most Cited Paper: DESI DR2 Results II: Measurements of Baryon Acoustic Oscillations and Cosmological Constraints
    \\
    Citation Count: 703
    \item Highest LCI Paper: Precise Measurement of Cosmic Ray Light and Helium Spectra above 0.1 Peta-electron-Volt
    \\
    LCI: 468
    \item Highest GCI Paper: GWTC-4.0: Population Properties of Merging Compact Binaries
    \\
    GCI: 2133
    \item Highest NGCI Paper: The CosmoVerse White Paper: Addressing observational tensions in cosmology with systematics and fundamental physics
    \\
    NGCI: 57

\end{itemize}

\section{Discussion} \label{sec:conclusions}

We have presented the most detailed and holistic overview to date for all astrophysics research over the past year in the form of papers uploaded to the arXiv in 2025. We have explored these papers through a series of different parameters and metrics to find and highlight both number statistics as well as trends and patterns in relation to the actual content in these papers.

Overall, the yearly increase in the number of research papers submitted to the arXiv is an indication that interest in astrophysics is growing. The community is producing papers at a rate of 52 articles a day. An astonishing 80\% of the papers uploaded to the arXiv end up being published in a journal. These articles cover a diverse range of subjects and utilise a variety of telescopes. We all work in our small research groups or large collaborations on both the hot topics and niche, ultra-specialised areas of the field to produce papers that are journal-worthy most of the time. We plan to include analysis across multiple years once we have sufficient data, from past years as well as the current year. This will provide a deeper insight into any trends or patterns amongst the various metrics, and enable predictions to be made for the future.

We believe the results in this paper can be useful for predicting future trends and projecting statistics, which could help better inform both students and professionals in their career planning as well as broader institutions and collaboration teams on expectations for their respective sub-fields and areas of interest. It will also provide a foundational basis for macro-decisions on the future of astrophysics globally, such as informing the level of grant-funding allocated to astrophysics research at the government and university level, the proposal and construction of new telescopes and observatories, and most importantly, informing critically, the direction of decadal surveys like those carried out by NASA -- to directly quote, ``NASA and its partners ask the National Research Council once each decade to look out 10 or more years into the future and prioritize research areas, observations, and notional missions to make those observations"\footnote{\url{https://science.nasa.gov/earth-science/decadal-surveys/}}.

\noindent Our hope for this paper is that it shines light on the different, and perhaps underappreciated, aspects of the field and community, which in turn will promote healthy collaboration between all scientists across the world and open new avenues of research. It has been an amazing year for Astrophysics, and we look forward to another wonderful, ground-breaking year.

\section*{Acknowledgments}
 We would like to thank Paul Woods and Josè Maria Diego for useful discussion, comments and feedback that greatly improved the quality of this paper. R.F.L. and A.A. would like to thank Jeremy Lim for supporting us at the University of Hong Kong and allowing us to take up interesting side-projects like this. We also thank Sung Kei Li and James Nianias for their insightful comments on the overall presentation of this paper. R.F.L. would like to thank Siddhant Dutta, Priyansh Shah and Yuvraj Bhandari for useful suggestions.
 
 We especially thank all the members of the community who reached out to us via email with extremely positive comments, feedback, suggestions and corrections. 
 
 We express our appreciation for the team at arXiv (\url{https://arxiv.org}) that made this paper possible in their continuous management and maintenance of the arXiv, which was the source for all our data. This work used the NASA ADS API, which in turn uses the Astrophysics Data System, funded by NASA under Cooperative Agreement 80NSSC25M7105. This work was not supported in any way by a specific organization and was simply a project taken up independently.\\

 We would like to stress that this paper is not meant to incite unhealthy competition among the community, nor is it meant to be a cheat-sheet to maximising metrics. Instead, we hope that this work reveals the many interesting statistics we would all like to know and helps us gauge what we as a community should be working towards in the long run. \\

We also include the credit for all images used in Figure \ref{fig:telescopes}:
\begin{itemize}
    \item JWST Image Credit: ESA
    \item ALMA Logo Credit: ESO
    \item TESS Image Credit: NASA
    \item SDSS Image Credit: SDSS
    \item HST Image Credit: NASA
    \item Fermi Image Credit: NASA
    \item Euclid Image Credit: ESA
    \item DESI Image Credit: NOIRLab
    \item Gaia Logo Credit: ESA
    \item LSST Image Design Credit: LSST Project Office
    
\end{itemize}

The maps in this paper were made using two Python packages- Folium which can be found at \url{https://python-visualization.github.io/folium/latest/}, and basemap which can be found at \url{https://matplotlib.org/basemap/stable/}. \\

Credit for Figure 1:
Leaflet: \url{https://leafletjs.com}, OpenStreetMap contributors: \url{https://www.openstreetmap.org/copyright}, CARTO: \url{https://carto.com/attribution}

\bibliographystyle{apsrev4-1}

\bibliography{oja_template}

@ARTICLE{trimble2006,
       author = {{Trimble}, Virginia and {Aschwanden}, Markus J. and {Hansen}, Carl J.},
        title = "{Astrophysics in 2006}",
      journal = {\ssr},
         year = 2007,
        month = oct,
       volume = {132},
       number = {1},
        pages = {1-182},
          doi = {10.1007/s11214-007-9224-0},
archivePrefix = {arXiv},
       eprint = {0705.1730},
 primaryClass = {astro-ph},
       adsurl = {https://ui.adsabs.harvard.edu/abs/2007SSRv..132....1T}
}

@ARTICLE{trimble2005,
       author = {{Trimble}, Virginia and {Aschwanden}, Markus J. and {Hansen}, Carl J.},
        title = "{Astrophysics in 2005}",
      journal = {\pasp},
         year = 2006,
        month = jul,
       volume = {118},
       number = {845},
        pages = {947-1047},
          doi = {10.1086/506157},
archivePrefix = {arXiv},
       eprint = {astro-ph/0606663},
 primaryClass = {astro-ph},
       adsurl = {https://ui.adsabs.harvard.edu/abs/2006PASP..118..947T}
}

@article{trimble2004,
author = {Trimble, Virginia and Aschwanden, Markus},
year = {2005},
month = {04},
pages = {311-394},
title = {Astrophysics in 2004},
volume = {117},
journal = {Publications of The Astronomical Society of The Pacific - PUBL ASTRON SOC PAC},
doi = {10.1086/429117}
}

@ARTICLE{trimble2003,
       author = {{Trimble}, Virginia and {Aschwanden}, Markus J.},
        title = "{Astrophysics in 2003}",
      journal = {\pasp},
         year = 2004,
        month = mar,
       volume = {116},
       number = {817},
        pages = {187-265},
          doi = {10.1086/383241},
       adsurl = {https://ui.adsabs.harvard.edu/abs/2004PASP..116..187T}
}

@article{trimble2002,
doi = {10.1086/374651},
url = {https://doi.org/10.1086/374651},
year = {2003},
month = {may},
publisher = {The University of Chicago Press},
volume = {115},
number = {807},
pages = {514},
author = {Trimble, Virginia and Aschwanden, Markus J.},
title = {Astrophysics in 2002},
journal = {Publications of the Astronomical Society of the Pacific}
}

@article{trimble2001,
doi = {10.1086/341673},
url = {https://doi.org/10.1086/341673},
year = {2002},
month = {may},
publisher = {The University of Chicago Press},
volume = {114},
number = {795},
pages = {475},
author = {Trimble, Virginia and Aschwanden, Markus J.},
title = {Astrophysics in 2001},
journal = {Publications of the Astronomical Society of the Pacific}
}

@article{trimble2000,
 ISSN = {00046280, 15383873},
 URL = {http://www.jstor.org/stable/10.1086/322844},
 author = {Virginia Trimble and Markus J. Aschwanden},
 journal = {Publications of the Astronomical Society of the Pacific},
 number = {787},
 pages = {1025--1114},
 publisher = {[The University of Chicago Press, Astronomical Society of the Pacific]},
 title = {Astrophysics in 2000},
 urldate = {2026-06-08},
 volume = {113},
 year = {2001}
}

@article{trimble1999,
doi = {10.1086/316546},
url = {https://doi.org/10.1086/316546},
year = {2000},
month = {apr},
publisher = {The University of Chicago Press},
volume = {112},
number = {770},
pages = {434},
author = {Trimble, Virginia and Aschwanden, Markus J.},
title = {Astrophysics in 1999},
journal = {Publications of the Astronomical Society of the Pacific}
}

@ARTICLE{trimble1998,
       author = {{Trimble}, Virginia and {Aschwanden}, Markus},
        title = "{Astrophysics in 1998}",
      journal = {\pasp},
         year = 1999,
        month = apr,
       volume = {111},
       number = {758},
        pages = {385-437},
          doi = {10.1086/316342},
       adsurl = {https://ui.adsabs.harvard.edu/abs/1999PASP..111..385T}
}

@article{trimble1997,
doi = {10.1086/316143},
url = {https://doi.org/10.1086/316143},
year = {1998},
month = {mar},
publisher = {The University of Chicago Press},
volume = {110},
number = {745},
pages = {223},
author = {Trimble, Virginia and McFadden, Lucy‐Ann},
title = {Astrophysics in 1997},
journal = {Publications of the Astronomical Society of the Pacific}
}

@article{trimble1996,
doi = {10.1086/133865},
url = {https://doi.org/10.1086/133865},
year = {1997},
month = {feb},
publisher = {The Astronomical Society of the Pacific},
volume = {109},
number = {732},
pages = {78},
author = {Trimble, Virginia and McFadden, Lucy Ann},
title = {ASTROPHYSICS IN 1996},
journal = {Publications of the Astronomical Society of the Pacific}
}

@misc{Coles_2025a, title={Like a million pounds: Published by The Open Journal of Astrophysics}, url={https://astro.theoj.org/post/3602-like-a-million-pounds}, journal={The Open Journal of Astrophysics}, publisher={Open Journal of Astrophysics}, author={Coles, Peter}, year={2025}, month={Dec}}

\begin{appendix} \label{appendix}

Here we present some plots that we found interesting and might be interesting to some of the readers. We also have the table of the top keyword and top sub-field for every country or region in our dataset at the very end.

\begin{figure}[H]
    \centering
	\includegraphics[width=0.7\linewidth]{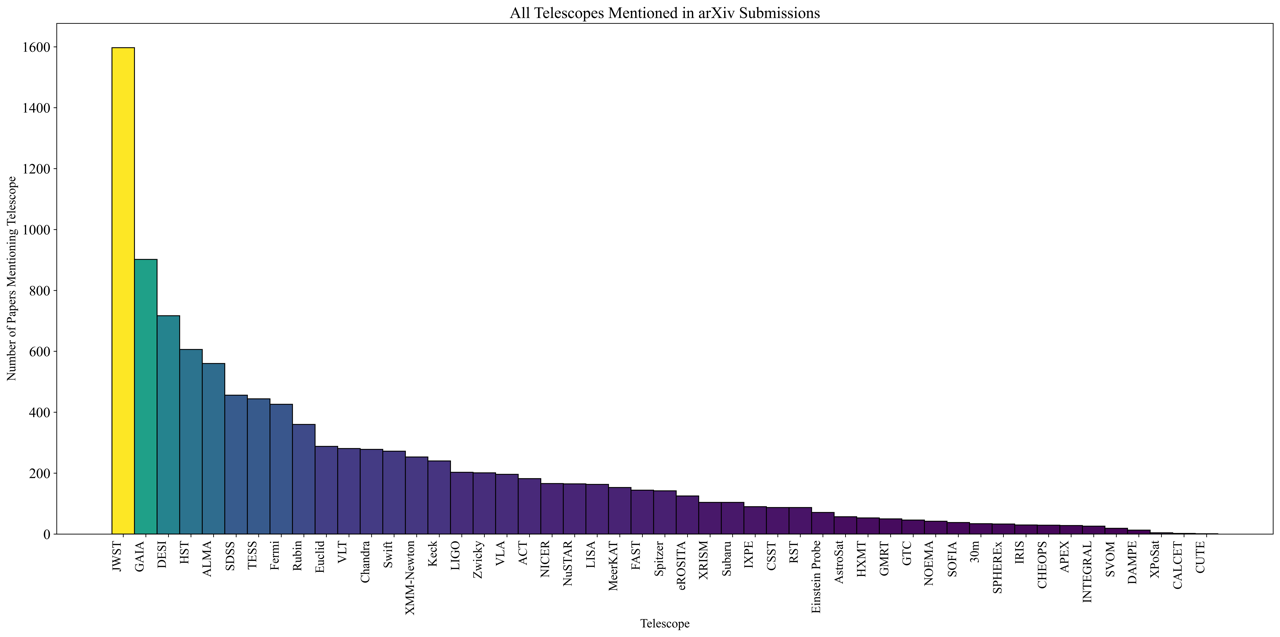}
    \caption{Distribution of the number of papers mentioning a telescope for all telescopes in our predefined search list for 2025}
    \label{fig:appen_telescopes}
\end{figure}

\begin{figure}[H]
    \vspace*{-0.5cm}
    \centering
	\includegraphics[width=0.9\columnwidth]{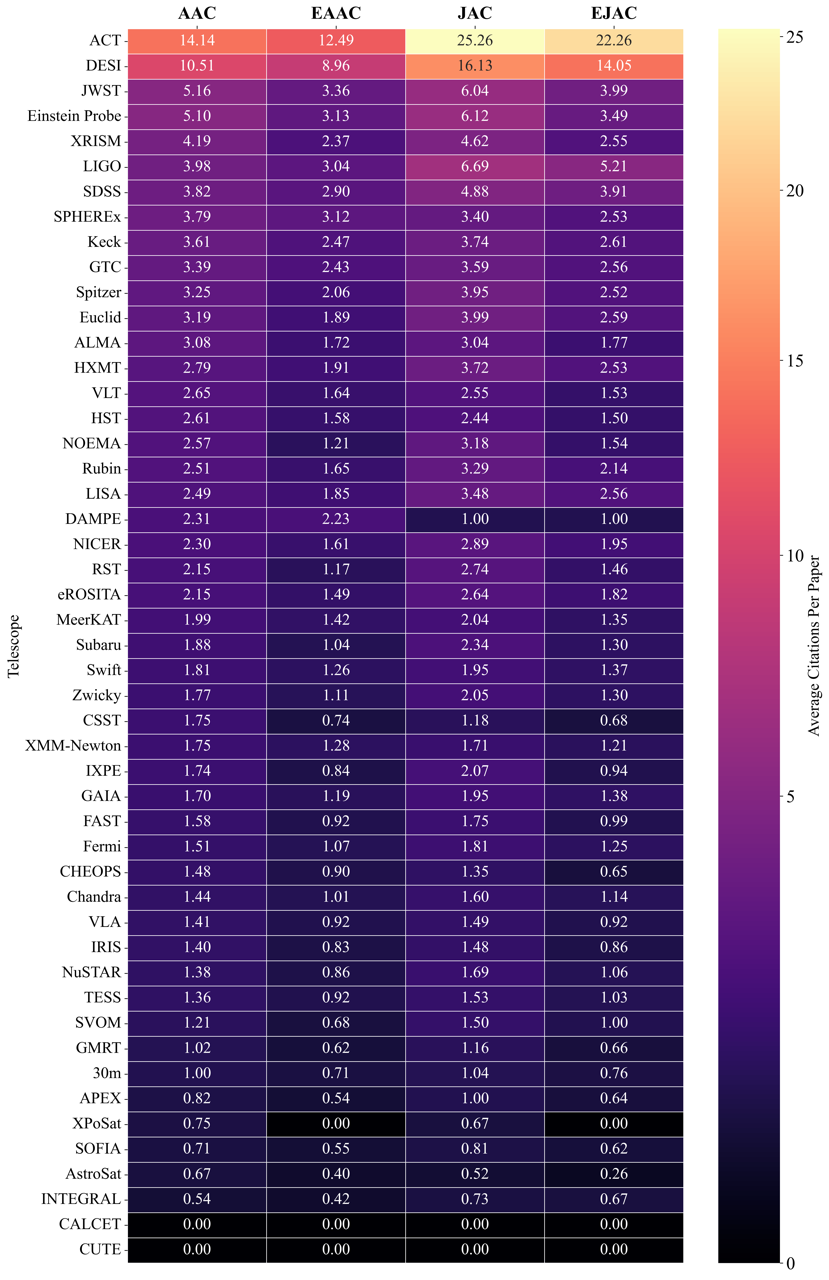}
    \caption{Heat-map of Citation Metrics for All Telescopes}
    \label{fig:telescope_heatmap_big}
\end{figure}

\begin{figure}[H]
    \centering

    \begin{subfigure}{0.48\columnwidth}
        \centering
        \includegraphics[width=\linewidth]{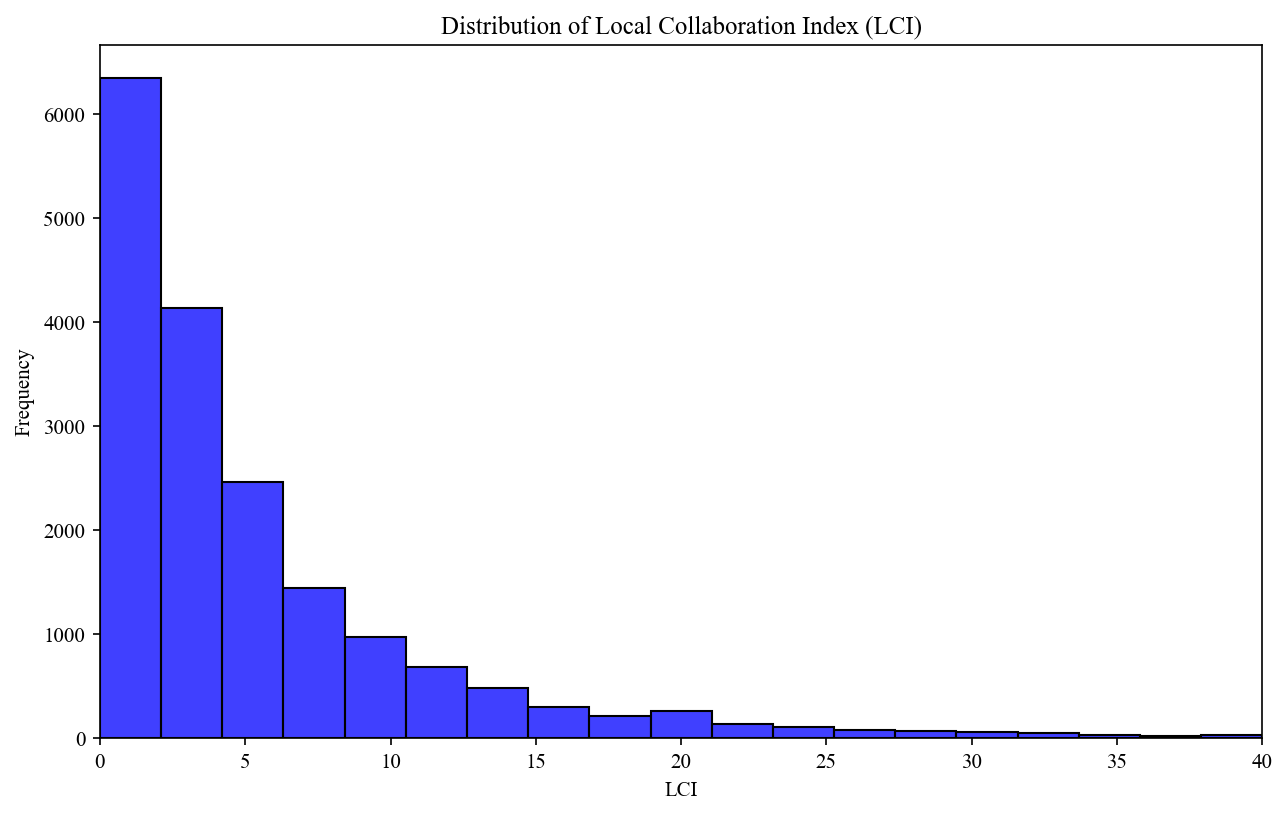}
        \caption{Histogram of LCI}
        \label{fig:appen_1}
    \end{subfigure}
    \hfill
    \begin{subfigure}{0.48\columnwidth}
        \centering
        \includegraphics[width=\linewidth]{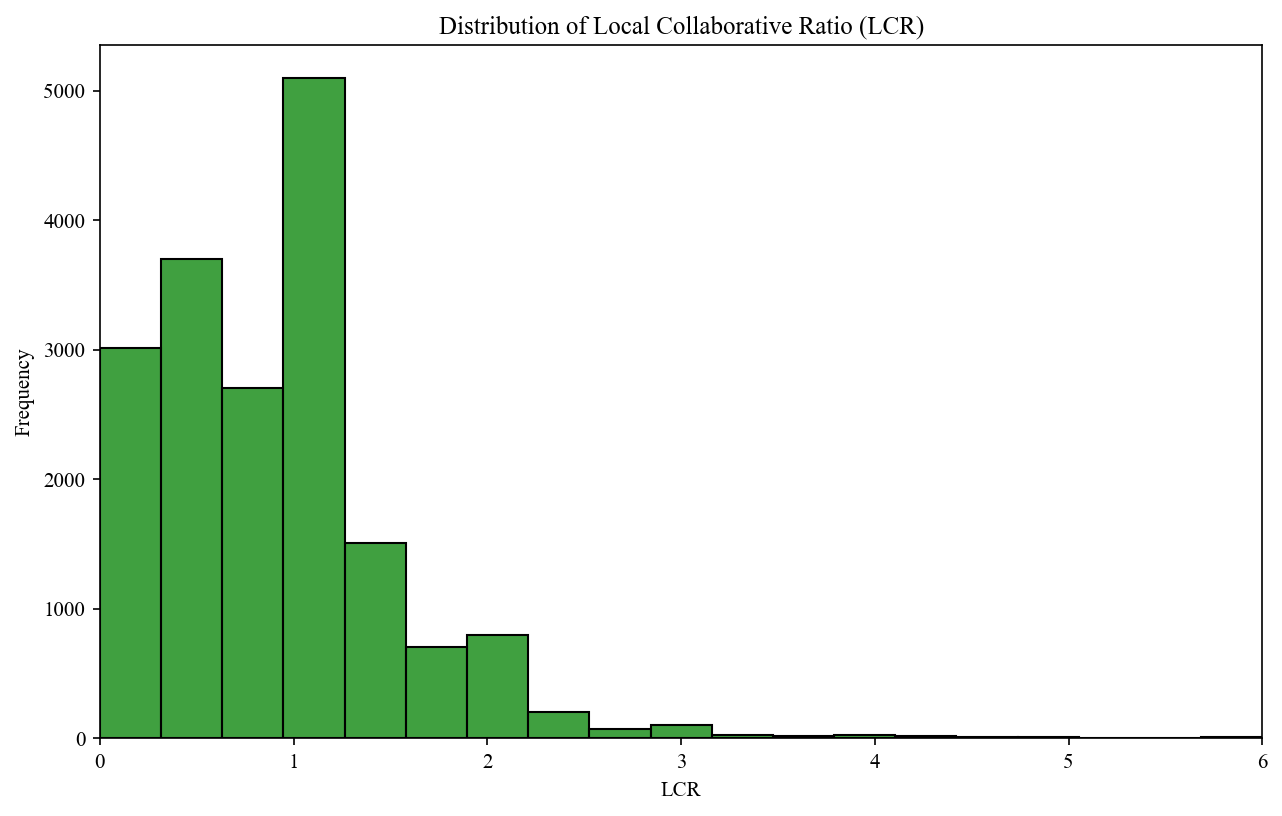}
        \caption{Histogram of LCR}
        \label{fig:appen_2}
    \end{subfigure}

    \vspace{0.5em}

    \begin{subfigure}{0.48\columnwidth}
        \centering
        \includegraphics[width=\linewidth]{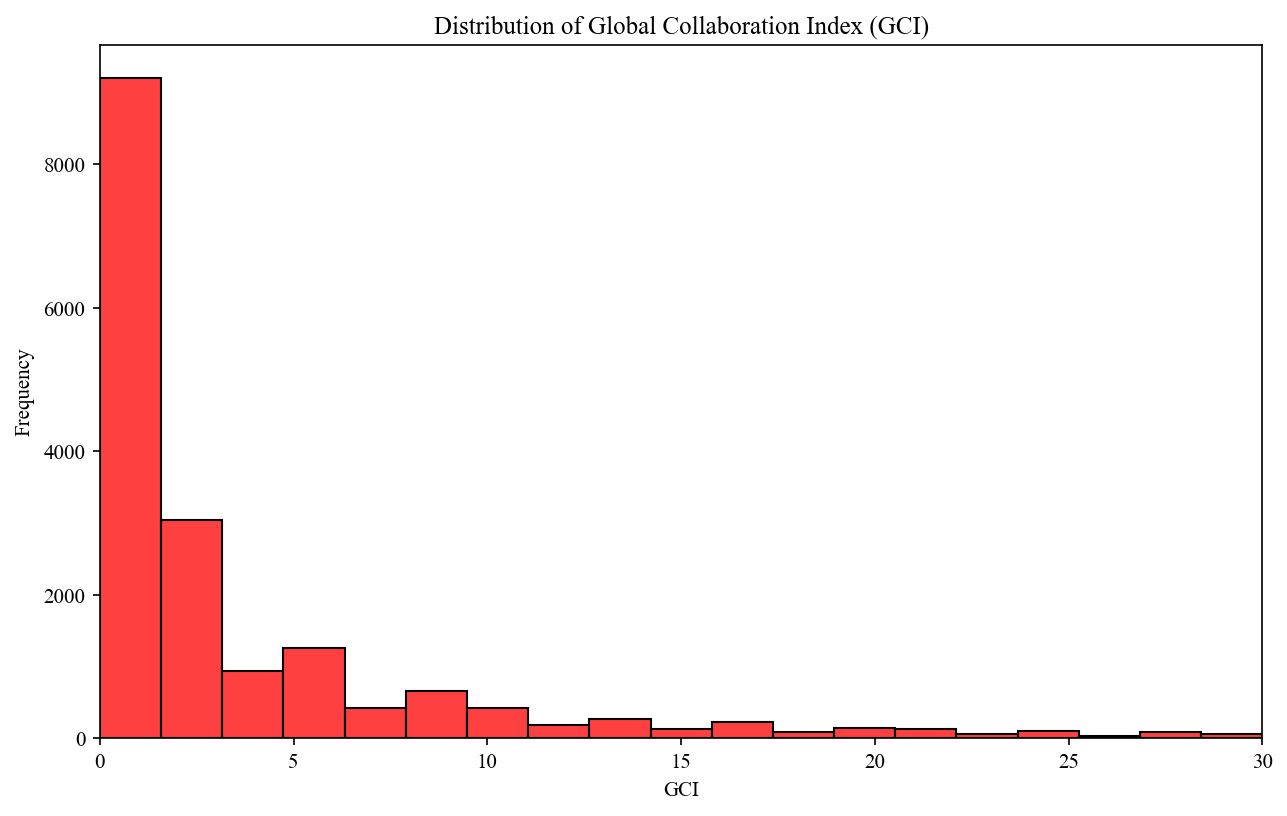}
        \caption{Histogram of GCI}
        \label{fig:appen_3}
    \end{subfigure}
    \hfill
    \begin{subfigure}{0.48\columnwidth}
        \centering
        \includegraphics[width=\linewidth]{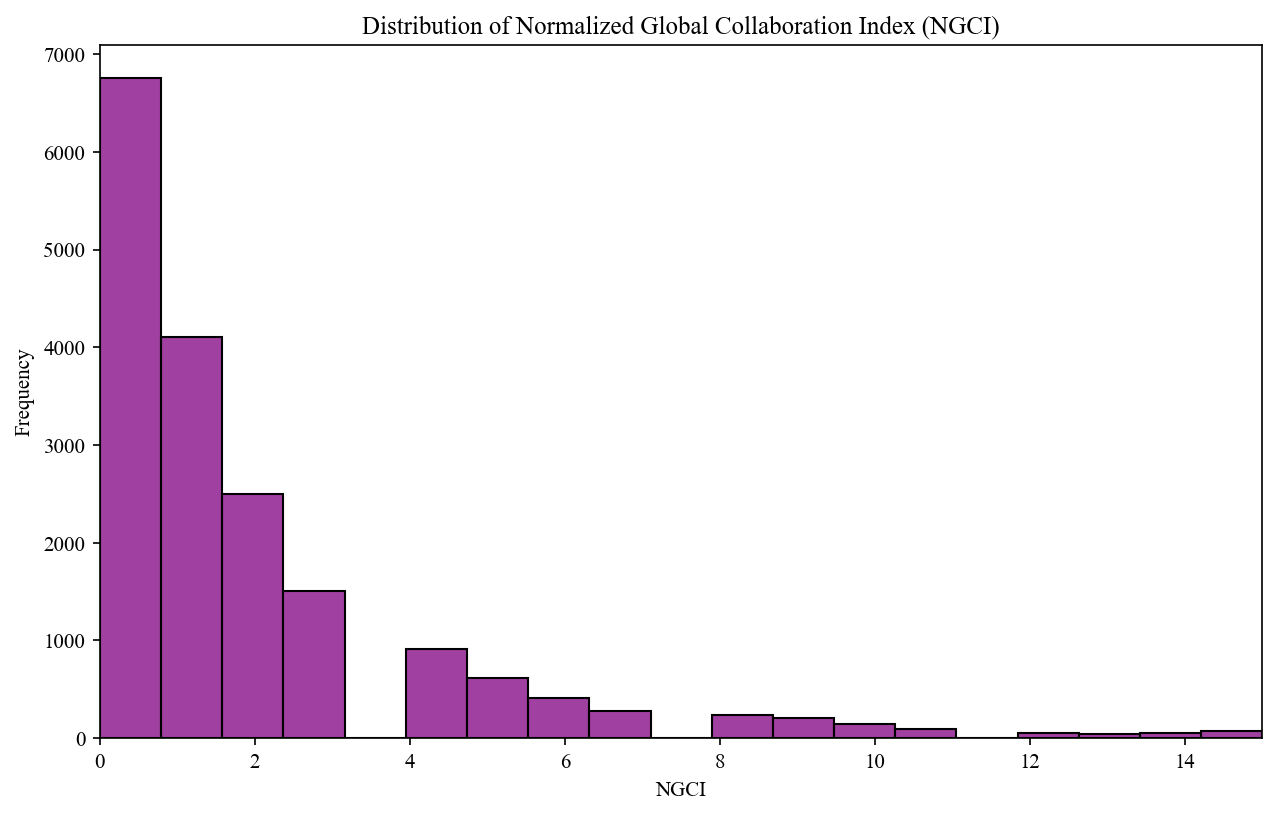}
        \caption{Histogram of NGCI}
        \label{fig:appen_4}
    \end{subfigure}

    \caption{Histograms for the collaboration indicies}
    \label{fig:appendix_histograms}
\end{figure}

\begin{table}[htbp]
\centering
\caption{Country or Region Statistics: Top Subfield and Keyword (Part 1)}
\label{tab:country_stats_1}
\begin{tabular}{lll}
\toprule
\textbf{Country or Region} & \textbf{Top Subfield} & \textbf{Top Keyword} \\
\midrule
United States & Simulations & Accretion \\
Germany & Simulations & Accretion \\
United Kingdom & Simulations & Accretion \\
Italy & Star Formation & Accretion \\
China & Simulations & Active Galactic Nuclei \\
France & Simulations & Accretion \\
Spain & Star Formation & Accretion \\
Japan & Simulations & Protoplanetary Disks \\
Canada & Star Formation & Galaxy Evolution \\
Australia & Star Formation & Galaxy Evolution \\
Netherlands & Simulations & Accretion \\
Chile & Star Formation & Protoplanetary Disks \\
Switzerland & Simulations & High-redshift Galaxies \\
Denmark & Star Formation & Galaxy Evolution \\
Sweden & Star Formation & Reionization \\
Belgium & Simulations & Dust \\
India & Simulations & Accretion \\
South Korea & Star Formation & Star Formation \\
Taiwan & Star Formation & Protoplanetary Disks \\
Austria & Star Formation & High-redshift Galaxies \\
Israel & Supernova & Reionization \\
Poland & Simulations & Accretion \\
Argentina & Cosmic Rays & Accretion \\
Brazil & Simulations & Cosmology \\
Czech Republic & Cosmic Rays & Accretion \\
Mexico & Star Formation & Accretion \\
Ireland & Supernova & Supernovae \\
Norway & Simulations & Cosmic Microwave Background \\
South Africa & Star Formation & Accretion \\
Colombia & Dark Energy & Spectroscopy \\
Finland & Simulations & Accretion \\
Russia & Pulsars & Accretion \\
Greece & Active Galactic Nuclei & Accretion \\
Portugal & Exoplanets & Accretion \\
Romania & Cosmic Rays & Minor Planets \\
Slovenia & Cosmic Rays & Galaxies High-redshift \\
Hungary & Supernova & Nucleosynthesis \\
Georgia & Simulations & Radio pulsars \\
Thailand & Supernova & Supernovae \\
Hong Kong & Dark Matter & Black Holes \\
\bottomrule
\end{tabular}
\end{table}

\begin{table}[htbp]
\centering
\caption{Country or Region Statistics: Top Subfield and Keyword (Part 2)}
\label{tab:country_stats_2}
\begin{tabular}{lll}
\toprule
\textbf{Country or Region} & \textbf{Top Subfield} & \textbf{Top Keyword} \\
\midrule
Iran & Star Formation & Machine Learning \\
Iceland & Gamma-Ray Bursts & Gamma-ray bursts \\
United Arab Emirates & Simulations & Accretion \\
Cyprus & Supernova & Supernovae General \\
Jersey & Simulations & Supernovae \\
New Zealand & Simulations & Gravitational Lensing Micro \\
Ukraine & Simulations & Solar Wind \\
Mayotte & Cosmic Rays & N/A \\
Croatia & Dark Matter & Exoplanets \\
Malta & Star Formation & High-redshift Galaxies \\
Serbia & Active Galactic Nuclei & Active Galactic Nuclei \\
Ecuador & Star Formation & Galaxies \\
Ethiopia & Star Formation & Pulsars General \\
Jordan & Inflation & Exoplanets \\
Kazakhstan & Star Formation & Dark Energy \\
Monaco & Star Formation & Surveys \\
Nigeria & Star Formation & Sun Solar Cycle \\
Oman & Dark Matter & N/A \\
Turkey & Pulsars & Open Clusters \\
Bangladesh & Strong Lensing & N/A \\
Botswana & Star Formation & N/A \\
Bulgaria & Dark Matter & Acceleration of Particles \\
Estonia & Simulations & N/A \\
Nepal & Cosmic Rays & Radio Detection \\
Puerto Rico & Pulsars & Millisecond Pulsars \\
Uzbekistan & Black Holes & Dark Energy \\
Bahamas & Simulations & Cosmology \\
Burkina Faso & Neutron Stars & Gravitational Wave (GWs) \\
Greenland & Simulations & Integral-field Spectroscopy \\
Kuwait & Black Holes & Dark Energy \\
Lebanon & Simulations & Gravitational Lensing Strong \\
Macau & Cosmic Microwave Background & X-ray Transient Sources \\
Uganda & Star Formation & N/A \\
Vietnam & Planet Formation & N/A \\
Antarctica & Simulations & Galaxies \\
Armenia & Supernova & Radiation Mechanisms Non-thermal \\
Costa Rica & Supernova & N/A \\
Egypt & Galaxy Clusters & N/A \\
Latvia & Fast Radio Bursts & N/A \\
Lithuania & Neutron Stars & N/A \\
\bottomrule
\end{tabular}
\end{table}

\begin{table}[htbp]
\centering
\caption{Country or Region Statistics: Top Subfield and Keyword (Part 3)}
\label{tab:country_stats_3}
\begin{tabular}{lll}
\toprule
\textbf{Country or Region} & \textbf{Top Subfield} & \textbf{Top Keyword} \\
\midrule
Malaysia & Active Galactic Nuclei & Accretion \\
Montenegro & Galaxy Clusters & N/A \\
Morocco & Dark Energy & Dark Energy \\
Namibia & Dark Matter & Stars Individual $\eta$Carinae \\
Pakistan & Simulations & pn-QRPA \\
Tanzania & Fast Radio Bursts & N/A \\
Azerbaijan & Simulations & Herbig Ae/Be \\
Cuba & Neutron Stars & Bose-Einstein Condensate Stars \\
Indonesia & Dark Matter & N/A \\
Mali & Neutron Stars & Gravitational Wave (GWs) \\
Palau & Simulations & N/A \\
Peru & Dark Matter & Gravitational Waves (GWs) \\
Saudi Arabia & Star Formation & N/A \\
Slovakia & Dark Matter & Meteors \\
Sudan & Gamma-Ray Bursts & Arizona \\
Tunisia & Interstellar Medium & N/A \\
Albania & Dark Matter & N/A \\
Algeria & Cosmic Rays & N/A \\
Bolivia & Stellar Streams & N/A \\
Bosnia and Herzegovina & Dark Matter & N/A \\
Cameroon & Astrochemistry & N/A \\
Chad & Neutron Stars & Nanofabrication \\
El Salvador & Cosmic Microwave Background & N/A \\
Faroe Islands & Black Holes & N/A \\
Ghana & Large-Scale Structure & Singular Value Decomposition \\
Honduras & Black Holes & N/A \\
Iraq & Galaxy Evolution & N/A \\
Kenya & Supernova & N/A \\
Luxembourg & Stellar Evolution & N/A \\
Madagascar & Active Galactic Nuclei & N/A \\
Mauritius & Early Universe & N/A \\
Montserrat & Star Formation & N/A \\
Nicaragua & Star Formation & Dwarf Galaxies \\
Philippines & Exoplanets & N/A \\
Qatar & Exoplanets & Techniques Photometric \\
Rwanda & Star Formation & N/A \\
Singapore & Simulations & Planets and Satellites Dynamical \\
Syria & Cosmic Microwave Background & N/A \\
Uruguay & Quasars & Resonances \\
Vatican City & Pulsars & N/A \\
Venezuela & Hubble Tension & ISM Clouds \\
Zambia & Pulsars & N/A \\
Guinea & N/A & N/A \\
Mongolia & N/A & Near-Earth Asteroids \\
Paraguay & N/A & MAS-CCD \\
Sri Lanka & N/A & N/A \\
\bottomrule
\end{tabular}
\end{table}

\end{appendix}

\end{document}